\documentclass[superscriptaddress,pre,aps,twocolumn,showpacs,showkeys]{revtex4-1}  

\usepackage{relsize}

 \usepackage{etoolbox}
\usepackage{lmodern}
\usepackage{footnote}
\usepackage{colordvi}
\usepackage{array}
 \usepackage{graphicx}

\usepackage{dcolumn}
\usepackage{longtable}
 \usepackage{epsfig}
\usepackage{eqnarray}

\usepackage{amsmath,amssymb,amsfonts, gensymb}
\usepackage{longtable} 
\usepackage{booktabs}
\usepackage{float}
\usepackage{xcolor}

\usepackage{etoolbox}
\usepackage{placeins}
\usepackage[utf8]{inputenc}
\usepackage[english]{babel}

\usepackage{hyperref}
\hypersetup{
    colorlinks=true,
    linkcolor=blue,
    filecolor=red,      
    urlcolor=blue,
    citecolor=blue,
}
\def\eqref#1{\textcolor{blue}{(\ref{#1})}}

\makeatletter
\renewcommand{\fnum@figure}{FIG. \thefigure}
\makeatother

\begin{document}

\title{Measurement and analysis of nuclear $\gamma$-ray production cross sections in proton interactions with Mg, Si and Fe nuclei abundant in astrophysical sites over the incident energy range $E=30-66$~MeV. \\ }

\author {W. Yahia-Cherif}
\email{escaflowne113@gmail.com}
\author {S. Ouichaoui}
\email {souichaoui@usthb.dz, souichaoui@gmail.com}

\affiliation{University of Sciences and Technology Houari Boumedienne (USTHB), Faculty of Physics, P.O. Box 32, EL Alia, 16111 Bab Ezzouar, Algiers, Algeria}

\author {J. Kiener}
\affiliation{Centre de Sciences Nucl\'{e}aires et de Sciences de la Mati\`{e}re (CSNSM), CNRS-IN2P3 et Universit\'{e} de Paris-Sud, 91405 Orsay Campus, France}
 
\author {E.A. Lawrie}
 \affiliation{iThemba LABS, National Research Foundation, P.O. Box 722, Somerset West 7129, South Africa}
\affiliation{Department of Physics, University of the Western Cape, Private Bag X17, Bellville 7535, South Africa}

\author {J.J. Lawrie}
 \affiliation{iThemba LABS, National Research Foundation, P.O. Box 722, Somerset West 7129, South Africa}

 \author {V. Tatischeff}
\affiliation{Centre de Sciences Nucl\'{e}aires et de Sciences de la Mati\`{e}re (CSNSM), CNRS-IN2P3 et Universit\'{e} de Paris-Sud, 91405 Orsay Campus, France}
 
\author {A. Belhout}
\author {D. Moussa}
\affiliation{University of Sciences and Technology Houari Boumedienne (USTHB), Faculty of Physics, P.O. Box 32, EL Alia, 16111 Bab Ezzouar, Algiers, Algeria}

\author {P. Papka}
\affiliation{iThemba LABS, National Research Foundation, P.O. Box 722, Somerset West 7129, South Africa}
\affiliation{Department of Physics, Stellenbosch University, Private Bag X1, Matieland 7602, South Africa}

\author {H. Benhabiles}
\affiliation{Universit\'{e} M'Hamed Bougara, Institut de G\'{e}nie Electrique et Electronique, 35 000 Boumerd\'{e}s, Algeria}

\author {T.D. Bucher}
\affiliation{iThemba LABS, National Research Foundation, P.O. Box 722, Somerset West 7129, South Africa}
\affiliation{Department of Physics, Stellenbosch University, Private Bag X1, Matieland 7602, South Africa}

\author {A. Chafa}
\affiliation{University of Sciences and Technology Houari Boumedienne (USTHB), Faculty of Physics, P.O. Box 32, EL Alia, 16111 Bab Ezzouar, Algiers, Algeria}

\author {J.L. Conradie}
\affiliation{iThemba LABS, National Research Foundation, P.O. Box 722, Somerset West 7129, South Africa}

\author {S. Damache}
\affiliation{CRNA, 02 Bd. Frantz Fanon, B.P. 399 Alger-gare, Algiers, Algeria}

\author {M. Debabi}
\affiliation{University of Sciences and Technology Houari Boumedienne (USTHB), Faculty of Physics, P.O. Box 32, EL Alia, 16111 Bab Ezzouar, Algiers, Algeria}

\author {I. Deloncle}
\affiliation{Centre de Sciences Nucl\'{e}aires et de Sciences de la Mati\`{e}re (CSNSM), CNRS-IN2P3 et Universit\'{e} de Paris-Sud, 91405 Orsay Campus, France}

\author {J.L. Easton}
\affiliation{iThemba LABS, National Research Foundation, P.O. Box 722, Somerset West 7129, South Africa}
\affiliation{Department of Physics, University of the Western Cape, Private Bag X17, Bellville 7535, South Africa}

\author {C. Hamadache}
\affiliation{Centre de Sciences Nucl\'{e}aires et de Sciences de la Mati\`{e}re (CSNSM), CNRS-IN2P3 et Universit\'{e} de Paris-Sud, 91405 Orsay Campus, France}

\author {F. Hammache}
\affiliation{Institut de Physique Nucl\'{e}aire (IPN), CNRS-IN2P3 et Universit\'{e} Paris-Sud, 91405 Orsay Campus, France}

\author {P. Jones}
\affiliation{iThemba LABS, National Research Foundation, P.O. Box 722, Somerset West 7129, South Africa}

\author {B.V. Kheswa}
\affiliation{iThemba LABS, National Research Foundation, P.O. Box 722, Somerset West 7129, South Africa}
\affiliation{Department of Physics, Stellenbosch University, Private Bag X1, Matieland 7602, South Africa}

\author {N.A. Khumalo}
\author {T. Lamula}
\affiliation{Department of Physics, University of the Western Cape, Private Bag X17, Bellville 7535, South Africa}

\author {S.N.T. Majola}
\affiliation{iThemba LABS, National Research Foundation, P.O. Box 722, Somerset West 7129, South Africa}
\affiliation{Department of Physics, University of Cape Town, Private Bag X3, 7701 Rondebosch, South Africa}
\affiliation{Department of Physics, University of Johannesburg, P.O. Box 524, Auckland Park 2006, South Africa}

\author {J. Ndayishimye}
\affiliation{iThemba LABS, National Research Foundation, P.O. Box 722, Somerset West 7129, South Africa}
\affiliation{Department of Physics, Stellenbosch University, Private Bag X1, Matieland 7602, South Africa}

\author {D. Negi}
\affiliation{iThemba LABS, National Research Foundation, P.O. Box 722, Somerset West 7129, South Africa}
\affiliation{Department of Nuclear and Atomic Physics, Tata Institute of Fundamental Research, Mumbai 400005, India}

\author {S.P. Noncolela}
\affiliation{iThemba LABS, National Research Foundation, P.O. Box 722, Somerset West 7129, South Africa}
\affiliation{Department of Physics, University of the Western Cape, Private Bag X17, Bellville 7535, South Africa}

\author {N. de S\'er\'eville}
\affiliation{Institut de Physique Nucl\'{e}aire (IPN), CNRS-IN2P3 et Universit\'{e} Paris-Sud, 91405 Orsay Campus, France}

\author {J.F. Sharpey-Schafer}
\affiliation{Department of Physics, University of the Western Cape, Private Bag X17, Bellville 7535, South Africa}

\author {O. Shirinda}
\affiliation{iThemba LABS, National Research Foundation, P.O. Box 722, Somerset West 7129, South Africa}
\affiliation{Department of Physics, Stellenbosch University, Private Bag X1, Matieland 7602, South Africa}

\author {M. Wiedeking}
\affiliation{iThemba LABS, National Research Foundation, P.O. Box 722, Somerset West 7129, South Africa}

\author {S. Wyngaardt}
\affiliation{University of Stellenbosch, Private Bag X1, 7602 Matieland, South Africa}

\begin{abstract}
Gamma-ray production cross section excitation functions have been measured for $30$, $42$, $54$ and $66$~MeV proton beams accelerated onto targets of astrophysical interest, $^{nat}$C, C + O (Mylar), $^{nat}$Mg, $^{nat}$Si and $^{56}$Fe, at the Sector Separated Cyclotron (SSC) of iThemba LABS (near Cape Town, South Africa). The AFRODITE array equipped with 8 Compton suppressed HPGe clover detectors was used to record $\gamma$-ray data. For known, intense $\gamma$-ray lines the previously reported experimental data measured up to $E_{p}\simeq$ $25$~MeV at the Washington and Orsay tandem accelerators were extended to higher proton energies. Our experimental data for the last 3 targets are reported here and discussed with respect to previous data and the Murphy \textit{et al.} compilation [ApJS 183, 142 (2009)], as well as to predictions of the nuclear reaction code TALYS. The overall agreement between theory and experiment obtained in first-approach calculations using default input parameters of TALYS has been appreciably improved by using modified optical model potential (OMP), deformation, and level density parameters. The OMP parameters have been extracted from theoretical fits to available experimental elastic/inelastic nucleon scattering angular distribution data by means of the coupled-channels reaction code OPTMAN. Experimental data for several new $\gamma$-ray lines are also reported and discussed. The astrophysical implications of our results are emphasised.

\end{abstract}



\maketitle

\volumeyear{year}
\volumenumber{number}
\issuenumber{number}
\eid{identifier}
\date[Date text]{date}
\received[Received text]{date}

\revised[Revised text]{date}

\accepted[Accepted text]{date}

\published[Published text]{date}

\startpage{1}
\endpage{2}


\section{\label{sec1} Introduction}

Gamma astronomy uses $\gamma-$ray lines produced in interactions of highly accelerated charged particles in astrophysical sites as a tool for probing non-thermal processes in the Universe~\cite{1, 2}. For example, in strong solar flares (SFs), electrons and ions (protons, $^{3}$He, $^{4}$He and heavier ions) are accelerated to energies of several hundreds~MeV. Their interactions with abundant nuclei in the solar atmosphere result in complex $\gamma-$ray emission spectra consisting of several components~\cite{3} including narrow, broad, and non-resolved weak lines.\qquad

The broad component of $\gamma-$ray emission spectra from SFs results from the interactions of accelerated heavy ions on hydrogen and helium nuclei and is subject to large Doppler effects making its analysis rather complicated. In contrast, the narrow $\gamma-$ray line component, primarily generated in nuclear reactions induced by accelerated protons and $\alpha-$particles on abundant heavier nuclei ($^{12}$C, $^{14}$N, $^{16}$O, $^{20}$Ne, $^{24}$Mg, $^{28}$Si and $^{56}$Fe), is less affected by Doppler line shifts and broadenings. A thorough analysis of the intensities and shapes of $\gamma-$ray lines~\cite{4} might reveal the properties of particle distributions that are related to the particle acceleration mechanism and to the magneto-hydrodynamic structure of the astrophysical, ambient medium~\cite{5, 5b}. It might also provide valuable information on the properties (density, temperature, ambient chemical elemental composition, etc.) of the emitting astrophysical sites (Sun, stars).\qquad

On the other hand, similar complex $\gamma-$ray spectra involving a large component of both narrow and broad $\gamma-$ray lines are generated in the interactions of low-energy cosmic rays (LECRs, of kinetic energies, $E_{c}$ $\lesssim$ $1$ GeV nucleon$^{-1}$) with abundant nuclei in the interstellar medium (ISM)~\cite{6}. Strong $\gamma-$ray lines produced in the ISM are emitted directly by the excited nuclei and/or by nuclear reaction products following fusion-evaporation, direct, charge-exchange or pre-equilibrium processes induced by energetic protons and $\alpha-$particles on these nuclei.\qquad

Among the strongest $\gamma-$ray lines are the line at $4439$~keV from the decay of the $2^{+}$ first excited state of $^{12}$C, the line at $E_{\gamma}$ $=6129$~keV emitted by the $3^{-}$ second excited state of $^{16}$O, and the lines of $^{14}$N, $^{20}$Ne, $^{24}$Mg, $^{28}$Si and $^{56}$Fe at energies of $E_{\gamma}$ $=1635$, $1634$, $1369$, $1779$ and $847$~keV, respectively. All these strong, main $\gamma-$ray lines are emitted in transitions from the first excited states of abundant nuclei while weaker lines, result from the deexcitation of higher-lying nuclear states.\qquad

To understand the observed $\gamma-$ray spectra from astrophysical sites, the interaction processes at work in these environments requires the knowledge of $\gamma-$ray line production cross sections over a wide energy range of the accelerated particles, extending from reaction thresholds up to several hundreds of~MeV. Gamma-ray line production cross sections for the strong lines at $E_{\gamma}$ $=$ $4.438$~MeV and $6.129$~MeV of $^{12}$C and $^{16}$O are available~\cite{6} up to $E_{p}$ $=$ $85$~MeV. Systematic $\gamma-$ray cross section measurements for the strongest lines have been carried out~\cite{7,8,9} at the Washington university tandem accelerator. They have later been extended (\cite{10, 10b} and Refs therein) to moderately strong lines at the $14$-MV tandem accelerator of Orsay. However, the particle energy range explored in those experiments was limited to $E_{p}$ $<25$~MeV and $E_{\alpha}$ $<40$~MeV, respectively. Production cross sections for some lines have also been measured (\cite{11} and Refs therein) at cyclotron facilities for higher particle energies (up to $E_{p}$ $=50$~MeV and $E_{\alpha}$ $=40$~MeV). In contrast, experimental data for a large number of other, less intense lines covering the lowest proton energy range of astrophysical interest are scarce or even
lacking. A database of $\gamma-$ray line cross section excitation functions has been established four decades ago by Ramaty \textit{et al.}~\cite{12} and was later successively updated by Kozlovsky \textit{et al.}~\cite{13} (in 2002) and by Murphy \textit{et al.}~\cite{14} (2009). This compilation reports cross section data for $\gamma-$ray lines produced in proton and $\alpha-$particle induced reactions on target nuclei from He to Fe for particle incident energies ranging from reaction thresholds up to several hundred~MeV, and also in $^{3}$He$^{+}$ ion induced reactions. Predicted trends for numerous weak lines over the high energy region not covered by experiment are exclusively based on TALYS code~\cite{15} extrapolations. Conversely, existing experimental cross section data can be used to check and improve the predictions of nuclear reaction theoretical models.\qquad

More recently, measurements of $\gamma-$ray line cross sections have been carried out~\cite{16, 16b} at the Orsay tandem accelerator for reactions induced by swift protons and $\alpha-$particles on N, O, Ne, and Si targets over the incident energy range extending up to $26$ and $39$~MeV, respectively. The measured cross section data completed earlier data sets taken at the same facility~\cite{10, 10b,17,18} for lower particle energies. However, the experimental $\gamma-$ray production cross sections remain absent for higher energies.\qquad

During particle-nucleus collisions various reaction mechanisms (compound nucleus formation, direct reactions, pre-equilibrium emission, etc.) occur with different probabilities depending on the projectile energy, the interaction time and the structure of the nuclear reaction partners. They should be quantitatively taken into account in calculations by nuclear reaction codes like TALYS~\cite{15} or EMPIRE II~\cite{19}. Calculations performed by members of our group using these codes assuming built-in default OMP and nuclear level structure parameters have led to $\gamma-$ray line production cross sections significantly lower than the experimental values for particle energies above $20-30$~MeV. However, the agreements between the calculated $\gamma-$ray line production cross sections and corresponding experimental data can be appreciably improved by using in TALYS modified nuclear level deformation parameters instead of the built-in default ones as shown in Ref.~\cite{16, 16b}, e.g., for the $^{14}$N, $^{22}$Ne or $^{28}$Si nuclei. Furthermore, these calculations revealed the presence of large structures at higher particle energies where no experimental data are available.\qquad

Consequently, measurements of $\gamma-$ray line production cross sections for projectile energies of $E\geqslant30$~MeV are necessary. Such experimental data are of crucial importance for understanding the nuclear interaction processes taking place in astrophysical sites and for reliably adjusting theoretical model parameters. Furthermore, these data are of great interest for various applications such as proton radiotherapy~\cite{20}, non-destructive analysis of archeaological materials~\cite{21}, etc. We aim to expand the existing cross section database~\cite{14}, to check and improve the predictions of modern nuclear reaction codes~\cite{15, 19}, and to simulate the nuclear collisions at work in SFs and the ISM in by modelling $\gamma-$ray emission fluxes from these sites~\cite{22,23,24}. In this context, we have undertaken a comprehensive experimental program for measuring nuclear $\gamma-$ray line production cross sections for protons and alpha-particles at energies $E_{lab} = 30 - 200$ MeV, on various target nuclei known to be abundant in the solar atmosphere and in the ISM.\qquad

In the present paper, we report and discuss our results obtained in an experiment carried out at the SSC facility of iThemba LABS for $30-66$~MeV proton beams interacting with Mg, Si, and Fe target nuclei.\qquad

In Sections~\ref{sec2} and~\ref{sec3} we describe the experiment and the data analysis, respectively. The experimental results are reported and discussed in Section~\ref{sec4}, where the cross section excitation functions for known, main $\gamma-$ray lines are compared to previous counterparts measured at proton energies, $E_{p}<$ 30~MeV~\cite{7,10,10b,11,16, 16b}. Results for new lines, measured for the first time in this work, are also reported in this section, where present and previous experimental cross section data are compared to the Murphy \textit{et al.} database~\cite{14} and to TALYS calculations in Section~\ref{sec5}. An astrophysical application of $\gamma-$ray line cross sections is presented in Section~\ref{sec6}. Finally, a summary and conclusion is provided in Section~\ref{sec7}.\newline

 \FloatBarrier

\begin{table*}[!hpbt]
\begin{tabular}
[c]{ccccccc}
 \hline
 \hline

{Target} & {Thickness} & \multicolumn{5}{c}{Composition} \\
          \cline{3-7}    \\
& {(mg~cm${^{-2}}$)}& {(This Work)} &{Ref. }\cite{7} & {Ref. }\cite{11} &
{Ref. }\cite{10} & {Ref. }\cite{16, 16b}\\
\hline
 & & & & & & \\
Mg & {9.80 $\pm$ 0.49} & {${^{24}}$Mg ($>$ 99\%)} & {${^{24}}$Mg ($>$ 99\%)} & {natural } & {natural } & \\
& {5.00 $\pm$ 0.02} & {natural } &  &  &  & \\ \hline

 & & & & & &\\
{Si} & {6.00 $\pm$ 0.30} & {natural } & {natural } & {natural } &  & {natural}\\ \hline

 & & & & & &\\
{Fe} & {7.93 $\pm$ 0.08} & {${^{56}}${Fe} ($>$ 99\%) } & {${^{56}}${Fe} ($>$99\%)} & {natural } & {natural } & \\
& {7.40 $\pm$ 0.37} & {natural } &  &  &  & \\ \hline

\end{tabular}
\caption{\label{tab:1}List and properties of targets. The isotopic composition of natural
targets are as follows: $^{nat}$Mg = $^{24}$Mg(78.99\%) + $^{25}$Mg(10\%) +
$^{26}$Mg(11.01\%), $^{nat}$Si = $^{28}$Si(92.22\%) + $^{29}$Si(4.68\%) + $^{30}$Si(3.09\%), $^{nat}$Fe = $^{54}$Fe(5.85\%) + $^{56}$Fe(91.75\%) + $^{57}$Fe(2.12\%) + $^{58}$Fe(0.28\%).}%
\end{table*} 

We supply in the \hyperlink{thesentence}{Appendix} the procedure used for extracting, within the framework of the coupled-channels nuclear reaction approach, improved OMP and nuclear level deformation parameters that prove to be more appropriate for $\gamma$-ray line production cross section calculations than the default input data of TALYS. In addition, we provide information on the experimental angular distribution data for nucleon scattering off the studied Mg, Si and Fe target nuclei, taken from the literature, that was used to derive new OMP and nuclear level structure parameters.\qquad

\section{\label{sec2} Experimental set up and method}

\subsection{Beam, targets, and detection system}

The experiment was carried out at the SSC facility of iThemba LABS.

Proton beams of incident energies $E_{p}$ $=30$, $42$, $54$ and $66$~MeV and current intensity, $I=3-5$ nA, were directed onto solid elemental targets. The incident proton beam was stopped in an electrically well isolated Faraday cup
located 3 m downstream of the reaction chamber and placed $1.5$ m deep inside a concrete wall. The integrated current on the Faraday cup was measured to an accuracy of $\sim1\%$, and the total collected beam charge for a typical $1$ hour run was of\ $\sim5$ $\mu$C.\qquad

Only solid targets (natural or isotopically enriched) of C, Mylar ($^{12}$C $+^{16}$O), Mg, Al, Si, Ca and Fe elements prepared mainly at iThemba LABS (some of them were imported from Orsay) were used in the experiment. The Mylar foils were $5$ $\mu$m thick while the thicknesses of the other explored targets varied in the range, $6-9.8$ mg~cm$^{-2}$; the isotopic compositions and thicknesses of the $^{nat}$Mg, $^{24}$Mg, $^{nat}$Si, $^{nat}$Fe and $^{56}$Fe targets for which results are presented in this paper are reported in Table~\ref{tab:1}. The thickness of each target was then converted in units of nuclei$\cdotp$cm$^{-2}$\qquad

The targets were mounted onto small rectangular Al frames and placed on a ladder allowing for $4$ target positions: one for a beam viewer, one for an empty frame and two for frames with targets. The viewer consisted of a rectangular-shaped, $3$ mm thick fluorescent Al$_{2}$O$_{3}$ layer with a $3$ mm diameter hole in its centre. During the experiment the proton beam spot was reduced down to less than $3$ mm diameter by focusing it through the hole in the Al$_{2}$O$_{3}$ viewer and was frequently checked and focused after target changes. Improved beam tuning was achieved by minimizing the $\gamma$ count rate from an empty target frame.\qquad

The detection of $\gamma$ rays was made by means of the AFRODITE array~\cite{25, 25b, 26} that consisted of 8 Compton suppressed clover detectors~\cite{27} of the EUROGAM phase II type. The clovers were placed in a fixed geometry with 4 detectors at $90\degree$ and 4 at $135\degree$ relative to the incident beam direction.\qquad

With a distance between the target and the front face of each clover of $17.6$ cm, the individual crystals of a clover were at $\pm5\degree$ relative to the clover centre. Hence, a total of $32$ HPGe crystals were used, which enabled us to measure $\gamma-$ray line energy spectra at four detection angles, i.e., $\theta_{lab}=85\degree$, $95\degree$, $130\degree$ and $140\degree$. The whole detection system subtended a solid angle of $\sim11\%$ of $4\pi$.\qquad

\subsection{Measurement of $\gamma-$ray energy spectra}

The $\gamma-$ray energy spectra from the individual HPGe crystals were recorded using a digital data acquisition system based on XIA modules of DGF Pixie$-16$ type. At each energy and for each target, data were acquired until a statistical error of better than $5\%$ in the less intense peaks of the $\gamma-$ray spectrum in a single clover crystal was reached. Irradiation times with the target exposed to the proton beam were typically $1$ hour. After each target irradiation spectra from activation of the target and the surrounding material were recorded for half an hour without beam but with the irradiated target in place. Thereafter beam-induced $\gamma-$ray background, with an empty target frame in place, was measured for half an hour. Finally, room background runs without beam and without a target in place were frequently measured for longer times of $\sim1.5$ h. For each run the duration and accumulated charge were recorded. \qquad 

\begin{figure*}[!htbp]

\centering
    \textbf{p + $^{nat}$Mg for $E_{protons} = 42$~MeV, $\theta${$_{lab}$} $= 85\degree$ }\par\medskip

\includegraphics[scale=0.75]{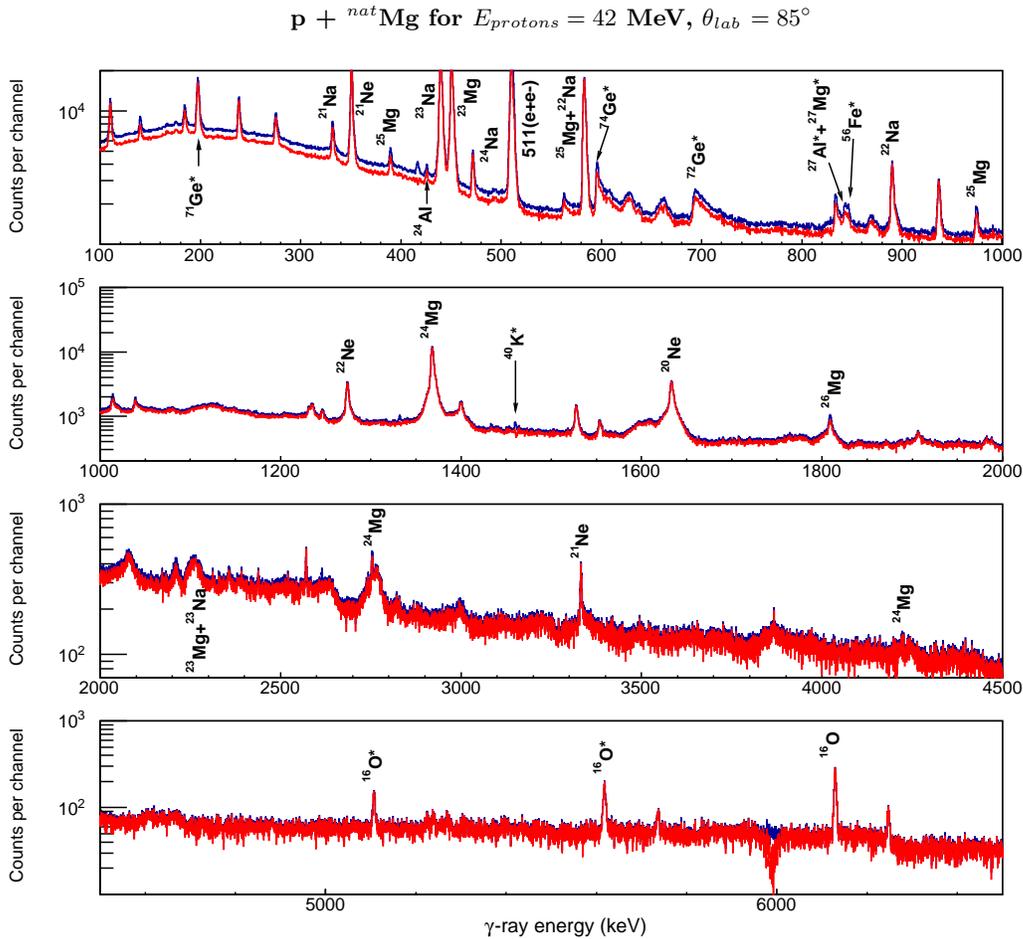}
\caption[c]{\label{fig1}Calibrated $\gamma$-ray spectra produced by the scattering of 42~MeV protons off the $^{nat}$Mg target. The blue histograms correspond to the raw spectrum of one clover crystal without background suppressions, while the red one is background subtracted.}

\end{figure*}

Detection of $\gamma-$ray lines with energies of at least up to $E_{\gamma}=$ $8$~MeV was expected in the measured energy spectra. The energy calibration of the individual HPGe crystals was performed by means of standard radioactive sources of $^{137}$Cs, $^{60}$Co and $^{152}$Eu covering the low $\gamma-$ray range up to 1408~keV, while the prominent line of $^{16}$O at $E_{\gamma}=6129$~keV (and associated escape lines at $5618$ and $5107$~keV) produced in the $^{16}$O(p, p$^{\prime}\gamma$) reaction from a Mylar target was used at high energies. The relative energy resolution $\Delta E_{_{\gamma}}/E_{_{\gamma}}$, determined at $E_{\gamma}=1332$~keV was less than $0.23\%$.\qquad

Detection efficiency measurements for the low-energy region were performed using the same standard radioactive sources. The obtained results were extrapolated to higher $\gamma-$ray energies of up to $E_{\gamma}\sim10$~MeV, by means of Monte-Carlo simulations of the detection system including a detailed description of the target chamber and all the clover detectors using GEANT4~\cite{28} (see section~\ref{sec3}).\qquad

Throughout the experiment, care was taken to minimize the neutron and $\gamma-$ray backgrounds. All the experimental runs were performed with relatively low proton beam intensities, $I\leq5$ nA, typical count rates through the empty frame of less than $400$ Hz, beam-on-target count rates limited to $5-6$ kHz per crystal, reduced dead time (see below) and minimum pulse pile-up.\qquad

A measurement of the acquisition dead time was made using a pulser signal inserted on the test line of one clover detector as well as directly into one channel of the data acquisition system. A comparison of the total pulser triggers and the counts in the pulser peaks in the spectra yielded a dead time of $8(\pm2)\%$ at an average count rate of $\sim6$ kHz per clover crystal. Since all the clovers were identical the dead time is assumed to be the same for all elements of the array.\qquad

Figure~\ref{fig1} reports typical Compton suppressed $\gamma-$ray spectra from a single HPGe crystal located at $\theta_{lab}=85\degree$, obtained in the irradiation of the $^{nat}$Mg target with $42$~MeV protons. The total spectrum is shown in blue, while the spectrum after subtraction of the normalised beam-related background is depicted in red.\qquad

\ \ \ \ \ \ \ \ \ \ \ \ 

\section{\label{sec3} Data analysis, determination of $\gamma-$ray cross sections}

Since only relatively thick solid targets were used in the experiment (see Table~\ref{tab:1}), the $\gamma-$ray lines from some targets, for instance the Fe targets, were not found to be dramatically affected by Doppler shifts and broadenings, although for other targets, for instance Mg and Si targets, the measured $\gamma-$ray spectra were rather complex.\qquad

In the data analysis the clover crystals were treated as individual detectors and, in addition to the normal Compton suppression from the BGO suppression shield, coincident events between elements in the same clover were rejected. The background-corrected spectra were obtained by subtracting the beam related background, recorded with beam on an empty target frame, after normalising to the same accumulated charge. The numbers of counts in the observed $\gamma-$ray peaks were extracted using the ROOT program~\cite{29} and/or the GF3 program included in the RADWARE package~\cite{30}. In our analysis we fitted the observed $\gamma-$ray peaks with symmetric Gaussian shapes on a linear background. For most of the clover detectors the resulting uncertainties in the identified $\gamma-$ray energies were less than $0.1$~keV.\qquad

In most cases, the analysis of the peaks associated with the $\gamma$ rays of interest was performed similarly to that in Ref.~\cite{10,10b}. For well-defined, symmetric peaks the corresponding areas were first extracted from Gaussian distribution fits. For each peak, a second estimation of its area was performed by summation over a region of interest after subtraction of a linear background. The final peak area was determined as the average of the two estimations and the systematic error as the difference. The total uncertainty in the peak area consisted of the systematic error, the statistical error in the peak area and the error from the fitting procedure. For most analysed $\gamma-$ray lines, the statistical error was lower than $2\%$, while the systematic error varied from $1\%$ for intense peaks (such as, e.g., the $1408$~keV peak) to $9\%$ (e.g., the $1303$~keV peak).\qquad

The experimental differential cross sections for the analysed $\gamma-$ray lines were determined from values of the target thickness, the total beam charge deposited in the Faraday-cup, the extracted peak area and the absolute detection efficiency, $\epsilon(E_{\gamma},\theta)$. We assumed that $\theta_{lab}\approx\theta_{c.m.}$ since one deals with reactions induced by light projectiles on appreciably heavier target nuclei. The efficiency as function of energy was derived from a fit to the experimental data from the radioactive sources, and the normalised data from a GEANT4 simulation using the following expression~\cite{30} 

\qquad%
\begin{equation}
\epsilon(E_{\gamma},\theta)=e^{[(A+BX)^{-H}+(D+EY+FY^{2}+GY^{3}%
)^{-H}]^{-1/H}}, \label{eq.1}%
\end{equation}

\noindent where $X=log(E_{\gamma}/100)$, $Y=log(E_{\gamma}/1000)$ and the other quantities are adjustable free parameters.\qquad

The obtained differential cross section values from all the HPGe crystals at each given observation angle were averaged, and Legendre polynomial expansions of the form,

\begin{equation}
W(\theta)=\sum\limits_{l=0}^{l_{max}}a_{l}Q_{l}P_{l}(cos(\theta)),
\label{eq. 2}%
\end{equation}

\noindent were fitted to the experimental angular distribution data. In this expression, the summation extends only over integer \textit{l-}values with $l_{max}$ taking on twice the $\gamma-$ray multipolarity and the $Q_{l}$ are energy-dependent geometrical attenuation coefficients, as described by Rose~\cite{31} (see also Ref.~\cite{32}). Except for the $6.129$~MeV line of $^{16}$O, the multipolarity of the $\gamma-$ray lines of interest in this work is at most $2$, which then fixes, $l_{max}\leq4$, in the above expansion. The $Q_{l}$-values were specifically calculated for the AFRODITE detection array via our GEANT4~\cite{28} simulations; they were found to remain almost constant over the photon energy range, $E_{\gamma }=0.1-10$~MeV, i.e., $Q_{l}\sim0.981-1$ with relative uncertainties of at most $1\%$.\qquad

The total relative uncertainties in the measured differential cross section values for all the analysed lines were taken as the relative errors in the above parameters, and were found to fall within the range $10-15\%$.\qquad

The angle-integrated production cross sections for the observed $\gamma-$ray lines were directly derived from the $a_{0}$ coefficients of eq.~\eqref{eq. 2}, using the relation, $\sigma_{int}=4\pi a_{0}$.\qquad 

The corresponding results are reported and discussed in the following section, where additional information on the analysis of the $\gamma-$ray energy spectra is provided.\qquad

\section{\label{sec4} Experimental results and discussion}

\subsection{\bigskip$\gamma-$ray energy spectra and transition properties}

As can be seen in Figure~\ref{fig1}, various more or less prominent $\gamma-$ray lines, whose origin is indicated in this figure, are observed below and above the intense line at $511$~keV from the ($e^{+}$, $e^{-}$) pair annihilation. One notes, e.g., the presence of asymmetric peaks for lines resulting from the inelasting scattering of secondary neutrons off the isotopic constituents of the HPGe crystals~\cite{33,34}, i.e., off the $^{74}$Ge isotope (line peak at $E_{\gamma}$ $=595.85$~keV) and off $^{72}$Ge (peaks at $E_{\gamma}$ $=689.6$~keV and $834.01$), along with the Gaussian shape peak for the line at $E_{\gamma}$ $=198.39$~keV emitted in the $^{70}$Ge(n, $\gamma$)$^{71m}$Ge radiative neutron capture reaction~\cite{33}. These neutron peaks affected the measured $\gamma-$ray energy spectra from all the explored targets. One can also observe the line at $E_{\gamma}$ $=846.76$~keV from the $^{56}$Fe(p,p$^{\prime}\gamma$) inelastic proton scattering and the $843.76$~keV line of $^{27}$Al, produced both in the $^{27}$Al(p, p$^{\prime}\gamma$) inelastic proton scattering and the $\beta$-decay of $^{27}$Mg following the $^{27}$Al(n, p)$^{27}$Mg$^{\ast}$ charge exchange reaction. These two lines result from reactions induced by scattered protons and secondary neutrons on the aluminium of the target chamber and the beam pipes. In the high energy part of the energy spectrum the characteristic peak associated with the $6.129$~MeV line of $^{16}$O appears (see Figure~\ref{fig1}), likely resulting from the energetically possible $^{24}$Mg(p, p2$\alpha\gamma$)$^{16}$O reaction. The properties (transition energy, emitting isotope, corresponding nuclear levels with their spin-parity assignements, branching ratio) of the $\gamma-$ray lines of interest identified in the collected energy spectra in the proton irradiations of the Mg, Si and Fe targets are listed in Table~\ref{tab:2}.\qquad

\begin{table*}
 \begin{tabular}{p{2cm}p{1.5cm}cp{1.5cm}p{1.5cm}cccc}


\hline
\hline
\centering{Main $\gamma$ ray} & \centering{Intruder} & \centering{emitting nucleus} &  \centering{$E_{i}$}  & \centering{$E_{f}$} & \centering {$J^{\pi}_i$}  & \centering {$J^{\pi}_f$} & \centering {$M\lambda$} & \centering{Branching ratio}  \tabularnewline 


\centering {keV} & \centering {keV} & \centering {} & \centering {keV} & \centering {keV} & \centering {$\%$} \tabularnewline 
\midrule
\hline
\multicolumn{9}{c}{\underline{Magnesium}} \tabularnewline 
\\

\centering {425.8} & {} & \centering {$^{24}$Al} & \centering {425.8} & \centering {$g.s.$} & \centering {$1^+_1$} & \centering {$4^+_1$} & \centering {\textit{M3}} & \centering {100}  \tabularnewline
\centering {1808.68} & {} & \centering {$^{26}$Mg} & \centering {1808.74} & \centering {$g.s.$} &  \centering {$2^+_1$} & \centering {$0^+_1$} & \centering {\textit{E2}} & \centering {100} \tabularnewline
\centering {585.03} & {} & \centering {$^{25}$Mg} & \centering {585.05} & \centering {$g.s.$} &  \centering {$\frac{1}{2}^+_1$} & \centering {$\frac{5}{2}^+_1$} & \centering {\textit{E2}}  & \centering {100} \tabularnewline
\centering {} & \centering  {583.04} & \centering {$^{22}$Na} & \centering {583.05} & \centering {$g.s.$} & \centering {$1^+_1$} & \centering {$3^+_1$} & \centering {\textit{E2}} & \centering {100}\tabularnewline
\centering {389.71} & {} & \centering {$^{25}$Mg} & \centering {974.76} & \centering {585.05} & \centering {$\frac{3}{2}^+_1$} & \centering {$\frac{1}{2}^+_1$} & \centering {\textit{M1+E2}} & \centering {45.89\textit{(81)}} \tabularnewline
\centering {974.74} & {} & \centering {$^{25}$Mg} & \centering {974.76} & \centering {$g.s.$} & \centering {$\frac{3}{2}^+_1$} & \centering {$\frac{5}{2}^+_1$} & \centering {\textit{M1+E2}} & \centering {54.11\textit{(81)}} \tabularnewline
\centering {1368.62} & {} & \centering {$^{24}$Mg} & \centering {1368.67} & \centering {$g.s.$} & \centering {$2^+_1$} & \centering {$0^+_1$} & \centering {\textit{E2}}  & \centering {100} \tabularnewline
\centering {} & \centering  {1368.7} & \centering {$^{22}$Na} & \centering {1951.8} & \centering {583.05} & \centering {$2^+_1$} & \centering {$1^+_1$} &  & \centering {99.71} \tabularnewline
\centering {2754.01} & {} & \centering {$^{24}$Mg} & \centering {4122.89} & \centering {1368.67} & \centering {$4^+_1$} & \centering {$2^+_1$} & \centering {\textit{E2}}  & \centering {100} \tabularnewline
\centering {4237.96} & {} & \centering {$^{24}$Mg} & \centering {4238.24} & \centering {$g.s.$} & \centering {$2^+_2$} & \centering {$0^+_1$} & \centering {\textit{E2}} & \centering {78.93\textit{(47)}}  \tabularnewline
\centering {450.70} & {} & \centering {$^{23}$Mg} & \centering {450.71} & \centering {$g.s.$} & \centering {$\frac{5}{2}^+_1$} & \centering {$\frac{3}{2}^+_1$} & \centering {\textit{M1+E2}} & \centering {100} \tabularnewline
\centering {472.20} & {} & \centering {$^{24}$Na} & \centering {472.21} & \centering {$g.s.$} & \centering {$1^+_1$} & \centering {$4^+_1$} & \centering {\textit{M3}}  & \centering {100}   \tabularnewline
\centering {439.99} & {} & \centering {$^{23}$Na} & \centering {439.99} & \centering {$g.s.$} &  \centering {$\frac{5}{2}^+_1$} & \centering {$\frac{3}{2}^+_1$} & \centering {\textit{M1+E2}} & \centering {100} \tabularnewline
\centering {331.91} & {} & \centering {$^{21}$Na} & \centering {331.90} & \centering {$g.s.$} &  \centering {$\frac{5}{2}^+_1$} & \centering {$\frac{3}{2}^+_1$} & \centering {\textit{M1+E2}} & \centering {100} \tabularnewline
\centering {350.73} & {} & \centering {$^{21}$Ne} & \centering {350.73} & \centering {$g.s.$} &  \centering {$\frac{5}{2}^+_1$} & \centering {$\frac{3}{2}^+_1$} & \centering {\textit{M1+E2}} & \centering {100} \tabularnewline

\multicolumn{9}{c}{\underline{Silicon}} \tabularnewline 
\\
\centering {1273.36} & {} & \centering {$^{29}$Si} & \centering {1273.39} & \centering {$g.s.$} &  \centering {$\frac{3}{2}^+_1$} & \centering {$\frac{1}{2}^+_1$} & \centering {\textit{M1+E2}}  & \centering {100} \tabularnewline
\centering {1778.97} & {} & \centering {$^{28}$Si} & \centering {1779.03} & \centering {$g.s.$} & \centering {$2^+_1$} & \centering {$0^+_1$} & \centering {\textit{E2}} & \centering {100} \tabularnewline
\centering {2838.29} & {} & \centering {$^{28}$Si} & \centering {4617.86} & \centering {1778.97} & \centering {$4^+_1$} & \centering {$2^+_1$} & \centering {\textit{E2}}  & \centering {100} \tabularnewline
\centering {780.8} & {} & \centering {$^{27}$Si} & \centering {780.9} & \centering {$g.s.$} &  \centering {$\frac{1}{2}^+_1$} & \centering {$\frac{5}{2}^+_1$} & \centering {\textit{E2}} & \centering {100}  \tabularnewline
\centering {957.3} & {} & \centering {$^{27}$Si} & \centering {957.3} & \centering {$g.s.$} &  \centering {$\frac{3}{2}^+_1$} & \centering {$\frac{5}{2}^+_1$} & \centering {\textit{M1+E2}} & \centering {93.97\textit{(21)}} \tabularnewline
\centering {843.76} & {} & \centering {$^{27}$Al} & \centering {843.76} & \centering {$g.s.$} &  \centering {$\frac{1}{2}^+_1$} & \centering {$\frac{5}{2}^+_1$} & \centering {\textit{E2}} & \centering {100} \tabularnewline
\centering {416.85} & {} & \centering {$^{26}$Al} & \centering {416.85} & \centering {$g.s.$} & \centering {$3^+_1$} & \centering {$5^+_1$} & \centering {\textit{[E2]}} & \centering {100} \tabularnewline
\centering {585.03} & {} & \centering {$^{25}$Mg} & \centering {585.05} & \centering {$g.s.$} &  \centering {$\frac{1}{2}^+_1$} & \centering {$\frac{5}{2}^+_1$} & \centering {\textit{E2}}  & \centering {100} \tabularnewline
\centering {} & \centering  {583.04} & \centering {$^{22}$Na} & \centering {583.05} & \centering {$g.s.$} & \centering {$1^+_1$} & \centering {$3^+_1$} & \centering {\textit{E2}} & \centering {100}\tabularnewline
\centering {389.71} & {} & \centering {$^{25}$Mg} & \centering {974.76} & \centering {585.05} & \centering {$\frac{3}{2}^+_1$} & \centering {$\frac{1}{2}^+_1$} & \centering {\textit{M1+E2}} & \centering {45.89\textit{(81)}} \tabularnewline
\centering {1368.62} & {} & \centering {$^{24}$Mg} & \centering {1368.67} & \centering {$g.s.$} & \centering {$2^+_1$} & \centering {$0^+_1$} & \centering {\textit{E2}} & \centering {100} \tabularnewline
\centering {} & \centering  {1368.7} & \centering {$^{22}$Na} & \centering {1951.8} & \centering {583.05} & \centering {$2^+_1$} & \centering {$1^+_1$} &  & \centering {99.71} \tabularnewline
\centering {450.70} & {} & \centering {$^{23}$Mg} & \centering {450.71} & \centering {$g.s.$} & \centering {$\frac{5}{2}^+_1$} & \centering {$\frac{3}{2}^+_1$} & \centering {\textit{M1+E2}}  & \centering {100} \tabularnewline
\centering {439.99} & {} & \centering {$^{23}$Na} & \centering {439.99} & \centering {$g.s.$} &  \centering {$\frac{5}{2}^+_1$} & \centering {$\frac{3}{2}^+_1$} & \centering {\textit{M1+E2}} & \centering {100} \tabularnewline
\centering {331.91} & {} & \centering {$^{21}$Na} & \centering {331.90} & \centering {$g.s.$} &  \centering {$\frac{5}{2}^+_1$} & \centering {$\frac{3}{2}^+_1$} & \centering {\textit{M1+E2}} & \centering {100} \tabularnewline
\centering {350.73} & {} & \centering {$^{21}$Ne} & \centering {350.73} & \centering {$g.s.$} &  \centering {$\frac{5}{2}^+_1$} & \centering {$\frac{3}{2}^+_1$} & \centering {\textit{M1+E2}} & \centering {100} \tabularnewline

\midrule
\multicolumn{9}{c}{\underline{Iron}} \tabularnewline 
\\
\centering {846.78} & {} & \centering {$^{56}$Fe} & \centering {846.76} & \centering {$g.s.$} & \centering {$2^+_1$} & \centering {$0^+_1$} & \centering {\textit{E2}} & \centering {100} \tabularnewline
\centering { } & \centering  {847} & \centering {$^{55}$Fe} & \centering {2255.5} & \centering {1408.45} & & &  & \centering {100} \tabularnewline
\centering {1238.27} & {} & \centering {$^{56}$Fe} & \centering {2085.11} & \centering {846.76} & \centering {$4^+_1$} & \centering {$2^+_1$} & \centering {\textit{E2}}  & \centering {100\textit{(2)}} \tabularnewline
\centering {1810.76} & {} & \centering {$^{56}$Fe} & \centering {2657.59} & \centering {846.76} & \centering {$2^+_2$} & \centering {$2^+_1$} & \centering {\textit{M1+E2}}  & \centering {96.99\textit{(29)}} \tabularnewline
\centering {1303.4} & {} & \centering {$^{56}$Fe} & \centering {3388.55} & \centering {2085.11} & \centering {$6^+_1$} & \centering {$4^+_1$} & \centering {\textit{E2}} & \centering {98.72\textit{(40)}} \tabularnewline
\centering {411.9} & {} & \centering {$^{55}$Fe} & \centering {411.42} & \centering {$g.s.$} & \centering {$\frac{1}{2}^-_1$} & \centering {$\frac{3}{2}^-_1$} & \centering {\textit{M1(+E2)}}  & \centering {100\textit{(6)}} \tabularnewline
\centering { } & \centering {411.4} & \centering {$^{54}$Fe} & \centering {2949.2} & \centering {2538.1} & \centering {$6^+_1$} & \centering {$4^+_1$} & \centering {\textit{E2}} & \centering {100} \tabularnewline
\centering {1316.4} & {} & \centering {$^{55}$Fe} & \centering {1316.54} & \centering {$g.s.$} & \centering {$\frac{7}{2}^-_1$} & \centering {$\frac{3}{2}^-_1$} & \centering {\textit{E2}}  & \centering {92.94\textit{(13)}} \tabularnewline
 {} & \centering{1312.6} & \centering {$^{56}$Fe} & \centering {4683.04} & \centering {3369.95} & \centering {$(2^{+}),3^{+}$} & \centering {$2^{+}$} & \centering {}  & \centering { $<$ 48} \tabularnewline

\centering {1408.4} & {} & \centering {$^{55}$Fe} & \centering {1408.45} & \centering {$g.s.$} & \centering {$\frac{7}{2}^-_2$} & \centering {$\frac{3}{2}^-_1$} & \centering {\textit{E2}}  & \centering {44.3\textit{(21)}} \tabularnewline
\centering { } &  \centering {1408.1} & \centering {$^{54}$Fe} & \centering {1408.19} & \centering {$g.s.$} & \centering {$2^+_1$} & \centering {$0^+_1$} & \centering {\textit{E2}} & \centering {100} \tabularnewline
\centering {274.8} & {} & \centering {$^{55}$Fe} & \centering {2813.8} & \centering {2539.11} &  \centering {$\frac{13}{2}^-_1$} & \centering {$\frac{11}{2}^-_1$} & \centering {\textit{E2}}  & \centering {100} \tabularnewline
\centering {3432.0} & {} & \centering {$^{54}$Fe} & \centering {6380.9} & \centering {2949.2} & \centering {$8^+_1$} & \centering {$6^+_1$} & \centering {\textit{E2}}  & \centering {100} \tabularnewline
\centering {156.27} & {} & \centering {$^{54}$Mn} & \centering {156.29} & \centering {$g.s.$}  & \centering {$4^+_1$} & \centering {$3^+_1$} & \centering {\textit{M1+E2}} & \centering {100} \tabularnewline
\centering {211.98} & {} & \centering {$^{54}$Mn} & \centering {368.22} & \centering {156.29} & \centering {$5^+_1$} & \centering {$4^+_1$} & \centering {\textit{M1}} & \centering {99.11} \tabularnewline
\centering {377.88} & {} & \centering {$^{53}$Mn} & \centering {377.89} & \centering {$g.s.$} & \centering {$\frac{5}{2}^-_1$} & \centering {$\frac{7}{2}^-_1$} & \centering {\textit{M1+E2}} & \centering {100} \tabularnewline
\centering {1441.2} & {} & \centering {$^{53}$Mn} & \centering {1441.15} & \centering {$g.s.$} & \centering {$\frac{11}{2}^-_1$} & \centering {$\frac{7}{2}^-_1$} & \centering {\textit{E2}} & \centering {100} \tabularnewline
\centering {1434.07} & {} & \centering {$^{52}$Cr} & \centering {1434.09} & \centering {$g.s.$} & \centering {$2^+_1$} & \centering {$0^+_1$} & \centering {\textit{E2}} & \centering {100} \tabularnewline

\hline
\hline

\end{tabular}

\caption{ \label{tab:2} Gamma-ray lines from the Mg, Si and Fe targets considered in this work. In the first, second and third columns are listed the transition energies of the main and overlapping peaks and the emitting nuclei. In the fourth to the seventh columns the characteristics of the involved nuclear states are listed, while the $\gamma$-ray multipolarities and branching ratios are presented in the eighth and ninth columns respectively.}

\end{table*}

\subsection{Absolute $\gamma-$ray efficiencies}

\begin{figure}[!htbp]
\includegraphics[scale=0.45]{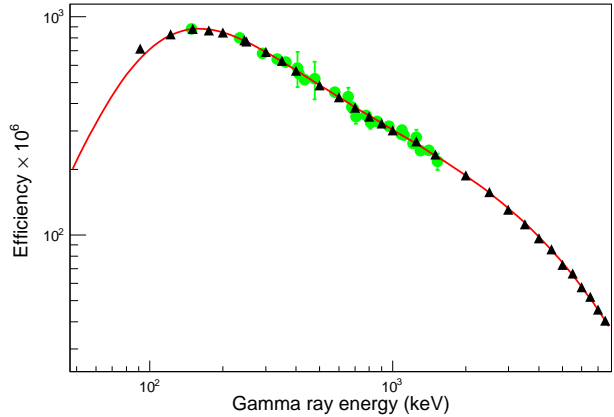}
\caption {\label{fig2} Absolute $\gamma-$ray efficiencies for a single HPGe crystal. The green data points correspond to experimental values measured using standard radioactive sources of $^{60}$Co, $^{137}$Cs and $^{152}$Eu. The black data points represent the results obtained via GEANT4 simulations at various $\gamma-$ray energies that were normalised to experimental data. The red solid line is the result of the fit of the efficiency function of equation~\eqref{eq.1}~\cite{30} to the experimental data and to the normalised GEANT4-simulated values.}
\end{figure}

Figure~\ref{fig2} reports an example of the experiment absolute efficiency data measured with the standard radioactive sources and GEANT4-simulated values, normalised to the former ones, for a single HPGe crystal. The fitted curve to these two data sets using eq.~\eqref{eq.1} is also shown.\qquad

The efficiency of the whole array in the addback mode amounts to $1.6\%$ at $E_{\gamma}=1.33$~MeV , but one expects it to be substantially lower in single-crystal mode. Indeed, the $\gamma-$ray full absorption peak efficiency for a single Ge crystal was found to amount to $\sim0.03\%$ at $E_{\gamma}=1.33$~MeV and to $\sim0.004\%$ at $E_{\gamma}=8$~MeV on average (see Figure~\ref{fig2}) which suggests a total efficiency of $0.96\%$ at $E_{\gamma}=1.33$~MeV for the full array. The relative uncertainty in the detection efficiency was estimated to be lower than $5\%$ over the whole $\gamma-$ray energy domain explored.\qquad

\subsection{$\gamma-$ray angular distributions}

\begin{figure}[!htbp]
\includegraphics[scale=0.5]{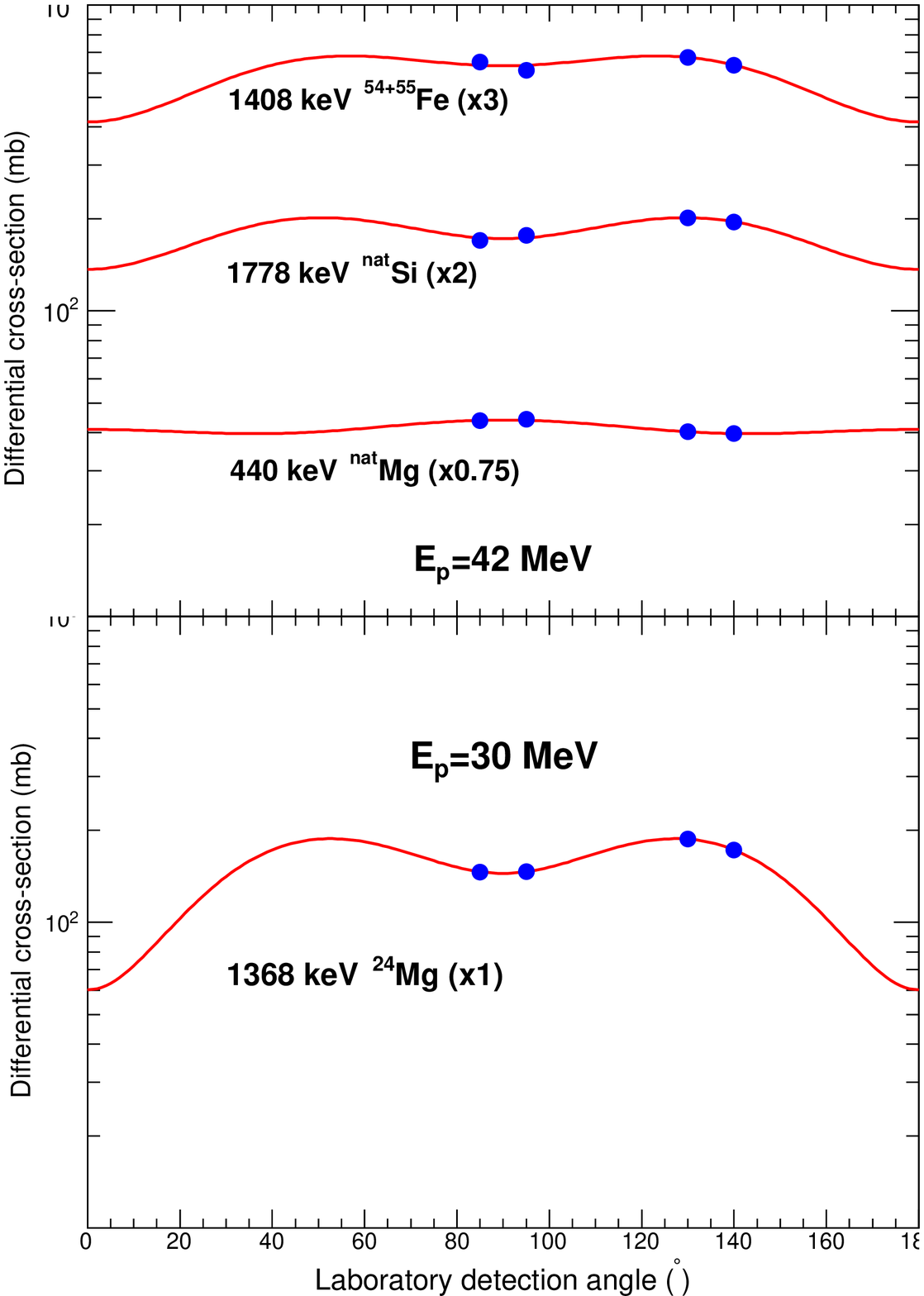}
\caption {\label{fig3} $\gamma-$ray angular distributions: (Top) for main lines produced in interactions of 42~MeV protons with the $^{nat}$Mg, $^{nat}$Si and $^{56}$Fe targets, (Bottom) for the line at $E_{\gamma}=1368$~keV emitted by the $^{24}$Mg isotope upon inelastic scattering of $30$~MeV protons.}
\end{figure}

The measured angular distributions of the observed $\gamma$ rays, produced mainly in (p, p$^{\prime}\gamma$) inelastic proton scattering off the Mg, Si and Fe targets, are dominated by $E2$ or ($M1+ E2$) transitions (see Table~\ref{tab:2}). Illustrative examples of experimental angular distributions for some lines induced by $30$ and $42$~MeV protons are shown in Figure~\ref{fig3}, together with the associated least-squares Legendre polynomial best-fit curves generated according to equation~\eqref{eq. 2}.\qquad

\subsection{Integrated $\gamma-$ray production cross section results}

The results obtained in this work are reported in Figures~\ref{fig4},~\ref{fig5} and~\ref{fig6} where they are compared to previous experimental data from Refs.~\cite{7,10,10b,11,16, 16b} and to the semi-empirical compilation of Murphy \textit{et al.}~\cite{14} when $\sigma$ ($E_{p}$) extrapolations from the latter database are proposed. We have thus determined production cross sections for a total number of $41$ $\gamma-$ray lines produced in proton induced reactions on the $^{nat}$Mg, $^{nat}$Si and $^{56}$Fe targets with $30$, $42$, $54$ and $66$~MeV proton beams, and on the $^{24}$Mg target for $30$ and $66$~MeV protons.\qquad

As pointed out in Ref.~\cite{10,10b}, previous $\gamma-$ray production cross section measurements have been performed for at most three excited states in inelastic particle scattering or in residual nuclei following the removal of one or more nucleons from the target nucleus. The Orsay nuclear astrophysics group in~\cite{10,10b} and~\cite{16, 16b} has extracted a large number of production cross sections for $\gamma$ rays generated in proton and $\alpha-$particle induced reactions on $^{12}$C, $^{16}$O, $^{24}$Mg, $^{nat}$Si and $^{nat}$Fe targets. Their results for incident protons accelerated up to $25$~MeV on the last three targets are plotted together with our experimental data in Figures~\ref{fig4},~\ref{fig5} and~\ref{fig6}. To allow a comparison of the cross section data sets from various experiments, data obtained with targets of different isotopic compositions (see Table~\ref{tab:1}) were normalised: for $^{56}$Fe nucleus, data from this work and Refs.~\cite{7} were normalised to the natural isotopic composition of Fe, and results for $^{24}$Mg from this work as well as Refs.~\cite{7,10,10b} were normalised to the natural abundance of this isotope.\qquad

One expects, in general, that the $\gamma-$ray production cross sections should decrease smoothly as the incident particle energy increases beyond the low-energy region of compound nucleus resonances since several low-energy reaction channels are then successively opened. However, this does not seem to be the case for all observed $\gamma-$ray lines, as can be seen in Figures~\ref{fig4},~\ref{fig5} and~\ref{fig6}. Below, we discuss the obtained $\gamma-$ray production cross section results with concentrating on the main $\gamma$ rays following the decay of the ground-state bands of $^{56}$Fe, $^{28}$Si and $^{24}$Mg isotopes.\qquad

\subsubsection{$\gamma$ rays in proton reactions with Mg}

Production cross sections have been determined for thirteen $\gamma$-ray lines observed in proton induced reactions on the $^{nat}$Mg and $^{24}$Mg targets, i.e., in (p, p$^{\prime}\gamma$) inelastic proton scattering and other binary reactions.\qquad

\begin{figure*}[!htbp]

\includegraphics[scale=0.95]{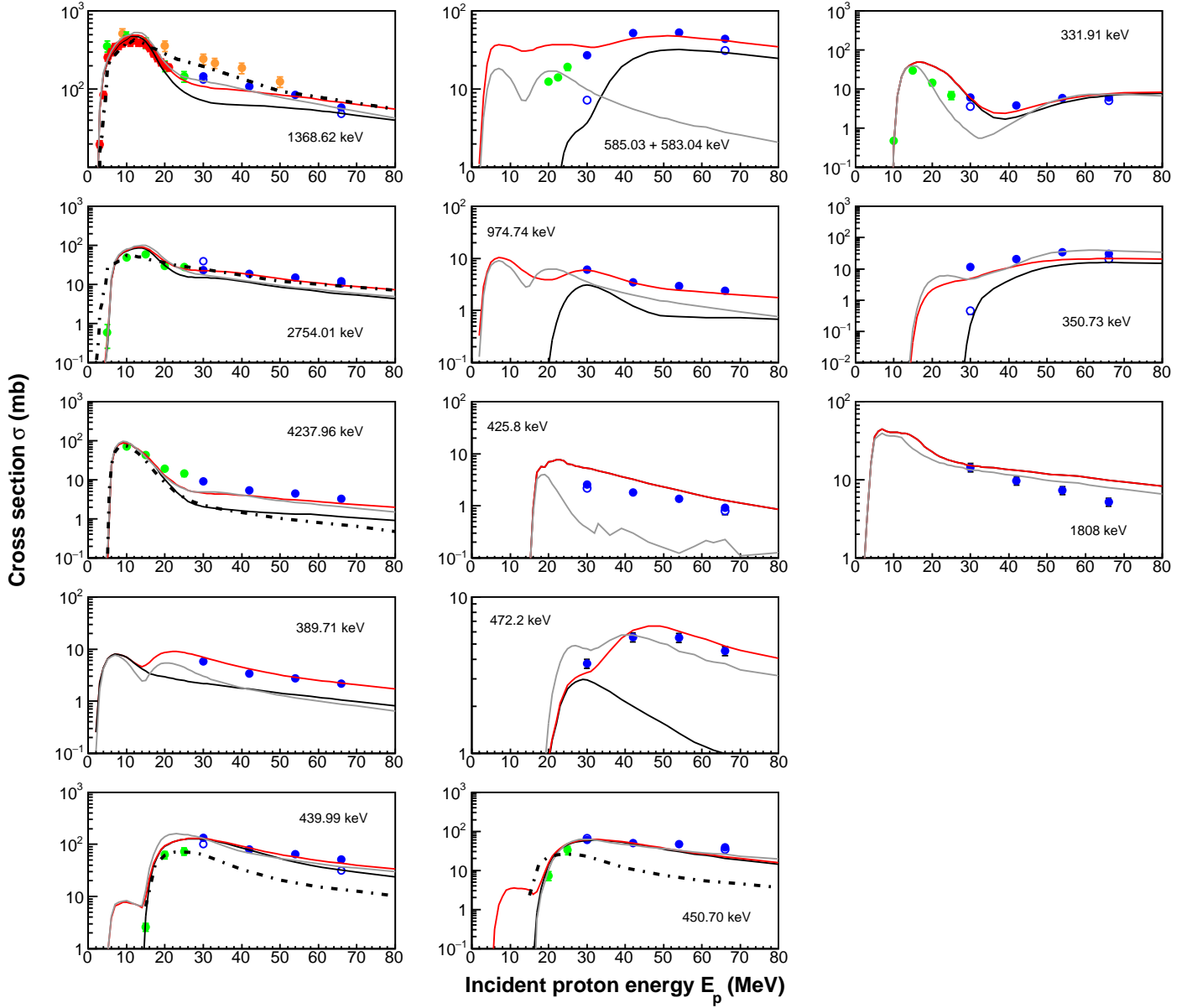}
\caption[c]{\label{fig4} Total production cross sections for $\gamma-$ray lines produced in proton reactions with the $^{24}$Mg (open circles) and $^{nat}$Mg (filled circles) targets (see Table~\ref{tab:2} for more details and Table~\ref{tab:1} for the properties of targets used in this and previous works). The experimental data are shown with full circles: in blue colour (this work), in red (Dyer \textit{et al.}~\cite{7}), in green (Belhout \textit{et al.}~\cite{10,10b}) and in orange (Lesko \textit{et al.}~\cite{11}). The dashed-dotted curves correspond to predictions of the Murpy \textit{et al.} semi-empirical compilation~\cite{14}. The results of TALYS calculations performed with default parameters are shown by grey curves; those performed with our modified OMP and level structure parameters are depicted by black curves for $\gamma-$ray lines emitted by the most abundant isotope, and by red curves for $\gamma-$ray lines from the target of natural composition.}

\end{figure*}

\begin{figure*}[!htbp]

\includegraphics[scale=0.95]{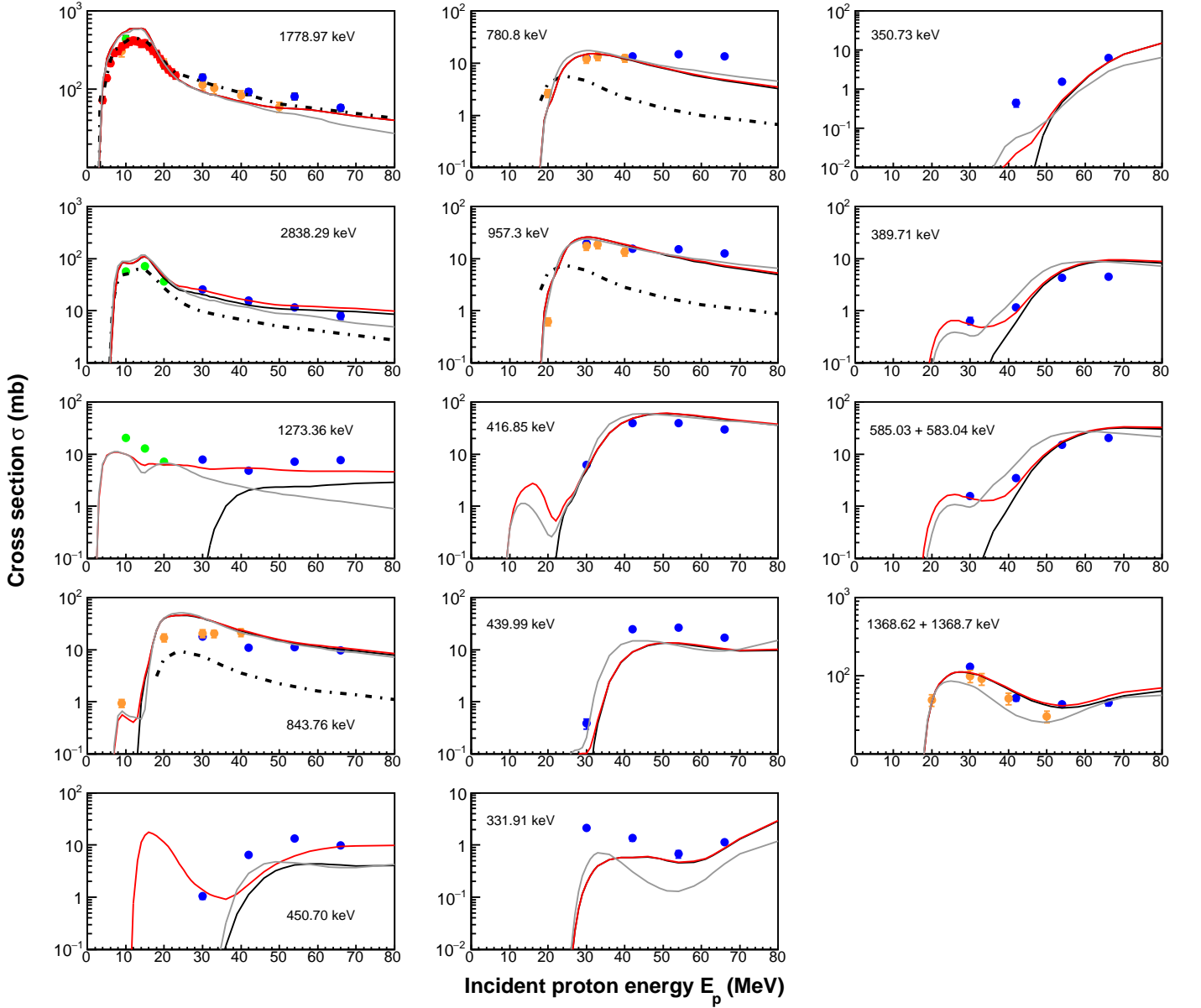}
\caption[c]{\label{fig5} Same as in Figure~\ref{fig4} but for $\gamma-$ray lines produced in proton reactions with the Si targets. The same symbols are used, except that here the green circles represent the experimental data of Benhabiles \textit{et al.}~\cite{16, 16b}.}

\end{figure*}

\begin{figure*}[!htbp]

\includegraphics[scale=0.95]{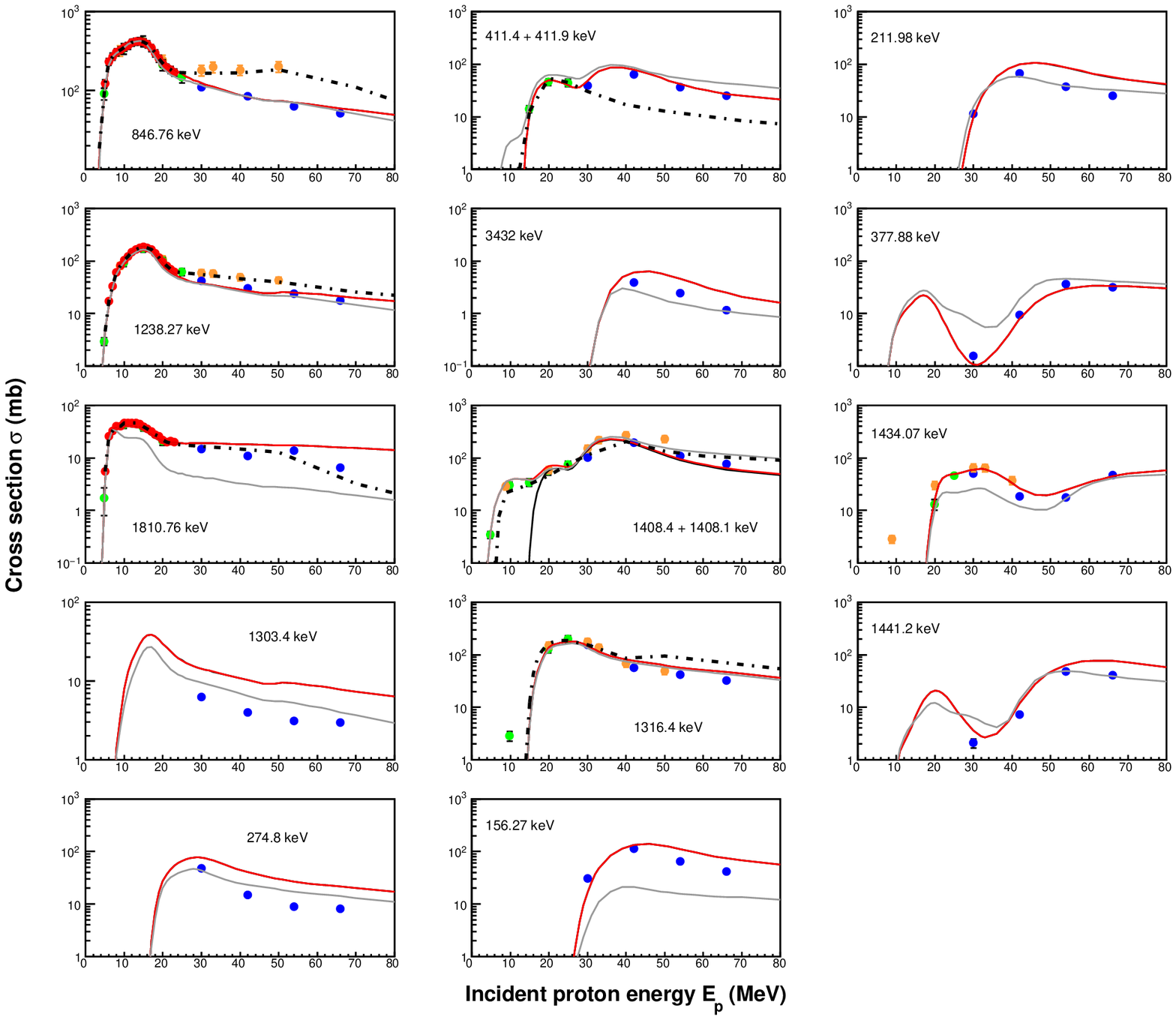}
\caption[c]{\label{fig6} Same as in Figure~\ref{fig4} but for $\gamma-$ray lines produced in proton reactions with the $^{56}$Fe target.}

\end{figure*}

In the $^{nat}$Mg target, $^{24}$Mg is the most abundant isotope in comparison to $^{25}$Mg and $^{26}$Mg (see Table~\ref{tab:1}). The measured production cross sections in the irradiation of the Mg targets are reported in Figure~\ref{fig4}. One observes that the cross section values determined for $30$ and $66$~MeV incident protons on the isotopically enriched $^{24}$Mg target lie slightly below those obtained from the natural Mg target, which is an indication of weak contributions from reactions on the $^{25}$Mg and $^{26}$Mg isotopes leading to the production of $^{24}$Mg. Three $\gamma-$ray lines emitted by $^{24}$Mg were analysed, namely the lines at $E_{\gamma} = 1368.6$~keV (with the possible overlapping line at $E_{\gamma} = 1368.7$~keV emitted by $^{22}$Na), $E_{\gamma} = 2754.01$~keV and $E_{\gamma} = 4237.96$~keV (see Table~\ref{tab:2} for the caracteristics of the transitions). It was not practically possible to extract the peak area for the line at $E_{\gamma} = 4237.96$~keV from the isotopically enriched $^{24}$Mg target due to low statistics and a very broad peak shape (see Figure~\ref{fig1}). Particular care was given to the analysis of these three lines due to their Doppler broadening and, in the case of the line at $4237.96$~keV, to peak splitting. A corresponding line shape calculation showed that, due to the very short lifetime of the decaying nuclear level and the recoil of the emitting nucleus, almost all $\gamma$ rays were emitted in flight. Comparing our $\gamma-$ray production cross section data to previous results from the literature, one observes a smooth extension to higher proton energies the data of Refs.~\cite{7} and Ref.~\cite{10,10b} at $E_{p}<30$~MeV. In contrast, the cross section data reported by Ref.~\cite{11} is significantly higher than our values.\qquad

\FloatBarrier

We provide in this work cross section data for the $\gamma$ rays produced in the de-excitation of the first excited state of $^{25}$Mg (see Figure~\ref{fig4}) at $E_{\gamma}=$ $585.03$~keV, alongside two $\gamma$ rays at $E_{\gamma}=$ $389.71$~keV and $974.74$~keV, resulting from the deexcitation of the second excited state of $^{25}$Mg (see Table~\ref{tab:2}). For the latter two lines cross sections are measured for the first time. The $585.03$~keV line overlaps with another line of very close energy at $E_{\gamma}=$ $585.04$~keV, emitted in the de-excitation process of $^{22}$Na. Considering the fact that the $^{25}$Al compound nucleus decays to the ground state of $^{25}$Mg (with branching ratio $>$ $99\%$), one can safely attribute the observed $E_{\gamma}=$ $585.04$ line from the $^{24}$Mg target to $^{22}$Na. In addition, the absence of lines emitted by the second excited state of $^{25}$Mg from the $^{24}$Mg enriched target corroborates this statement. Previous (unpublished) experimental data for the line at $E_{\gamma}=$ $585.03$~keV from Ref.~\cite{10,10b} for $E_{p}<$ 30~MeV are reported in Figure~\ref{fig4}, and are consistent with our cross section values measured at $E_{p}=30-66$~MeV.\qquad

Production cross sections for the line of $^{26}$Mg at $E_{\gamma}=1808.7$~keV are also reported. There are previous data from an experiment at the Orsay tandem accelerator~\cite{10,10b} for $E_{p}<27$~MeV, that were obtained with a MgO target of natural isotopic composition on a thin Al foil. But the $1808.7$~keV line produced in this experiment contained an important contribution from reactions with the Al backing~\cite{10b}, and the corresponding cross section data are therefore not reported in Figure~\ref{fig4}.\qquad

Three lines at $E_{\gamma}=331.91$, $439.99$ and $450.71$~keV, emitted respectively by the $^{21}$Na, $^{23}$Na and $^{23}$Mg isotopes (see Table~\ref{tab:2}) and for which previous measurements have been reported in Ref.~\cite{10,10b}, were analysed as well. As can be seen in Figure~\ref{fig4}, our cross section data for these lines are in overall agreement with the data from Ref.~\cite{10,10b} over the right hand side of the apparent bump (broad compound resonance) at $E_{p}\sim20$~MeV, although the trend of our data goes slightly higher than theirs.\qquad

Three other lines at $E_{\gamma}=$ $425.8$, $472.20$ and $350.73$~keV for which no experimental cross section data are available in the literature, were observed and analysed (see Figures~\ref{fig1} and~\ref{fig4}), for which no experimental cross section data are avaible in the previous literature. They can be attributed to the de-excitations of the first excited states of $^{24}$Al, $^{24}$Na and $^{21}$Ne, respectively. Concerning the $472.20$~keV line, no cross section measurement was performed with the $^{24}$Mg enriched target since the production of $^{24}$Na from this target is not energetically possible. The measured cross section data from the $^{nat}$Mg target come from the reactions induced by protons on the $^{25}$Mg and $^{26}$Mg isotopes. The $425.8$~keV line of $^{24}$Al shows a small contribution from reactions induced on the less abundant isotopes of Mg. Finally, one can easily see that the production cross section for the $350.73$~keV line of $^{21}$Ne is dominated (to an order of magnitude) at $E_{p}=30$~MeV by contributions from reactions on $^{25+26}$Mg isotopes, while for $66$~MeV incident protons the contribution from the $^{24}$Mg isotope dominates.

\subsubsection{$\gamma$ rays in proton reactions with Si}

A total of fourteen $\gamma-$ray lines have been observed in the irradiations of the $^{nat}$Si target for which we have measured the production cross sections reported in Figure~\ref{fig5}. Most of these lines were produced in binary reactions on the $^{28,29,30}$Si isotopes (see Table~\ref{tab:2}).\qquad

The two main lines emitted in the deexcitation of $^{28}$Si are at $1778.97$ and $2838.29$~keV. Experimental cross section data have been reported for the $1778.97$~keV line by Refs.~\cite{7},~\cite{16, 16b} and~\cite{11}. As can be seen in Figure~\ref{fig5}, our data are consistent with lower proton energy data from Refs.~\cite{7,16, 16b} but their values are slightly higher than the data of Ref.~\cite{11} in proportions of $23\%$ at $E_{p}=30$~MeV and $36\%$ at $E_{p}=54$~MeV. For the $2838.29$~keV line, the only available previous data are those reported by Ref.~\cite{16, 16b} that seem to be consistently extended to higher proton energies by our data. Production cross sections were also measured for the line at $E_{\gamma}=1273.4$~keV emitted in the deexcitation of the first excited state of $^{29}$Si.\qquad

We have also performed production cross section measurements for $\gamma-$ray lines emitted in the deexcitation of the $^{27}$Si, $^{27}$Al and $^{24}$Mg nuclei (see Table~\ref{tab:2}), for which only Ref.~\cite{11} have reported cross section data. As can be seen in Figure~\ref{fig5}, an overall agreement between the two cross section data sets for these lines is observed. Indeed, our cross section values for the two lines of $^{27}$Si at $E_{\gamma}=780.8$~keV and $957.3$~keV are consistent with the data of these authors~\cite{11}. For the line of $^{27}$Al at $843.76$~keV, one observes a very good agreement between our values and the data of Ref.~\cite{11} around $E_{p}$ $=30$~MeV, while at higher proton energies the cross sections measured by these authors seem to be higher, evolving with different energy dependence than our data. A $\gamma-$ray line at $E_{\gamma}=1368.62$~keV corresponding to $^{24}$Mg was observed. As already mentioned in the case of the Mg target, another line of very close energy, $E_{\gamma}=1368.7$~keV (see Table~\ref{tab:2}), emitted by $^{22}$Na very likely overlaps with this line. Our experimental cross section data for this doublet is consistent with those reported by Ref.~\cite{11}, with moderate differences of $\sim30\%$ and $\sim40\%$ for proton energies $E_{p}=30$ and $50$~MeV, respectively.\qquad

Finally, six $\gamma-$ray lines for which no previous cross section data are available in the literature, to our knowledge, were analysed. These are the lines at $E_{\gamma}=$ $416.85$~keV of $^{26}$Al, $389.71$~keV of $^{25}$Mg, $450.70$~keV of $^{23}$Mg, $439.99$~keV of $^{23}$Na and the two lines at $331.91$~keV and $350.73$~keV both coming from the deexcitation of $^{21}$Na and $^{21}$Ne, respectively (see Table~\ref{tab:2}). The measured cross sections are shown in Figure~\ref{fig5}.\qquad

\subsubsection{$\gamma$ rays in proton reactions with Fe}

The iron target used in the experiment was highly enriched in $^{56}$Fe, to better than 99\%. Production cross sections have been determined for fourteen $\gamma-$ray lines emitted from this target. The obtained experimental results are reported in Figure~\ref{fig6}. Previous cross section experimental data, for the known seven lines at $E_{\gamma}=411.9$, $846.78$, $1238.27$, $1316.4$, $1408.4$, $1434.07$ and $1810.76$~keV, have been reported in Refs.~\cite{7,10,10b,11} and are also plotted in Figure~\ref{fig6} for comparison. In contrast, cross sections for the remaining seven lines at $E_{\gamma}=156.27$, $211.98$, $274.8$, $377.88$, $1303.4$, $1441.2$ and $3432.0$~keV (see Table~\ref{tab:2}) are measured in this work for the first time, to our knowledge. One can notice the conservation of the inelastic character (decreasing cross section with increasing proton energy) for all reported $\gamma-$ray lines with the exception of the lines at $E_{\gamma}=411.9$, $1408.4$ and $1434.07$~keV for which a second bump appears at around $E_{p}=42$~MeV.\qquad

Among the first group of known lines, the three main intense lines at $E_{\gamma}=846.78$ ($2_{1}^{+}$ to $0_{1}^{+}$), $1238.27$ ($4_{1}^{+}$ to $2_{1}^{+}$) and $1810.76$~keV ($2_{2}^{+}$ to $2_{1}^{+}$) are produced in the $^{56}$Fe(p, p$^{\prime}\gamma$) reaction following the deexcitation of the first three excited states of $^{56}$Fe. For the first time, cross sections are measured for the line at $1303.4$~keV ($6_{1}^{+}$ to $4_{1}^{+}$, see Table~\ref{tab:2}) which sheds light on the $\gamma$-ray production cross section for the three first levels of the $g.s.$ band of $^{56}$Fe. The $846.78$~keV line needed a careful analysis because it lies on a neutron background (see Section~\ref{sec2}). In addition, the presence of two small peaks at $E_{\gamma}=843.76$~keV (Al background) and $E_{\gamma}\sim848$~keV (possibly due to the deexcitation of the $5^{+}_{1}$ excited state of $^{52}$Cr to its $4^{+}_{2}$ excited state) had to be considered assuming Gaussian shapes. Similarly, for the fit of the $1238.27$~keV $\gamma$-ray line, a small peak at $E_{\gamma}=1241.7$~keV (presumably associated with the deexcitation of the fourth excited state of the $^{53}$Mn isotope located at $E_{x}=1620.12$~keV) was added. The areas of the peaks were extracted by considering the aforementioned fits with the RADWARE package~\cite{30}. As can be seen in Figure~\ref{fig6}, our cross section data for these lines appear to be consistent with data measured at lower energy in Refs.~\cite{7, 10,10b}. In the case of the $E_{\gamma}=846.78$~keV and $1238.27$~keV lines there are marked differences between our experimental data and those of Ref.~\cite{11}, the latter ones being higher, by $\sim50\%$ to $\sim190\%$ and by $\sim30\%$ up to $\sim70\%$ at $E_{p}=30$~MeV and $54$~MeV, respectively. For the $E_{\gamma}=1810.76$~keV line, our data are fairly consistent with the previous measurements of Refs.~\cite{7, 10}.\qquad

Lines from the $^{54,55}$Fe and $^{53,54}$Mn isotopes have also been observed in $\gamma-$ray energy spectra with the iron target, and the production cross sections have been established.\qquad

Gamma-ray production cross sections have been determined for lines in $^{54,55}$Fe produced in binary reactions on the $^{56}$Fe target, namely for the $E_{\gamma}= 1316.4$~keV line and for the first time for the $274.8$~keV line from $^{55}$Fe (see Table~\ref{tab:2}), and for the $E_{\gamma}= 3432.0$~keV ($8_{1}^{+}$ to $6_{1}^{+}$) line from $^{54}$Fe (see Table~\ref{tab:2}). In addition, cross sections were measured for two doublet lines at $E_{\gamma}=1408.4$~keV and $411.9$~keV, both from $^{54+55}$Fe (see Table~\ref{tab:2}). The $\gamma$ ray at $1316.4$~keV was affected by a bump structure in the low-energy part, assigned to a line of $E_{\gamma}=1312.2$~keV, possibly coming from the deexcitation of a state of $^{56}$Fe (see Table~\ref{tab:2}). Since no shape separation was possible due to the Doppler broadening of both lines, their areas were considered as one. Finally, our cross section data seem to describe a resonance structure which has a maximum at $E_{p}\sim40$~MeV for the doublet at $E_{\gamma}=411.9$~keV. Apart from the newly measured production cross sections, our measurements show a continuity with the data of Refs.~\cite{10,10b, 11}. However, discrepancies are noticed between our data and the data of Ref.~\cite{11} for the line at $E_{\gamma}= 1408$~keV prominent at $E_{p}=54$~MeV ($\sim93\%$).\qquad

Gamma-ray production cross sections for four lines produced in the deexcitations of the $^{53,54}$Mn isotopes, for which no cross section measurements have been carried out previously to our knowledge, are presented (see Table~\ref{tab:2}). These are $E_{\gamma}= 156.27$~keV and $211.98$~keV from $^{54}$Mn, and $E_{\gamma}=377.88$~keV and $1441.2$~keV from $^{53}$Mn.\qquad

Finally, a line at $E_{\gamma}=1434.07$~keV produced in the deexcitation of the first excited state of $^{52}$Cr has been observed and analysed. Our experimental cross sectiondata for this line seem to be consistent with the data of Refs.~\cite{10,10b, 11}.\qquad

\subsection{ Comparison to the Murphy \textit{et al.} compilation} 

In this subsection our measured $\gamma-$ray line cross sections are compared to the Murphy \textit{et al.} database~\cite{14}, which includes cross section data on about 140 intense $\gamma-$ray lines, produced in the interaction of protons and $\alpha-$particles with abundant nuclei in astrophysical sites. \qquad

The nuclear reaction codes TALYS~\cite{15} and EMPIRE~\cite{19} have been used previously to calculate prompt and delayed $\gamma-$ray production cross sections. These results were compared to experimental data~\cite{10,10b,16, 16b} for several strong lines, emitted in proton and $\alpha -$particle induced reactions on various stable and radioactive nuclei synthesised in SFs~\cite{35}. \qquad

Murphy \textit{et al.}~\cite{14} updated earlier databases for (p, p$^{\prime}$), ($\alpha$, $\alpha^{\prime}$), (p, x) and ($\alpha$, x) reactions and extended the existing low-energy data up to several hundred MeV/nucleon by using energy dependences obtained in TALYS calculations. When available, the excitation function curves derived by Ref.~\cite{14} are also reported in Figures~\ref{fig4},~\ref{fig5}, and~\ref{fig6} (dashed-dotted lines).\qquad

These figures show that for the majority of lines the excitation curves from Ref.~\cite{14} have similar qualitative trends as our experimental data. However, they overestimate our cross sections when data from Ref.~\cite{11} are available, i.e. for $E_{\gamma}= 1368.62$~keV (Mg) and $E_{\gamma}= 846.78$, $1238.27$ and $1316.4$~keV (Fe); and underestimate them when data from Ref.~\cite{11} are lacking, i.e. for $E_{\gamma}= 4237.96$, $439.99$ and $450.70$~keV (Mg), $E_{\gamma}= 2838.29$, $843.76$, $780.8$ and $957.3$~keV (Si) and $E_{\gamma}= 411.4+411.9$~keV (Fe). In the remaining cases a very good agreement is observed.\qquad

\section{\label{sec5} Comparison to TALYS code predictions} 

The comparison of experimental cross sections to those predicted by nuclear reaction models using modern computer codes is crucial. The TALYS code~\cite{15} allows the calculation of theoretical cross sections for nuclear reactions induced by a variety of projectiles ($\gamma$, n, p, d, t, $^{3}$He and $\alpha$) on atomic nuclei over the energy range of $E_{lab}$ $=1$~keV up to $250$~MeV, with contributions from the main nuclear reaction mechanisms (compound nucleus, direct reactions, pre-equilibrium, etc.) using built-in parameter values from either phenomenological or microscopic models. It comprises libraries of nuclear data like masses, level densities, discrete states, OMP, level deformation parameters, etc. Alternatively, one is allowed to introduce modified nuclear data derived from an analysis of experimental data.\qquad

Gamma-ray production cross sections were initially calculated using default parameters in TALYS~\cite{15}. While the calculations reproduced the energy dependence of the measured cross sections reasonably well for most of the $\gamma-$ray lines, they deviated significantly in some cases. In this section, we will discuss the modifications brought to the TALYS source code, as well as the adjustement of OMP and nuclear level deformation parameters (from nucleon angular distribution analyses, see the \hyperlink{thesentence}{Appendix}) utilising the coupled-channels reactions code OPTMAN~\cite{36}. The optimised OMP parameters used in the TALYS calculations are presented in Table~\ref{tab:3} while the procedure for obtaining these is described in the \hyperlink{thesentence}{Appendix}. Additional adjustments of nuclear level density (LD) parameters and the $^{54}$Fe coupling scheme are also presented and described below. The results of the calculations were compared with the experimental data from this work and Refs.~\cite{7,10,11,16, 16b}.\qquad

\begin{figure}[!htbp]
\includegraphics[scale=0.45]{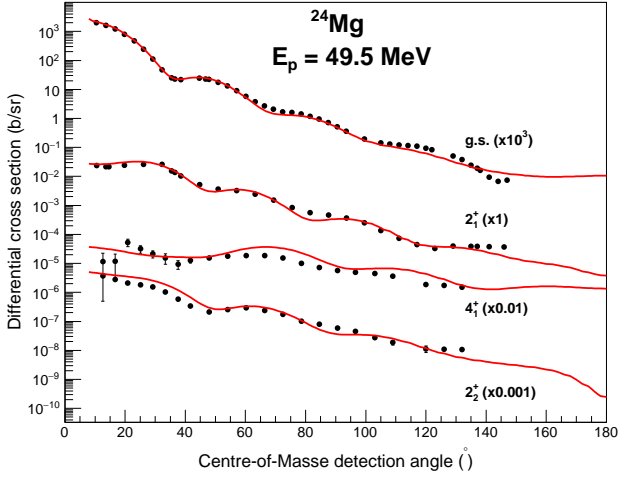}
\caption {\label{fig7} Illustrative examples of results from our theoretical analyses of experimental angular distribution cross sectiondata for elastically scattered protons off $^{24}$Mg taken from the literature and considered in the OPTMAN code~\cite{36} for extracting new OMP and nuclear level deformation parameters used in our TALYS code calculations of $\gamma-$ray line production cross sections.}
\end{figure}

\begin{figure}[!htbp]
\includegraphics[scale=0.45]{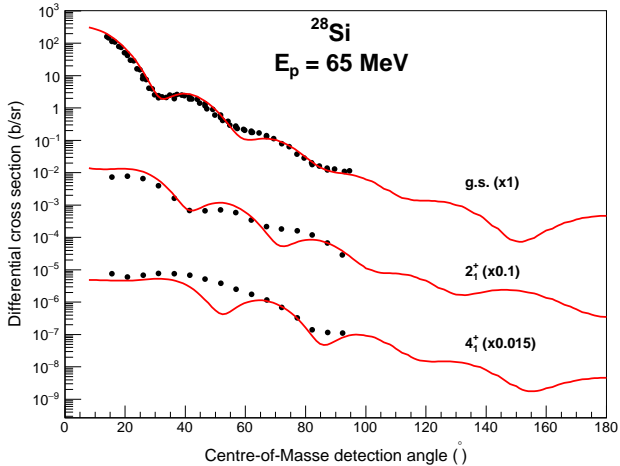}
\caption {\label{fig8} Same as in Figure~\ref{fig7} but for scattered protons off the $^{28}$Si target}
\end{figure}

\begin{figure}[!htbp]
\includegraphics[scale=0.45]{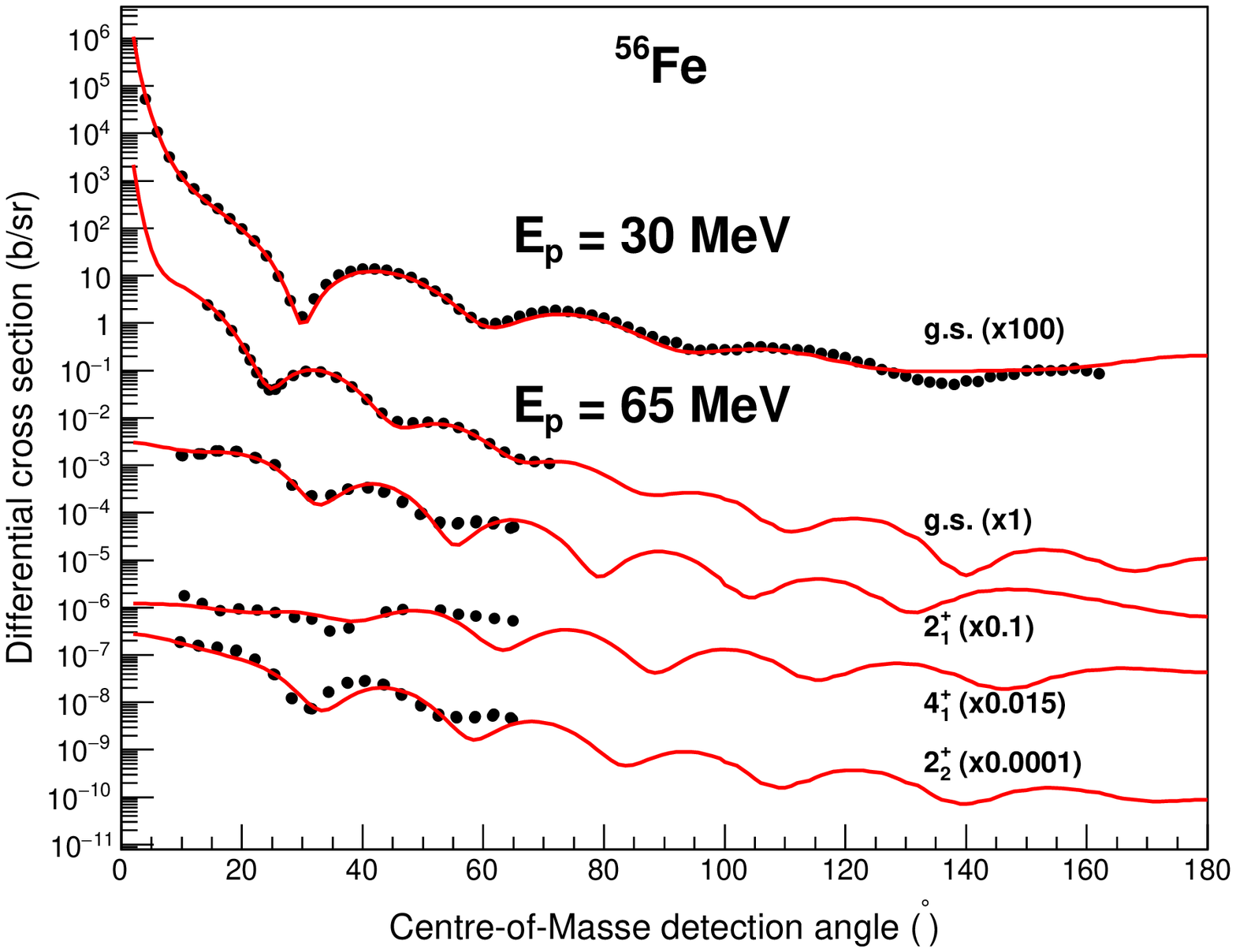}
\caption {\label{fig9} Same as in Figure~\ref{fig7} for scattered protons off the $^{56}$Fe target.}
\end{figure}

Over two hundred elastic and inelastic scattering angular distributions of protons and neutrons on $^{24,25,26}$Mg, $^{28,29,30}$Si and $^{54,56}$Fe have been analysed in order to extract OMP and deformation parameters. A nucleon OMP allows to better adjust the values for the Coulomb correction, radius, and diffusivity. Theoretical fits to the experimental data sets have been systematically performed using the coupled-channels reactions code OPTMAN. Examples of elastic and inelastic proton scattering angular distribution adjustments for $^{24}$Mg, $^{28}$Si and $^{56}$Fe are presented in Figures~\ref{fig7},~\ref{fig8}, and~\ref{fig9}. Analysing power studies are out of the scope of this paper.


\begin{table}[!htbp]

\begin{tabular}[c]{lccccc}%
\hline
 \hline
            &  &   &   &   &    \\

Nuclei & {$^{24,26}$Mg} & {$^{25}$Mg} &  $^{28,30}$Si & $^{29}$Si & $^{54,56}$Fe  \\
 Models & DF & ASR & DF & ASR & DF  \\
\hline
\hline

Parameters  &  &   &   &   &    \\
            &  &   &   &   &    \\

\textit{V$_{R}$} (MeV) & {50.29} & {50.29} & {51.17} & {51.17} & 50.84  \\
\textit{$\lambda${$_{R}$}} (MeV$^{-1}$) & {0.00530} & {0.00530} & {0.00557} &  {0.00557} & 0.00517  \\
\textit{W$_{S}$} (MeV) & {7.38} &  {7.38} & {8.00} & {8.00} & 7.32  \\
\textit{WID$_{S}$} & {9.96} & {9.96} & {11.88} & {11.88} & 11.56  \\
\textit{$\lambda${$_{S}$}} (MeV$^{-1}$) & {0.00335} & {0.00335} & {0.00334} & {0.00334} & 0.00203 \\
\textit{W$_{V}$} (MeV) & {3.08} & {3.08} & {2.14} & {2.14} & 7.01  \\
\textit{WID$_{D}$ } & {106.77} & {106.77} & {192.89 }& {192.89} & 91.58  \\
\textit{V$_{SO}$} (MeV) & {7.69} & {7.69} & {7.96} & {7.96} & 7.04  \\
\textit{$\lambda${$_{SO}$}} (MeV$^{-1}$) & {0.00235} & {0.00235} & {0.00350} & {0.00350} & 0.00119  \\
\textit{W$_{SO}$} (MeV) & {-4.19} & {-4.19} & {-4.54} & {-4.54} & -3.57  \\
\textit{WID$_{SO}$} & {223.80} & {223.80} & {332.02} & {332.02} & 208.57  \\
\textit{n}          & 4 & 4 & 4 & 4 & 4 \\
\textit{r$_{R}$/r$_{V}$}(fm) & 1.184 & 1.201   & 1.187 & 1.160    & 1.183 \\
\textit{r$_{S}$} (fm)        & 1.070 & 1.097   & 1.206 & 1.007    & 1.077 \\
\textit{r$_{SO}$} (fm)       & {1.078} &  {1.078} & {1.123} & {1.123} & 1.194 \\
\textit{r$_{C}$} (fm) & {1.294} & {1.294} & {1.234} & {1.234} & 1.375  \\
\textit{a$_{R}$/a$_{V}$} (fm)& 0.562 & 0.675   & 0.522 & 0.666    & 0.543  \\
\textit{a$_{S}$} (fm)        & 0.764 & 0.641   & 0.592 & 0.746    & 0.758 \\
\textit{a$_{SO}$} (fm)&  {0.815} & {0.815} & {0.757}  & {0.757}  & 0.686 \\
\textit{a$_{C}$} (fm) & {0.257} &  {0.257} & {0.179} &  {0.179} & 0.241 \\
$C_{Coul}$ & {0.74}  & {0.74} &  {0.68}   &  {0.68}   & 1.22 \\
$C_{viso}$ & {4.03}  & {4.03}  &  {46.62}  & {46.62}    & 0.82  \\
$C_{wiso}$ & {17.48} & {17.48} &  {1.49 }  &  {1.49 }   & 0.96  \\

\hline
\hline
\end{tabular}
\caption{\label{tab:3} Optimised OMP parameters used in the TALYS code calculations, extracted from the analysis of elastic nucleon scattering angular distribution data sets.
The DF label refers to the Davydov-Filipov Model, while the ASR label stands for the axially symmetric rigid rotor model.\qquad}
\end{table}

\squeezetable
\begin{table}[!htbp]
\begin{tabular}[c]{ccccccccc}
 \hline
 \hline
  & & & & & & & & \\
              & $^{24}$Mg & $^{25}$Mg & $^{26}$Mg & $^{28}$Si & $^{29}$Si &  $^{30}$Si & $^{54}$Fe  & $^{56}$Fe   \\
 \hline
  & & & & & & & & \\

$\beta_{2}$   & 0.578  & 0.554 (0.454)   & 0.499 & -0.419 & -0.332 & 0.310 & 0.174 & 0.237  \\
$\beta_{4}$   & -0.037 &         & 0.182 & 0.108  &        &       &-0.056 & 0.001  \\
$\gamma$ (deg)& 21.89  &         & 27    & 35.69  &        & 20.85 & 26.10 & 21.75  \\
 \hline
 \hline
\end{tabular}
\caption{\label{tab:4} Quadrupole and hexadecapole deformations, and $\gamma$-asymmetry obtained from the coupled-channels analysis of the inelastically scattered neutrons and protons off the nuclei listed in Table~\ref{tab:2}. The value between parentheses for $^{25}$Mg corresponds to the quadrupole deformation of the $K^{\pi}=\frac{1}{2}^{+}$ side band with a band head at $E = 585.045$~keV.}
\end{table}

\begin{center}
\begin{table*}
\begin{tabular}
[c]{cccccccccccc}%
\hline\hline
Nucleus & $a$ & $E_{sh}$ & $\tilde{a}$ & $P_{shift}$ & Reference & Nucleus &
$a$ & $E_{sh}$ & $\tilde{a}$ & $P_{shift}$ & Reference\\
\hline
$^{21}$Ne &  &  & 1.7866 &  &~\cite{38} & $^{21}$Na &  &  & 1.8449 &  &~\cite{38}\\
$^{22}$Ne &  &  & 2.3886 &  &~\cite{38} & $^{22}$Na &  &  & 1.7003 & 0.4180 &~\cite{38}\\
$^{23}$Na &  &  & 1.9302 &  &~\cite{38} & $^{23}$Mg &  &  & 1.9922 &  &~\cite{38}\\
$^{22}$Mg &  &  & 2.5089 &  &~\cite{38} & $^{24}$Mg &  &  & 1.8731 & 0.4160 &~\cite{38}\\
$^{25}$Mg & 1.9659 & -8.500 &  &  &~\cite{37} & $^{26}$Mg & 2.3630 & 0.671 &
&  &~\cite{37}\\
$^{24}$Al &  &  & 2.1026 & 0.3902 &~\cite{38} & $^{25}$Al &  &  & 2.1472 &  &~\cite{38}\\
$^{28}$Si &  &  & 2.2116 & 0.1300 &~\cite{38} & $^{29}$Si & 2.2116 & -2.4670 &
&  &~\cite{37}\\
$^{30}$Si & 1.7383 & -0.930 &  &  &~\cite{37} & $^{29}$P &  &  & 2.4756 &  &~\cite{38}\\
$^{30}$P &  &  & 2.4079 & 0.3564 &~\cite{38} & $^{31}$P &  &  & 2.5719 &  &~\cite{38}\\
\hline\hline
\end{tabular}
\caption{\label{tab:5} List of parameters used in the GSM.}

\end{table*}

\end{center}

As mentioned above, much better agreements between our experimental $\gamma-$ray line production cross section data and theoretical values calculated by means of the TALYS code were obtained when using the built-in generalised superfluid model (GSM) together with experimental level density parameters, instead of the default constant temperature + Fermi gas model (CT+FGM). For the nuclei appearing in the outgoing reaction channels, available experimental data of the level density, $a$, and values of the shell correction, $E_{sh}$, were taken from the RIPL-3 database~\cite{37}. In the case of nuclei for which no experimental data are available, values of the asymptotic level density, $\tilde{a}$, and the pairing energy shift, $P_{shift}$, were derived from the systematics given in Ref.~\cite{38}. All the parameters used in the GSM model are reported in Table~\ref{tab:5} below.\qquad

Since the spin-orbit potential is treated as spherical in OPTMAN but deformed in TALYS, the TALYS source code was modified accordingly. Furthermore, TALYS considers the Coulomb potential to be non-diffuse, contrary to that in OPTMAN. Consequently, Coulomb diffusivity had to be included in the TALYS source code. A final modification, for full compatibility, involves the Fermi energy. Values derived from neutron and proton separation energies for each studied nucleus were adopted instead of using the values calculated by TALYS (see TALYS manual).\qquad

Finally, for $^{54}$Fe we did not use the coupling scheme given in the RIPL-3 library~\cite{37}, where the first member of the K$^{\pi}=2^{+}$ $\gamma-$band is considered to be the level at $E_{x} = 2959.0$~keV ($2_{2}^{+}$), while the levels at $E_{x} = 2561.3$~keV ($0_{2} ^{+}$) and $3166.0$~keV ($2_{3}^{+}$) are regarded as the first two members of a $\beta$-band. The OPTMAN fit of the inelastic scattering data following this level scheme was not successful. A much better fit was obtained by considering the levels at $E_{x} = 3166.0$~keV and $2561.3$~keV as the first members of the $\gamma-$band, while the level at $2959.0$~keV was excluded since the $2959.0$~keV and $2949.2$~keV states ($6_{1}^{+}$) are not resolved in the experimental data.\qquad

The total $\gamma-$ray production cross sections obtained in our calculations, using our modified OMP and level density parameters in GSM, are plotted as solid curves in Figures~\ref{fig4},~\ref{fig5} and~\ref{fig6} together with the experimental data from the present and previous works~\cite{7,10,10b,11,16, 16b}. Apart from some Mg and Si lines, a good agreement between experiment and theory is achieved, with a deviation of at most $\sim50\%$.\qquad

\subsection{Reactions with magnesium}

In Figure~\ref{fig4}, the solid black curves represent the calculated cross sections for the $\gamma-$ray lines produced on $^{24}$Mg, while the solid red curves depict the sum of the contributions from all the isotopes in the $^{nat}$Mg target. For the lines from $^{25,26}$Mg, the curves represent the calculated cross sections from reactions on these isotopes in the $^{nat}$Mg target.\qquad

As can be seen in Figure~\ref{fig4}, the calculated cross sections employing our modified OMP and deformation parameters exhibit remarkably good agreement with our experimental data, both in absolute value and in energy dependence, over the whole explored energy range. This is particularly the case for the lines of $^{24}$Mg ($E_{\gamma}=1368.6$, $2754.01$ and especially $4237.96$~keV), $^{21}$Na ($E_{\gamma} = 331.91$~keV), $^{23}$Na ($E_{\gamma}=439.99$~keV) and $^{23}$Mg isotope ($E_{\gamma}=450.71$~keV). However, both the default and our adjusted OMP fail to reproduce the cross sections of the line at $E_{\gamma}=1808.7$~keV ($^{26}$Mg).\qquad

Furthermore, for the lines from the deexcitation of $^{25}$Mg ($E_{\gamma}=585.03$~keV, and the two new lines at $389.71$ and $974.74$~keV), the calculated cross sections using the TALYS default parameters lie significantly below those calculated with our modified parameters and below our experimental data. One can note the important contribution from $^{26}$Mg to the $^{25}$Mg lines at $E_{\gamma}=389.71$ and $974.74$~keV, as well as the noteworthy contribution of the $^{24}$Mg(p,x)$^{22}$Na reaction to the line at $E_{\gamma}=585.03$~keV.\qquad

Regarding the new lines observed at $E_{\gamma}=350.73$~keV ($^{21}$Ne), $425.8$~keV ($^{24}$Al) and $472.20$~keV ($^{24}$Na) (see Table~\ref{tab:2}), there is a very good agreement between the experimental data and the calculations done with our modified parameters while for the $425.8$~keV line the TALYS code with default parameters fails considerably.\qquad

\subsection{Reactions with silicon}

The excitation functions for the two main lines of $^{28}$Si at $E_{\gamma}=1778.97$ and $2838.29$~keV calculated using our modified input parameters and the TALYS default ones are in excellent agreement, as shown in Figure~\ref{fig5}, both fitting the experimental data sets very well.\qquad

For the line of $^{29}$Si at $E_{\gamma}=1273.4$~keV, the excitation curve generated by TALYS with default parameters deviates significantly from the experimental data for energies larger than 30~MeV while the calculations with our modified parameters describe very well our experimental data and are in reasonable agreement with previous data measured at lower proton energies~\cite{16, 16b}.\qquad

Concerning the two $\gamma-$ray lines of $^{27}$Si at $E_{\gamma}=780.8$ and $957.3$~keV, and the lines of $^{27}$Al and $^{24}$Mg at $E_{\gamma}=843.76$ and $1368.62$~keV, respectively, the calculated excitation functions using our modified input parameters are in overall agreement with those derived using the TALYS code default parameters, and account well in absolute values for all the experimental data sets.\qquad

In the case of the six new lines from the Si target, at $E_{\gamma}=331.91$ and $350.73$~keV ($^{21}$Na), $389.71$~keV ($^{25}$Mg), $416.85$~keV ($^{26}$Al), $439.99$~keV ($^{23}$Na) and $450.70$~keV ($^{23}$Mg) the calculated excitation functions using our modified parameters appear to account reasonably well for our experimental cross section data, both regarding the absolute values and the energy dependence over the explored proton energy range. The calculations performed with the TALYS default parameters show somewhat worse agreement with the experimental data sets. \qquad

Notice also that the bumps associated with compound resonances are predicted well by our calculations using modified parameters, see for instance the lines at $E_{\gamma}=450.7$, $1368.62$ and $331.91$~keV.\qquad

\subsection{Reactions with iron}

As can be seen in Figure~\ref{fig6}, the calculations using our modified OMP and nuclear level deformation parameters yield theoretical excitation functions that describe very well the experimental data sets for the majority of $\gamma-$ray lines generated in proton reactions with the Fe target.\qquad

Excellent agreements between theoretical values and experimental data are in particular observed over the whole proton energy range, $E_{p}=5-66$~MeV, in the cases of the two main lines of $^{56}$Fe at $E_{\gamma}=846.78$ and $1238.27$~keV produced in the $^{56}$Fe(p, p$^{\prime}\gamma$) reaction. The calculated $\gamma-$ray production cross sections for these two lines agree very well both at high energy (our data) and at low-energy (data from~\cite{7, 10,10b}). Note the discrepancy between our experimental data and those reported by Ref.~\cite{11}. Our experimental cross sectiondata for the line at $E_{\gamma}=1810.76$~keV, produced in the deexcitation of the band head of the $K^{\pi}=2^{+}$ $\gamma$-band, are slightly overestimated by the calculated values using our modified input parameters, see Figure~\ref{fig6}, while the calculations performed with default parameters, considerably underestimate the data. An excellent agreement between nuclear reaction theory and experiment is also observed in the case of the lines at energies, $E_{\gamma}=411.9$, $1408.4$, $1316.4$, $156.27$, $377.88$ and $1441.2$~keV.\qquad

The observed change in trend near 30~MeV observed in the experimental data of Ref.~\cite{10,10b} and our data for the line at $E_{\gamma} =$ $411.9$~keV is due to the overlapping of two distinct lines of the same energy, as mentioned in Table~\ref{tab:2}. In this case, performed calculations indicate that the component from $^{55}$Fe dominates below the proton energy of $30$~MeV, while the contribution from $^{54}$Fe is noticeable only above it. For the line at $1408.4$~keV, calculations show that for incident protons with $E_{p}\gtrsim15$~MeV, reactions of the types $^{56}$Fe(p, n) and $^{56}$Fe(p, pn) mostly dominate. In the case of the $^{nat}$Fe target, a small contribution from the $^{54}$Fe(p, p$^{\prime}$) reaction is predicted for proton energies $E_{p}$ $\leq15$~MeV, which explains the observed experimental cross sections of Ref.~\cite{10,10b} in this energy range. This contribution is also observed in the case of the line at $E_{\gamma}=1316.4$~keV. \qquad

For the lines at $274.8$, $1303.4$ and $3432.0$~keV the calculated cross sections using our modified input parameters overestimate our experimental data, especially for the second line, while calculations using the default input parameters of TALYS yield better agreement. \qquad

Thus, in the case of proton reactions with the Fe target, both calculations yield generally good agreement with the experimental data sets in both absolute value and energy dependence for most observed $\gamma-$ray lines, as can be seen in Figure~\ref{fig6}. In contrast, for proton induced reactions on the Mg and Si targets, substantially improved agreement between TALYS calculated cross sections and experimental data can only be obtained by using modified OMP and nuclear level structure parameters.

\section{\label{sec6} Calculation of the total nuclear $\gamma-$ray fluxes in interactions of LECRs in the inner Galaxy}


\begin{figure*}[!htbp]

 \includegraphics[scale=0.95]{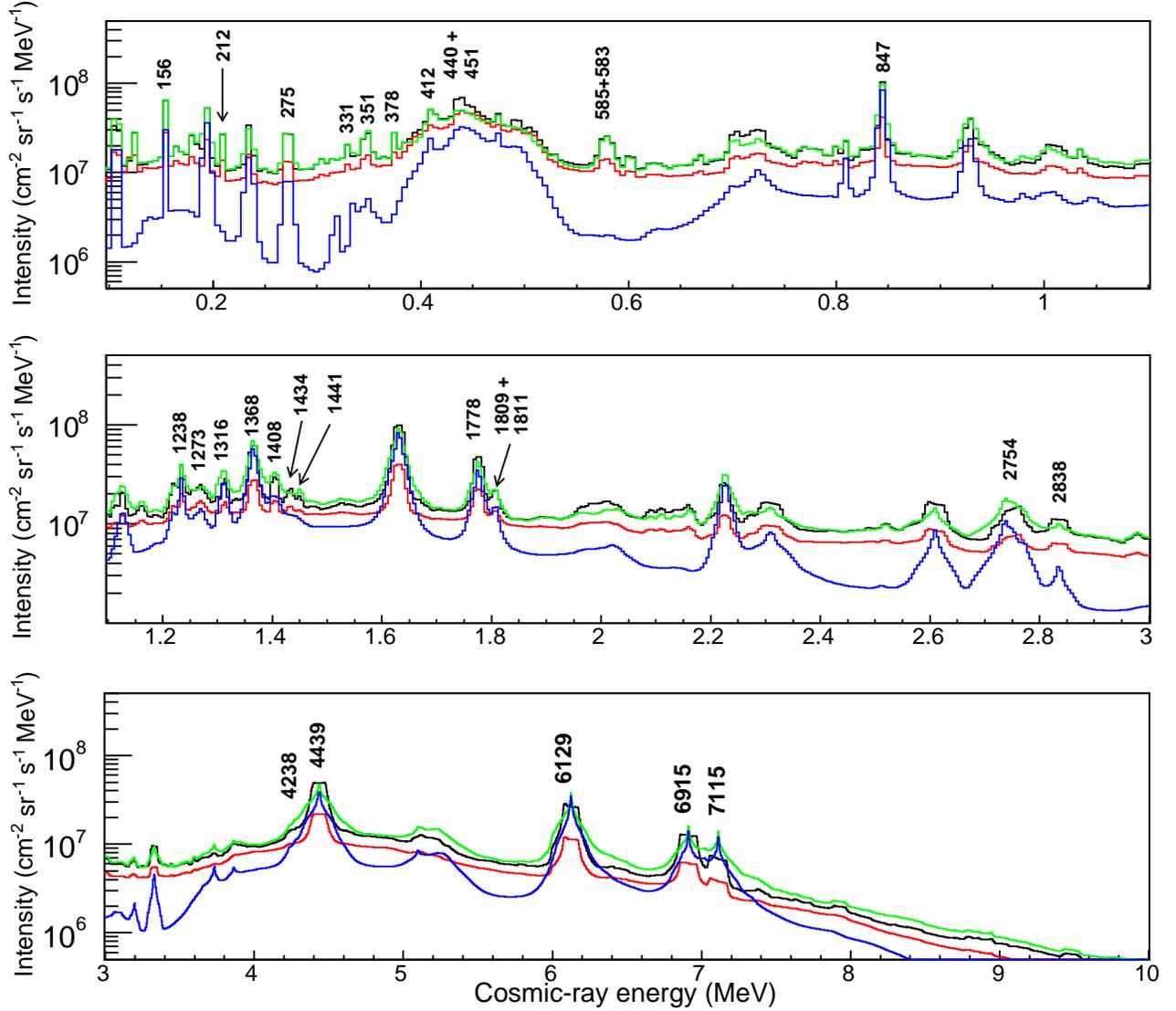}

\caption[c]{\label{fig10} Calculated nuclear $\gamma$-ray line emission spectra for LECRs produced by shock acceleration with the parameter $E_c=60$~MeV [26] interacting with interstellar matter in the inner Galaxy. The energies of $\gamma$-ray lines analysed in this work are indicated by vertical labels in units of~keV. The four spectra are calculated for various assumptions for the cross section data and the metallicity of the ambient medium. Black: cross section data derived with our modified parameters in the TALYS code and $M=3$, i.e. three-times solar metallicity; red: same, but $M=1$; green: cross section data derived with the default parameters of TALYS and $M=3$; blue: cross section data from the compilation of Murphy et al.~\cite{14} and $M=3$. 
}
\end{figure*}

The understanding of nuclear processes at work in astrophysical sites requires a good knowledge of $\gamma-$ray line production cross sections over a wide energy range of the accelerated particles, extending from nuclear reaction thresholds up to several hundreds~MeV per nucleon. However, only a limited number of experimental cross section data sets generated in proton and $\alpha-$particle interactions with abundant heavier nuclei in SFs and the ISM was available in the literature about two decades ago. They concerned the strongest lines produced in these reactions, but many of the data sets were limited to energies below $\sim25$~MeV for proton and $\alpha$-particle interactions (see e.g.~\cite{13}). Therefore, nuclear reaction code calculations were needed for estimating their values at higher particle energies, to model the $\gamma-$ray line emissions from these astophysical sites. In the latest cross section compilation, Murphy \textit{et al.}~\cite{14} used the nuclear reaction code TALYS for this purpose. \qquad

Murphy \textit{et al.}~\cite{14} also made for the first time extensive nuclear reaction calculations to assess the quasi-continuum $\gamma-$ray line emission with photon energies, $E_{\gamma} \sim 0.1-10$~MeV, composed of a large number of weaker lines, in the complete absence of corresponding experimental cross section data. This weak-line component is dominated by interactions with abundant nuclei, where the reaction products have an important number of excited states below the particle emission thresholds. In SFs and ISM the most important interactions are proton and $\alpha$-particle reactions with $^{20}$Ne, $^{24}$Mg, $^{28}$Si and $^{56}$Fe. This quasi-continuum component is in both astrophysical sites merged with another important quasi-continuum component of overlapping broad lines from interactions of accelerated heavier ions with ambient H and He.\qquad

An important, still unresolved and presently debated question concerns the energy spectrum of galactic LECRs (of energies below about $1$ GeV per nucleon) that were early assumed to be responsible for the nucleosynthesis of light elements in our Galaxy (the LiBeB problem)~\cite{39, 39b}. This component of cosmic rays of energy density similar to that of the interstellar photon and magnetic fields is assumed to play a crucial role in the dynamics and the chemical evolution of the Galaxy, including the processes of cosmic ray transport and star formation~\cite{40, 40b}.\qquad

Evidence of their existence rely only on indirect observations of marked ionisation rates in diffuse interstellar molecular clouds~\cite{42,43}, such as those inferred from the abundances of the H$_{3}^{+}$ molecular ion, and on the approximately linear increase of the Be abundance with metallicity~\cite{23}. It is expected that the interactions of the LECRs with the ISM matter in the inner Galaxy give rise to an important nuclear $\gamma$-ray line emission, whose intensity exceeds considerably the emission due to standard CRs (see e.g.~\cite{24}). In particular, the prominent nuclear lines emitted in the deexcitation of the first few excited states of most abundant nuclei, ($^{12}$C, $^{14}$N, $^{16}$O, $^{20}$Ne, $^{24}$Mg, $^{28}$Si and $^{56}$Fe~\cite{12}) have cross section maxima at LECR energies and are therefore strongly produced.\qquad

In this context, we have performed calculations of nuclear $\gamma-$ray line emission spectra ($\gamma-$ray emission fluxes versus the photon energy, $E_{\gamma}$) based on the present $\gamma-$ray production experimental cross sectiondata, that have been satisfactorily accounted for by nuclear reaction models with introducing in TALYS our own OMP, nuclear level structure and level density parameters reported in the previous section. We have also calculated $\gamma-$ray line emission spectra from the inner Galaxy assuming different values for the unknown metallicity of this medium, taken to be about twice the metallicity of the Sun.\qquad

For the energy spectra and the composition of the low-energy component of cosmic rays, we use source spectra from shock acceleration with an energy cutoff $E_c$ that are propagated with a simple leaky-box model, the so-called SA-LECR in Benhabiles \textit{et al.}~\cite{24}. The obtained results for SA-LECR with energy cutoff $E_c=60$~MeV are reported in Figure~\ref{fig10} showing, in particular, dominating contributions of main $\gamma-$ray lines emitted in the deexcitation of first low-lying states of the studied Mg, Si and Fe nuclei (see table~\ref{tab:2} and Figures~\ref{fig4},~\ref{fig5} and~\ref{fig6}), and of the strong lines at $E_{\gamma}=4439$~keV (from $^{12}$C) and $E_{\gamma}=6129$~keV (from $^{16}$O). \qquad

Differences between calculations based on the present cross sections and those from the TALYS calculations with default parameters (black and green curves in Figure~\ref{fig10}) can be seen for some of the moderately strong lines and in the quasi-continuum at higher energies, reaching or exceeding 30\%. The nuclear $\gamma$-ray line emission for lower metallicity (red line) shows, as expected, a decrease in the prominent narrow line fluxes, while the underlying quasi-continuum component is less affected, the broad-line component not being dependent on the metallicity. Finally, the blue line shows the component from the strong lines of the compilation of Murphy \textit{et al.}~\cite{14}, without the weak-line component. Strong differences in some energy ranges illustrate the importance of this latter emission component.\qquad

Space telescopes of improved technology with higher energy resolution and better sensitivity are now projected for near future satellite missions, like e-ASTROGAM (and All-Sky ASTROGAM) or COSI~\cite{48}. The possible observation of the predicted low-energy nuclear $\gamma-$ray line emission spectrum with $E_{\gamma}\sim0.1-10$~MeV should provide the most compelling signature of the LECRs interactions within the inner Galaxy and may help to elucidate the puzzling composition and energy spectrum of the latter component of cosmic rays. With the present new data and the improvements obtained in TALYS calculations, more reliable and accurate predictions are available for comparison with eventual future observations.\qquad

\section{\label{sec7} Summary and conclusion}

In the present work, we report experimental cross section excitation functions for 41 $\gamma$-ray lines produced in interactions of 30, 42, 54 and 66~MeV proton beams delivered by the SSC facility of iThemba LABS with Mg, Si and Fe, that are abundant nuclei in the SFs and ISM astrophysical sites. While cross section data for half of these lines are reported for the first time, the other half consists of lines resulting from the deexcitation of low-lying excited nuclear states for which data at lower energies already exist. Our experimental cross section data for these lines are found to be fairly consistent with previous data sets measured at the Washington~\cite{7} and Orsay~\cite{10,10b, 16, 16b} tandem accelerators for $E_p<27$~MeV, that are thus extended to higher proton energies. The observed small differences are likely due to variations in the experimental conditions, such as the properties of the targets used (see Table~\ref{tab:1}). In two cases, however, our experimental cross sections for known lines are significantly lower than previous data measured at the LBL cyclotron facility for proton energies up to 50~MeV~\cite{11}.\qquad

We have also compared our experimental data to the semi-empirical compilation of Murphy \textit{et al.}~\cite{14} in the case of lines for which previously measured values have been extrapolated to higher proton energies on the basis of TALYS code  calculations. The current experimental results thus improve this existing unique database for nuclear $\gamma$-ray production cross sections and allow for a more reliable extrapolation to higher energies.\qquad

Our experimental cross section data have also been compared with the predictions of nuclear reaction models via TALYS code calculations. First, the integral cross sections calculated using the built-in default OMP and nuclear level structure parameters of TALYS showed to be in overall agreement in terms of energy dependence, but exhibited noticeable differences in absolute values with experimental data for most observed $\gamma$-ray lines. Appreciable improvements have been obtained by introducing modified OMP parameters, as well as our coupling scheme and level density parameters of nuclear collective levels as input data in the TALYS code. These parameters were determined via our theoretical analysis of a large number of experimental nucleon elastic/inelastic scattering differential cross section data sets available in the literature using the coupled- channels code OPTMAN (see the \hyperlink{thesentence}{Appendix}). We have obtained substantially improved agreements between the calculated integral cross sections and the corresponding experimental data mainly for the Mg and Si targets, while the experimental results for the Fe targets appeared to be also satisfactorily accounted for by TALYS calculation using the default input parameters.\qquad

Experimental nuclear $\gamma$-ray line production cross sections are of great importance in
various scientific research fields and practical applications, notably in nuclear physics and nuclear astrophysics where they are crucially needed for diverse purposes, e.g. :\qquad

(i) Testing/improving the ability of reaction codes~\cite{15, 19} to predict~\cite{14} the cross sections for nuclear reactions where experimental data are lacking.\qquad

(ii) Modeling, analysing and interpreting nuclear processes, such as the $\gamma$-ray line emissions from astrophysical sites like solar flares and the interactions of cosmic rays in the ISM, and in particular the low-energy cosmic-ray component thought to be responsible for high ionisation rates in diffuse clouds towards the inner Galaxy.\qquad

On the basis of the present and previous~\cite{7, 10,10b, 16, 16b} experimental $\gamma$-ray line production cross section data and results of the TALYS code, we have performed calculations of $\gamma$-ray line emission spectra over the photon energy range, $E_{\gamma}=0.1-10$~MeV, expected to be generated in the interactions of the LECRs in the inner Galaxy. The accuracy of predictions for the $\gamma$-ray line emission in solar flares will likewise profit from the present extension of the reaction cross section data to higher energies. The obtained results should allow reliable comparisons (see Ref.~\cite{24}) with observational data from new generation space telescopes of higher energy resolution and better sensitivity. The combined progress may lead to accurate determinations of the accelerated particle populations and the interaction medium in solar flares and the presently unknown properties of the LECRs within the galactic Center could then be considerably constrained.\newline

\section*{ACKNOWLEDGMENTS}
The authors are indebted to the technical staff of the iThemba LABS SSC accelerator for their kind help and friendly cooperation. This work has been carried out in the framework of a joint scientific cooperation agreement between the USTHB university of Algiers and iThemba LABS of Cape Town. The work was partially supported by the General Direction of Scientific Research and Technological Development of Algeria (project code A/AS-2013-003), and  by the National Research Foundation of South Africa under grants GUN: 109134 and UID87454. Besides, travel support was granted to the collaborating French researchers by the CSNSM and the IPN of Orsay (CNRS/IN2P3 and University of Paris-Sud). Thanks are due to all persons from these institutions who helped in the realisation of this project. W. Yahia-Cherif would like to adress particular thanks to Dr. P. Adsley (iThemba LABS), Dr. B. M. Rebeiro (IPN Lyon) and Mr. K. C. W. Li (Stellenbosch University) for all their help and support in this work.

\hypertarget{thesentence}{\section*{ Appendix: determination of the OMP and collective level deformation parameters}}

The potential as used in the OPTMAN code~\cite{36} is expressed by:

\begin{equation}
  \begin{array} {l c l}
   
 V(E,r) & = & -V_{R}(E)f_{R}(r,r_{R}) \\
    {}    & {}  & -[ {\Delta}V_{V}(E) + iW_{V}(E) ]f_{V}(r,r_{R}) \\
    {}    & {}  & -[ {\Delta}V_{D}(E) - 4ia_{D}W_{D}(E) ]\dfrac{d}{dr}f_{D}(r,r_{D}) \\
    {}    & {}  & + (\dfrac{\hbar}{m_{\pi}c})^2[ {\Delta}V_{SO}(E) + V_{SO}(E) + iW_{SO}(E) ] \\
    {}    & {}  &  \times \dfrac{1}{r}\dfrac{d}{dr}f_{SO}(r,r_{SO})(\overrightarrow{\sigma}{\cdot}{\overrightarrow{L}}) \\
    {}    & {}  & + V_{Coul}(r,r_{c}), \\
  
\end{array}
\label{eq. 3} 
\end{equation}

\noindent where the real central potential is given by:

\begin{equation}
\begin{array} [c]{lcl}
V_{R}(E) & = & (V_{R}^{disp}+(-1)^{Z^{\prime}+1}C_{viso}(\tfrac
{N-2Z}{A}))e^{(-\lambda_{R}(E-E_{F}))}\\
  &  & + C_{coul}\tfrac{ZZ^{\prime}}{A^{1/3}}(\lambda_{R}V_{R}^{disp}e^{(-\lambda_{R}(E-E_{F}))}),
\end{array}
\label{eq. 4}
\end{equation}

\noindent and the imaginary surface and volume potentials are taken as~\cite{49, 50}:

\begin{equation}
\begin{array}[c]{lcl}
W_{V}(E) & = & (W_{V}^{disp}+(-1)^{Z^{\prime}+1}C_{wviso}
(\tfrac{N-2Z}{A}))\\
 & & \times \frac{(E-E_{F})^{2}}{(E-E_{F})^{2}+WID_{V}^{2}}\\
W_{D}(E) & = & (W_{D}^{disp}+(-1)^{Z^{\prime}+1}C_{wdiso}(\tfrac{N-2Z}{A})\\
 & & \times\frac{(E-E_{F})^{2}}{(E-E_{F})^{2}+WID_{D}^{2}}
e^{(-\lambda_{D}(E-E_{F}))},
\end{array}
\label{eq. 5}
\end{equation}

\noindent where $Z$ and $Z^{ \prime}$ are, respectively, the atomic numbers of the
target nucleus and the projectile.

The real and imaginary spin-orbit potentials were taken under the standard
forms of Koning and Delaroche~\cite{49}:

\begin{equation}
\begin{array}[c]{lcl}
V_{SO}(E) & = & V_{SO}^{disp}e^{(-\lambda_{SO}(E-E_{F}))}\\
W_{SO}(E) & = & W_{SO}^{disp}\frac{(E-E_{F})^{2}}{(E-E_{F}
)^{2}+WID_{SO}^{2}}.
\end{array}
\label{eq. 6}
\end{equation}

The positive quantities $C_{viso}$, $C_{wviso}$ and $C_{wdiso}$ in the above expressions are the constants of the isospin terms, $C_{coul}$ is the Coulomb correction constant and $E_{F}$ is the Fermi energy for neutrons and protons. The dynamic terms, $\Delta_{i}(E)$ (with index $i=V, S, SO$), are dispersive components calculated from the following integral~\cite{Quesada}: \ \ \ 

\begin{equation}
\begin{array}[c]{lcl}
\Delta V(r,E) & = & \dfrac{P}{\pi}\mathlarger{\int_{-\infty}^{+\infty}}\dfrac
{W(r,E^{\prime})}{E^{\prime}-E}dE^{\prime}.
\end{array}
\label{eq. 7}
\end{equation}

The adjustment of the OMP parameters was made in the framework of the CC formalism. For even-A nuclei, the Davydov-Filipov (DF) model~\cite{51} accounting for the $\gamma$ deformation of the nuclei was used in order to describe the collective states. The value of the $\gamma$ deformation was calculated according to the DF approach~\cite{51} using the ratio:

\begin{equation}
\begin{array}
[c]{lcl}%
R_{22} & = & \dfrac{E_{2_{2}^{+}}}{E_{2_{1}^{+}}}=\dfrac{3+\sqrt
{9-8sin^{2}(3\gamma)}}{3-\sqrt{9-8sin^{2}(3\gamma)}}
\end{array}
, \label{eq. 8}
\end{equation}

\noindent where $E_{2_{1}^{+}}$ and $E_{2_{2}^{+}}$ are, respectively, the energies of the first and second $2^{+}$ excited states, taken from the available experimental level schemes. The $\gamma$ value calculated from this ratio varies from 30\degree for $R_{22}=2$ down to $0\degree$ for $R_{22}$ = $\infty$. For $^{28}$Si, however, the second $2^{+}$ state is located at $E_{x}$ $=7380.59$~keV above the first $3^{+}$ state at $E_{x}=6276.20$~keV. This unnatural parity state is the second state of the $\gamma-$vibrational band. The $2^{+}$ state at $7380.59$~keV is a better candidate to form with the $0^{+}_2$ state at $E_{x}$ $=4979.92$~keV the first two members of a $\beta$-vibrational band~\cite{50}. Thus, to calculate the value of $\gamma$ for this nucleus, we have used the ratio

\begin{equation}
\begin{array}[c]{lcl}
R_{32} & = & \dfrac{E_{3_{1}^{+}}}{E_{2_{1}^{+}}}=\dfrac{18}{3-\sqrt
{9-8sin^{2}(3\gamma)}},
\end{array}
\label{eq. 9}
\end{equation}

\noindent where $E_{3_{1}^{+}}$ is the first $3^{+}$ state at $E_{x}= 6276.20$~keV. The $^{26}$Mg nucleus is another exception. For this nucleus, one has $R_{22}<2$. The value of $\gamma$ can be determined, instead, from the ratio of the reduced transition probbabilities:

\begin{equation}
\begin{array}[c]{lcl}
R_{b} & = & \dfrac{B(E2:2_{2}^{+}\rightarrow2_{1}^{+})}{{B}(E2:2_{2}^{+}\rightarrow0_{1}^{+})},
\end{array}
\label{eq. 10}
\end{equation}

\noindent in terms of the $B(E2)$ reduced electric transition probabilities, expressed (see~\cite{52} and references therein) as:

\begin{equation}
\begin{array}[c]{lcl}
B(E2:2_{2}^{+}\rightarrow2_{1}^{+}) & = & \dfrac{10}{7}\left(\dfrac{e^{2}Q_{0}^{2}
}{16\pi}\right)\dfrac{sin^{2}(3\gamma)}{9-8sin^{2}(3\gamma)}\\
B(E2:2_{2}^{+}\rightarrow0_{1}^{+}) & = & \dfrac{1}{2}\left(\dfrac{e^{2}Q_{0}^{2}
}{16\pi}\right)\left(1-\dfrac{3-2sin^{2}(3\gamma)}{\sqrt{9-8sin^{2}(3\gamma)}}\right)
\end{array}
\label{eq. 11}
\end{equation}

\noindent where $Q_{0}$ is the electric quadrupole moment. Glatz~\cite{53}, Alons~\cite{54} and Dybdal~\cite{55} have reported experimental reduced transition probabilities for the $2_{2}^{+}$ $\rightarrow$ $2_{1}^{+}$ and $2_{2}^{+}$ $\rightarrow$ $0_{1}^{+}$ ($g.s.$) transitions. Average $B(E2)$ values, weighted by the experimental errors estimated by these authors, have been calculated for these transitions and are, respectively: $B(E2:2_{2}^{+}\rightarrow 2_{1}^{+}) = 27.78 (619)$ $e^{2}fm^{4}$ and $B(E2: 2_{2}^{+} \rightarrow 0_{1}^{+}) =1.83 (16)$ $e^{2}fm^{4}$. Thus, the value of the $\gamma$ deformation parameter for $^{26}$Mg, calculated by equation~\eqref{eq. 11}, is $\gamma=27.0(1)\degree$.\qquad

For odd-A nuclei like $^{25}$Mg and $^{29}$Si, we have used the Axially-Symmetric Rigid Rotor (ASR) model~\cite{56}. In this case, the radii and diffusivities of the imaginary surface and volume potentials have been readjusted but with assuming the same parameters for the potential depths, as in the case of the even-A isotopes.\qquad

The derived results (potential depths, geometrical parameters, $\beta$ and $\gamma$ deformations) are reported in Section~\ref{sec5} (Tables~\ref{tab:3} and~\ref{tab:4}, respectively).\qquad

The validity of our OMP for each isotopic chain ranges from $E_p=0-250$~MeV for $^{24}$Mg, $E_p=0-200$~MeV for $^{28}$Si and $E_p=0-160$~MeV for $^{56}$Fe. The proton and neutron  experimental angular distribution cross sectiondata can be found in Refs~\cite{59, 60, 78, 79, 80, 81, 82, 83, 84, 85, 86, 87, 88, 89, 90, 91, 64, 91, 92} for $^{24,25,26}$Mg, Refs.~\cite{59, 61, 62, 63, 64, 65, 66, 67, 68, 69, 93, 79, 91, 94, 95, 96, 118, 97, 84, 98, 85, 86, 99, 100, 101, 93, 89, 102, 103} for $^{28,29,30}$Si and Refs.~\cite{70, 71, 72, 73, 74, 75, 77, 104, 105, 106, 107, 108, 109, 110, 111, 112, 113, 114, 115, 79, 116, 117, 118, 119, 120, 121, 122, 123, 124, 86, 125, 126, 127} for $^{54,56}$Fe. In addition, some data was taken from Ref.~\cite{57}: A. Virdis, Th. Schweitzer and W.N. Wang for Mg; A. Virdis, G. Boerker, S. Kliczewski, M. Baba and Y.Yamanouti for Si; I.A. Korzh, W.E. Kinney, Th. Schweitzer and R. Varner for Fe.

\bibliographystyle{apsrev4-1}

 \bibliography{PRC-Teleftaios}

\begin{thebibliography}{127}%
\makeatletter
\providecommand \@ifxundefined [1]{%
 \@ifx{#1\undefined}
}%
\providecommand \@ifnum [1]{%
 \ifnum #1\expandafter \@firstoftwo
 \else \expandafter \@secondoftwo
 \fi
}%
\providecommand \@ifx [1]{%
 \ifx #1\expandafter \@firstoftwo
 \else \expandafter \@secondoftwo
 \fi
}%
\providecommand \natexlab [1]{#1}%
\providecommand \enquote  [1]{``#1''}%
\providecommand \bibnamefont  [1]{#1}%
\providecommand \bibfnamefont [1]{#1}%
\providecommand \citenamefont [1]{#1}%
\providecommand \href@noop [0]{\@secondoftwo}%
\providecommand \href [0]{\begingroup \@sanitize@url \@href}%
\providecommand \@href[1]{\@@startlink{#1}\@@href}%
\providecommand \@@href[1]{\endgroup#1\@@endlink}%
\providecommand \@sanitize@url [0]{\catcode `\\12\catcode `\$12\catcode
  `\&12\catcode `\#12\catcode `\^12\catcode `\_12\catcode `\%12\relax}%
\providecommand \@@startlink[1]{}%
\providecommand \@@endlink[0]{}%
\providecommand \url  [0]{\begingroup\@sanitize@url \@url }%
\providecommand \@url [1]{\endgroup\@href {#1}{\urlprefix }}%
\providecommand \urlprefix  [0]{URL }%
\providecommand \Eprint [0]{\href }%
\providecommand \doibase [0]{http://dx.doi.org/}%
\providecommand \selectlanguage [0]{\@gobble}%
\providecommand \bibinfo  [0]{\@secondoftwo}%
\providecommand \bibfield  [0]{\@secondoftwo}%
\providecommand \translation [1]{[#1]}%
\providecommand \BibitemOpen [0]{}%
\providecommand \bibitemStop [0]{}%
\providecommand \bibitemNoStop [0]{.\EOS\space}%
\providecommand \EOS [0]{\spacefactor3000\relax}%
\providecommand \BibitemShut  [1]{\csname bibitem#1\endcsname}%
\let\auto@bib@innerbib\@empty
\bibitem [{\citenamefont {{Ramana Murthy}}\ and\ \citenamefont
  {{Wolfendale}}(1993)}]{1}%
  \BibitemOpen
  \bibfield  {author} {\bibinfo {author} {\bibfnamefont {P.~V.}\ \bibnamefont
  {{Ramana Murthy}}}\ and\ \bibinfo {author} {\bibfnamefont {A.~W.}\
  \bibnamefont {{Wolfendale}}},\ }\href@noop {} {\emph {\bibinfo {title}
  {Gamma-ray astronomy}}},\ Vol.~\bibinfo {volume} {22}\ (\bibinfo  {publisher}
  {Cambridge Astrophysics Series},\ \bibinfo {year} {1993})\BibitemShut
  {NoStop}%
\bibitem [{\citenamefont {{Cheng}}\ and\ \citenamefont {{Romero}}(2004)}]{2}%
  \BibitemOpen
  \bibfield  {author} {\bibinfo {author} {\bibfnamefont {K.~S.}\ \bibnamefont
  {{Cheng}}}\ and\ \bibinfo {author} {\bibfnamefont {G.~E.}\ \bibnamefont
  {{Romero}}},\ }\href@noop {} {\emph {\bibinfo {title} {Cosmic Gamma-Ray
  Sources}}},\ \bibinfo {edition} {1st}\ ed.,\ Astrophysics and Space Science
  Library 304\ (\bibinfo  {publisher} {Springer Netherlands},\ \bibinfo {year}
  {2004})\BibitemShut {NoStop}%
\bibitem [{\citenamefont {Vestrand}\ \emph {et~al.}(1999)\citenamefont
  {Vestrand}, \citenamefont {Share}, \citenamefont {Murphy}, \citenamefont
  {Forrest}, \citenamefont {Rieger}, \citenamefont {Chupp},\ and\ \citenamefont
  {Kanbach}}]{3}%
  \BibitemOpen
  \bibfield  {author} {\bibinfo {author} {\bibfnamefont {W.~T.}\ \bibnamefont
  {Vestrand}}, \bibinfo {author} {\bibfnamefont {G.~H.}\ \bibnamefont {Share}},
  \bibinfo {author} {\bibfnamefont {R.~J.}\ \bibnamefont {Murphy}}, \bibinfo
  {author} {\bibfnamefont {D.~J.}\ \bibnamefont {Forrest}}, \bibinfo {author}
  {\bibfnamefont {E.}~\bibnamefont {Rieger}}, \bibinfo {author} {\bibfnamefont
  {E.~L.}\ \bibnamefont {Chupp}}, \ and\ \bibinfo {author} {\bibfnamefont
  {G.}~\bibnamefont {Kanbach}},\ }\href {\doibase 10.1086/313180} {\bibfield
  {journal} {\bibinfo  {journal} {Astrophys. J. Supp. Ser.}\ }\textbf {\bibinfo
  {volume} {120}},\ \bibinfo {pages} {409} (\bibinfo {year}
  {1999})}\BibitemShut {NoStop}%
\bibitem [{\citenamefont {Kiener}(2019)}]{4}%
  \BibitemOpen
  \bibfield  {author} {\bibinfo {author} {\bibfnamefont {J.}~\bibnamefont
  {Kiener}},\ }\href {\doibase 10.1103/PhysRevC.99.014605} {\bibfield
  {journal} {\bibinfo  {journal} {Phys. Rev. C}\ }\textbf {\bibinfo {volume}
  {99}},\ \bibinfo {pages} {014605} (\bibinfo {year} {2019})}\BibitemShut
  {NoStop}%
\bibitem [{\citenamefont {Cassé}\ \emph {et~al.}(1995)\citenamefont {Cassé},
  \citenamefont {Lehoucq},\ and\ \citenamefont {Vangloni-Flam}}]{5}%
  \BibitemOpen
  \bibfield  {author} {\bibinfo {author} {\bibfnamefont {M.}~\bibnamefont
  {Cassé}}, \bibinfo {author} {\bibfnamefont {R.}~\bibnamefont {Lehoucq}}, \
  and\ \bibinfo {author} {\bibfnamefont {E.}~\bibnamefont {Vangloni-Flam}},\
  }\href {https://doi.org/10.1038/373318a0} {\bibfield  {journal} {\bibinfo
  {journal} {Nature}\ }\textbf {\bibinfo {volume} {373}},\ \bibinfo {pages}
  {318–319} (\bibinfo {year} {1995})}\BibitemShut {NoStop}%
\bibitem [{\citenamefont {Parizot}\ and\ \citenamefont {Drury}(1999)}]{5b}%
  \BibitemOpen
  \bibfield  {author} {\bibinfo {author} {\bibfnamefont {E.}~\bibnamefont
  {Parizot}}\ and\ \bibinfo {author} {\bibfnamefont {L.}~\bibnamefont
  {Drury}},\ }\href {http://aa.springer.de/papers/9349002/2300673.pdf}
  {\bibfield  {journal} {\bibinfo  {journal} {Astron. Astrophys.}\ ,\ \bibinfo
  {pages} {673–384}} (\bibinfo {year} {1999})}\BibitemShut {NoStop}%
\bibitem [{\citenamefont {Lang}\ \emph {et~al.}(1987)\citenamefont {Lang},
  \citenamefont {Werntz}, \citenamefont {Crannell}, \citenamefont {Trombka},\
  and\ \citenamefont {Chang}}]{6}%
  \BibitemOpen
  \bibfield  {author} {\bibinfo {author} {\bibfnamefont {F.~L.}\ \bibnamefont
  {Lang}}, \bibinfo {author} {\bibfnamefont {C.~W.}\ \bibnamefont {Werntz}},
  \bibinfo {author} {\bibfnamefont {C.~J.}\ \bibnamefont {Crannell}}, \bibinfo
  {author} {\bibfnamefont {J.~I.}\ \bibnamefont {Trombka}}, \ and\ \bibinfo
  {author} {\bibfnamefont {C.~C.}\ \bibnamefont {Chang}},\ }\href {\doibase
  10.1103/PhysRevC.35.1214} {\bibfield  {journal} {\bibinfo  {journal} {Phys.
  Rev. C}\ }\textbf {\bibinfo {volume} {35}},\ \bibinfo {pages} {1214}
  (\bibinfo {year} {1987})}\BibitemShut {NoStop}%
\bibitem [{\citenamefont {Dyer}\ \emph {et~al.}(1981)\citenamefont {Dyer},
  \citenamefont {Bodansky}, \citenamefont {Seamster}, \citenamefont {Norman},\
  and\ \citenamefont {Maxson}}]{7}%
  \BibitemOpen
  \bibfield  {author} {\bibinfo {author} {\bibfnamefont {P.}~\bibnamefont
  {Dyer}}, \bibinfo {author} {\bibfnamefont {D.}~\bibnamefont {Bodansky}},
  \bibinfo {author} {\bibfnamefont {A.~G.}\ \bibnamefont {Seamster}}, \bibinfo
  {author} {\bibfnamefont {E.~B.}\ \bibnamefont {Norman}}, \ and\ \bibinfo
  {author} {\bibfnamefont {D.~R.}\ \bibnamefont {Maxson}},\ }\href {\doibase
  10.1103/PhysRevC.23.1865} {\bibfield  {journal} {\bibinfo  {journal} {Phys.
  Rev. C}\ }\textbf {\bibinfo {volume} {23}},\ \bibinfo {pages} {1865}
  (\bibinfo {year} {1981})}\BibitemShut {NoStop}%
\bibitem [{\citenamefont {Seamster}\ \emph {et~al.}(1984)\citenamefont
  {Seamster}, \citenamefont {Norman}, \citenamefont {Leach}, \citenamefont
  {Dyer},\ and\ \citenamefont {Bodansky}}]{8}%
  \BibitemOpen
  \bibfield  {author} {\bibinfo {author} {\bibfnamefont {A.~G.}\ \bibnamefont
  {Seamster}}, \bibinfo {author} {\bibfnamefont {E.~B.}\ \bibnamefont
  {Norman}}, \bibinfo {author} {\bibfnamefont {D.~D.}\ \bibnamefont {Leach}},
  \bibinfo {author} {\bibfnamefont {P.}~\bibnamefont {Dyer}}, \ and\ \bibinfo
  {author} {\bibfnamefont {D.}~\bibnamefont {Bodansky}},\ }\href {\doibase
  10.1103/PhysRevC.29.394} {\bibfield  {journal} {\bibinfo  {journal} {Phys.
  Rev. C}\ }\textbf {\bibinfo {volume} {29}},\ \bibinfo {pages} {394} (\bibinfo
  {year} {1984})}\BibitemShut {NoStop}%
\bibitem [{\citenamefont {Dyer}\ \emph {et~al.}(1985)\citenamefont {Dyer},
  \citenamefont {Bodansky}, \citenamefont {Leach}, \citenamefont {Norman},\
  and\ \citenamefont {Seamster}}]{9}%
  \BibitemOpen
  \bibfield  {author} {\bibinfo {author} {\bibfnamefont {P.}~\bibnamefont
  {Dyer}}, \bibinfo {author} {\bibfnamefont {D.}~\bibnamefont {Bodansky}},
  \bibinfo {author} {\bibfnamefont {D.~D.}\ \bibnamefont {Leach}}, \bibinfo
  {author} {\bibfnamefont {E.~B.}\ \bibnamefont {Norman}}, \ and\ \bibinfo
  {author} {\bibfnamefont {A.~G.}\ \bibnamefont {Seamster}},\ }\href {\doibase
  10.1103/PhysRevC.32.1873} {\bibfield  {journal} {\bibinfo  {journal} {Phys.
  Rev. C}\ }\textbf {\bibinfo {volume} {32}},\ \bibinfo {pages} {1873}
  (\bibinfo {year} {1985})}\BibitemShut {NoStop}%
\bibitem [{\citenamefont {Belhout}\ \emph {et~al.}(2007)\citenamefont
  {Belhout}, \citenamefont {Kiener}, \citenamefont {Coc}, \citenamefont
  {Duprat}, \citenamefont {Engrand}, \citenamefont {Fitoussi}, \citenamefont
  {Gounelle}, \citenamefont {Lefebvre-Schuhl}, \citenamefont
  {S{\'e}r{\'e}ville}, \citenamefont {Tatischeff}, \citenamefont {Thibaud},
  \citenamefont {Chabot}, \citenamefont {Hammache},\ and\ \citenamefont
  {Benhabiles-Mezhoud}}]{10}%
  \BibitemOpen
  \bibfield  {author} {\bibinfo {author} {\bibfnamefont {A.}~\bibnamefont
  {Belhout}}, \bibinfo {author} {\bibfnamefont {J.}~\bibnamefont {Kiener}},
  \bibinfo {author} {\bibfnamefont {A.}~\bibnamefont {Coc}}, \bibinfo {author}
  {\bibfnamefont {J.}~\bibnamefont {Duprat}}, \bibinfo {author} {\bibfnamefont
  {C.}~\bibnamefont {Engrand}}, \bibinfo {author} {\bibfnamefont
  {C.}~\bibnamefont {Fitoussi}}, \bibinfo {author} {\bibfnamefont
  {M.}~\bibnamefont {Gounelle}}, \bibinfo {author} {\bibfnamefont
  {A.}~\bibnamefont {Lefebvre-Schuhl}}, \bibinfo {author} {\bibfnamefont
  {N.~d.}\ \bibnamefont {S{\'e}r{\'e}ville}}, \bibinfo {author} {\bibfnamefont
  {V.}~\bibnamefont {Tatischeff}}, \bibinfo {author} {\bibfnamefont {J.-P.}\
  \bibnamefont {Thibaud}}, \bibinfo {author} {\bibfnamefont {M.}~\bibnamefont
  {Chabot}}, \bibinfo {author} {\bibfnamefont {F.}~\bibnamefont {Hammache}}, \
  and\ \bibinfo {author} {\bibfnamefont {H.}~\bibnamefont
  {Benhabiles-Mezhoud}},\ }\href {\doibase 10.1103/PhysRevC.76.034607}
  {\bibfield  {journal} {\bibinfo  {journal} {Phys. Rev. C}\ }\textbf {\bibinfo
  {volume} {76}},\ \bibinfo {pages} {034607} (\bibinfo {year}
  {2007})}\BibitemShut {NoStop}%
\bibitem [{\citenamefont {Belhout}\ \emph {et~al.}(2009)\citenamefont
  {Belhout}, \citenamefont {Kiener}, \citenamefont {Coc}, \citenamefont
  {Duprat}, \citenamefont {Engrand}, \citenamefont {Fitoussi}, \citenamefont
  {Gounelle}, \citenamefont {Lefebvre-Schuhl}, \citenamefont {S\'er\'eville},
  \citenamefont {Tatischeff}, \citenamefont {Thibaud}, \citenamefont {Chabot},
  \citenamefont {Hammache},\ and\ \citenamefont {Benhabiles-Mezhoud}}]{10b}%
  \BibitemOpen
  \bibfield  {author} {\bibinfo {author} {\bibfnamefont {A.}~\bibnamefont
  {Belhout}}, \bibinfo {author} {\bibfnamefont {J.}~\bibnamefont {Kiener}},
  \bibinfo {author} {\bibfnamefont {A.}~\bibnamefont {Coc}}, \bibinfo {author}
  {\bibfnamefont {J.}~\bibnamefont {Duprat}}, \bibinfo {author} {\bibfnamefont
  {C.}~\bibnamefont {Engrand}}, \bibinfo {author} {\bibfnamefont
  {C.}~\bibnamefont {Fitoussi}}, \bibinfo {author} {\bibfnamefont
  {M.}~\bibnamefont {Gounelle}}, \bibinfo {author} {\bibfnamefont
  {A.}~\bibnamefont {Lefebvre-Schuhl}}, \bibinfo {author} {\bibfnamefont
  {N.~d.}\ \bibnamefont {S\'er\'eville}}, \bibinfo {author} {\bibfnamefont
  {V.}~\bibnamefont {Tatischeff}}, \bibinfo {author} {\bibfnamefont {J.~P.}\
  \bibnamefont {Thibaud}}, \bibinfo {author} {\bibfnamefont {M.}~\bibnamefont
  {Chabot}}, \bibinfo {author} {\bibfnamefont {F.}~\bibnamefont {Hammache}}, \
  and\ \bibinfo {author} {\bibfnamefont {H.}~\bibnamefont
  {Benhabiles-Mezhoud}},\ }\href {\doibase 10.1103/PhysRevC.80.029902}
  {\bibfield  {journal} {\bibinfo  {journal} {Phys. Rev. C}\ }\textbf {\bibinfo
  {volume} {80}},\ \bibinfo {pages} {029902} (\bibinfo {year}
  {2009})}\BibitemShut {NoStop}%
\bibitem [{\citenamefont {Lesko}\ \emph {et~al.}(1988)\citenamefont {Lesko},
  \citenamefont {Norman}, \citenamefont {Larimer}, \citenamefont {Kuhn},
  \citenamefont {Meekhof}, \citenamefont {Crane},\ and\ \citenamefont
  {Bussell}}]{11}%
  \BibitemOpen
  \bibfield  {author} {\bibinfo {author} {\bibfnamefont {K.~T.}\ \bibnamefont
  {Lesko}}, \bibinfo {author} {\bibfnamefont {E.~B.}\ \bibnamefont {Norman}},
  \bibinfo {author} {\bibfnamefont {R.-M.}\ \bibnamefont {Larimer}}, \bibinfo
  {author} {\bibfnamefont {S.}~\bibnamefont {Kuhn}}, \bibinfo {author}
  {\bibfnamefont {D.~M.}\ \bibnamefont {Meekhof}}, \bibinfo {author}
  {\bibfnamefont {S.~G.}\ \bibnamefont {Crane}}, \ and\ \bibinfo {author}
  {\bibfnamefont {H.~G.}\ \bibnamefont {Bussell}},\ }\href {\doibase
  10.1103/PhysRevC.37.1808} {\bibfield  {journal} {\bibinfo  {journal} {Phys.
  Rev. C}\ }\textbf {\bibinfo {volume} {37}},\ \bibinfo {pages} {1808}
  (\bibinfo {year} {1988})}\BibitemShut {NoStop}%
\bibitem [{\citenamefont {{Ramaty}}\ \emph {et~al.}(1979)\citenamefont
  {{Ramaty}}, \citenamefont {{Kozlovsky}},\ and\ \citenamefont
  {{Lingenfelter}}}]{12}%
  \BibitemOpen
  \bibfield  {author} {\bibinfo {author} {\bibfnamefont {R.}~\bibnamefont
  {{Ramaty}}}, \bibinfo {author} {\bibfnamefont {B.}~\bibnamefont
  {{Kozlovsky}}}, \ and\ \bibinfo {author} {\bibfnamefont {R.~E.}\ \bibnamefont
  {{Lingenfelter}}},\ }\href {\doibase 10.1086/190596} {\bibfield  {journal}
  {\bibinfo  {journal} {{IOP} Publishing}\ }\textbf {\bibinfo {volume} {40}},\
  \bibinfo {pages} {487} (\bibinfo {year} {1979})}\BibitemShut {NoStop}%
\bibitem [{\citenamefont {Kozlovsky}\ \emph {et~al.}(2002)\citenamefont
  {Kozlovsky}, \citenamefont {Murphy},\ and\ \citenamefont {Ramaty}}]{13}%
  \BibitemOpen
  \bibfield  {author} {\bibinfo {author} {\bibfnamefont {B.}~\bibnamefont
  {Kozlovsky}}, \bibinfo {author} {\bibfnamefont {R.~J.}\ \bibnamefont
  {Murphy}}, \ and\ \bibinfo {author} {\bibfnamefont {R.}~\bibnamefont
  {Ramaty}},\ }\href {\doibase 10.1086/340545} {\bibfield  {journal} {\bibinfo
  {journal} {Astrophys. J. Supp. Ser.}\ }\textbf {\bibinfo {volume} {141}},\
  \bibinfo {pages} {523} (\bibinfo {year} {2002})}\BibitemShut {NoStop}%
\bibitem [{\citenamefont {Murphy}\ \emph {et~al.}(2009)\citenamefont {Murphy},
  \citenamefont {Kozlovsky}, \citenamefont {Kiener},\ and\ \citenamefont
  {Share}}]{14}%
  \BibitemOpen
  \bibfield  {author} {\bibinfo {author} {\bibfnamefont {R.~J.}\ \bibnamefont
  {Murphy}}, \bibinfo {author} {\bibfnamefont {B.}~\bibnamefont {Kozlovsky}},
  \bibinfo {author} {\bibfnamefont {J.}~\bibnamefont {Kiener}}, \ and\ \bibinfo
  {author} {\bibfnamefont {G.~H.}\ \bibnamefont {Share}},\ }\href
  {http://stacks.iop.org/0067-0049/183/i=1/a=142} {\bibfield  {journal}
  {\bibinfo  {journal} {Astrophys. J. Supp. Ser.}\ }\textbf {\bibinfo {volume}
  {183}},\ \bibinfo {pages} {142} (\bibinfo {year} {2009})}\BibitemShut
  {NoStop}%
\bibitem [{\citenamefont {Koning}\ \emph {et~al.}(2008)\citenamefont {Koning},
  \citenamefont {Hilaire},\ and\ \citenamefont {Duijvestijn}}]{15}%
  \BibitemOpen
  \bibfield  {author} {\bibinfo {author} {\bibfnamefont {A.~J.}\ \bibnamefont
  {Koning}}, \bibinfo {author} {\bibfnamefont {S.}~\bibnamefont {Hilaire}}, \
  and\ \bibinfo {author} {\bibfnamefont {M.~C.}\ \bibnamefont {Duijvestijn}},\
  }in\ \href {\doibase https://doi.org/10.1051/ndata:07767} {\emph {\bibinfo
  {booktitle} {{International Conference on Nuclear Data for Science and
  Technology, Nice, France}}}},\ \bibinfo {editor} {edited by\ \bibinfo
  {editor} {\bibfnamefont {E.~R.}\ \bibnamefont {O.Bersillon}, \bibfnamefont
  {F.Gunsing}}\ and\ \bibinfo {editor} {\bibnamefont {S.Leray}}}\ (\bibinfo
  {organization} {EDP Sciences},\ \bibinfo {year} {2008})\ pp.\ \bibinfo
  {pages} {211--214}\BibitemShut {NoStop}%
\bibitem [{\citenamefont {Benhabiles-Mezhoud}\ \emph {et~al.}()\citenamefont
  {Benhabiles-Mezhoud}, \citenamefont {Kiener}, \citenamefont {Thibaud},
  \citenamefont {Tatischeff}, \citenamefont {Deloncle}, \citenamefont {Coc},
  \citenamefont {Duprat}, \citenamefont {Hamadache}, \citenamefont
  {Lefebvre-Schuhl}, \citenamefont {Dalouzy} \emph {et~al.}}]{16}%
  \BibitemOpen
  \bibfield  {author} {\bibinfo {author} {\bibfnamefont {H.}~\bibnamefont
  {Benhabiles-Mezhoud}}, \bibinfo {author} {\bibfnamefont {J.}~\bibnamefont
  {Kiener}}, \bibinfo {author} {\bibfnamefont {J.-P.}\ \bibnamefont {Thibaud}},
  \bibinfo {author} {\bibfnamefont {V.}~\bibnamefont {Tatischeff}}, \bibinfo
  {author} {\bibfnamefont {I.}~\bibnamefont {Deloncle}}, \bibinfo {author}
  {\bibfnamefont {A.}~\bibnamefont {Coc}}, \bibinfo {author} {\bibfnamefont
  {J.}~\bibnamefont {Duprat}}, \bibinfo {author} {\bibfnamefont
  {C.}~\bibnamefont {Hamadache}}, \bibinfo {author} {\bibfnamefont
  {A.}~\bibnamefont {Lefebvre-Schuhl}}, \bibinfo {author} {\bibfnamefont
  {J.-C.}\ \bibnamefont {Dalouzy}},  \emph {et~al.},\ }\href {\doibase
  https://doi.org/10.1103/PhysRevC.83.024603} {\bibfield  {journal} {\bibinfo
  {journal} {Phys. Rev. C}\ }\textbf {\bibinfo {volume} {83}},\ \bibinfo
  {pages} {024603}}\BibitemShut {NoStop}%
\bibitem [{\citenamefont {Benhabiles-Mezhoud}(2010)}]{16b}%
  \BibitemOpen
  \bibfield  {author} {\bibinfo {author} {\bibfnamefont {H.}~\bibnamefont
  {Benhabiles-Mezhoud}},\ }\emph {\bibinfo {title} {Calcul du spectre total de
  l’émission gamma induite par interactions nucléaires des particules du
  rayonnement cosmique avec le milieu interstellaire et comparaison avec les
  observations de l’astronomie gamma}},\ \href@noop {} {Ph.D. thesis},\
  \bibinfo  {school} {Université Paris-Sud.} (\bibinfo {year}
  {2010})\BibitemShut {NoStop}%
\bibitem [{\citenamefont {Kiener}\ \emph {et~al.}()\citenamefont {Kiener},
  \citenamefont {Belhout}, \citenamefont {Tatischeff},\ and\ \citenamefont
  {Benhabiles-Mezhoud}}]{17}%
  \BibitemOpen
  \bibfield  {author} {\bibinfo {author} {\bibfnamefont {J.}~\bibnamefont
  {Kiener}}, \bibinfo {author} {\bibfnamefont {A.}~\bibnamefont {Belhout}},
  \bibinfo {author} {\bibfnamefont {V.}~\bibnamefont {Tatischeff}}, \ and\
  \bibinfo {author} {\bibfnamefont {H.}~\bibnamefont {Benhabiles-Mezhoud}},\
  }in\ \href@noop {} {\emph {\bibinfo {booktitle} {{Proceedings of the DAE
  Symposium on Nuclear Physics V. 53}}}}\ (\bibinfo {organization} {{Bhabha
  Atomic Research Centre. Roorkee (India) }})\BibitemShut {NoStop}%
\bibitem [{\citenamefont {Kiener}\ \emph {et~al.}(1998)\citenamefont {Kiener},
  \citenamefont {Berheide}, \citenamefont {Achouri}, \citenamefont {Boughrara},
  \citenamefont {Coc}, \citenamefont {Lefebvre}, \citenamefont
  {de~Oliveira~Santos},\ and\ \citenamefont {Vieu}}]{18}%
  \BibitemOpen
  \bibfield  {author} {\bibinfo {author} {\bibfnamefont {J.}~\bibnamefont
  {Kiener}}, \bibinfo {author} {\bibfnamefont {M.}~\bibnamefont {Berheide}},
  \bibinfo {author} {\bibfnamefont {N.~L.}\ \bibnamefont {Achouri}}, \bibinfo
  {author} {\bibfnamefont {A.}~\bibnamefont {Boughrara}}, \bibinfo {author}
  {\bibfnamefont {A.}~\bibnamefont {Coc}}, \bibinfo {author} {\bibfnamefont
  {A.}~\bibnamefont {Lefebvre}}, \bibinfo {author} {\bibfnamefont
  {F.}~\bibnamefont {de~Oliveira~Santos}}, \ and\ \bibinfo {author}
  {\bibfnamefont {C.}~\bibnamefont {Vieu}},\ }\href {\doibase
  10.1103/PhysRevC.58.2174} {\bibfield  {journal} {\bibinfo  {journal} {Phys.
  Rev. C}\ }\textbf {\bibinfo {volume} {58}},\ \bibinfo {pages} {2174}
  (\bibinfo {year} {1998})}\BibitemShut {NoStop}%
\bibitem [{\citenamefont {{Herman}}\ \emph {et~al.}(2007)\citenamefont
  {{Herman}}, \citenamefont {{Capote}}, \citenamefont {{Carlson}},
  \citenamefont {{Oblo\v{z}insk\'{y}}}, \citenamefont {{Sin}}, \citenamefont
  {{Trkov}}, \citenamefont {{Wienke}},\ and\ \citenamefont {{Zerkin}}}]{19}%
  \BibitemOpen
  \bibfield  {author} {\bibinfo {author} {\bibfnamefont {M.}~\bibnamefont
  {{Herman}}}, \bibinfo {author} {\bibfnamefont {R.}~\bibnamefont {{Capote}}},
  \bibinfo {author} {\bibfnamefont {B.}~\bibnamefont {{Carlson}}}, \bibinfo
  {author} {\bibfnamefont {P.}~\bibnamefont {{Oblo\v{z}insk\'{y}}}}, \bibinfo
  {author} {\bibfnamefont {M.}~\bibnamefont {{Sin}}}, \bibinfo {author}
  {\bibfnamefont {A.}~\bibnamefont {{Trkov}}}, \bibinfo {author} {\bibfnamefont
  {H.}~\bibnamefont {{Wienke}}}, \ and\ \bibinfo {author} {\bibfnamefont
  {V.}~\bibnamefont {{Zerkin}}},\ }\href
  {https://www-nds.iaea.org/empire/index.html} {\bibfield  {journal} {\bibinfo
  {journal} {Nucl. Data Sheets}\ }\textbf {\bibinfo {volume} {108}},\ \bibinfo
  {pages} {2655} (\bibinfo {year} {2007})}\BibitemShut {NoStop}%
\bibitem [{\citenamefont {Bloemen}(1989)}]{20}%
  \BibitemOpen
  \bibfield  {author} {\bibinfo {author} {\bibfnamefont {H.}~\bibnamefont
  {Bloemen}},\ }\href {\doibase 10.1146/annurev.aa.27.090189.002345} {\bibfield
   {journal} {\bibinfo  {journal} {Annu. Rev. Astron. Astrophys.}\ }\textbf
  {\bibinfo {volume} {27}},\ \bibinfo {pages} {469} (\bibinfo {year}
  {1989})}\BibitemShut {NoStop}%
\bibitem [{\citenamefont {Dasari}\ \emph {et~al.}(2014)\citenamefont {Dasari},
  \citenamefont {Chhillar}, \citenamefont {Acharya}, \citenamefont {Ray},
  \citenamefont {Behera}, \citenamefont {Das},\ and\ \citenamefont
  {Pujari}}]{21}%
  \BibitemOpen
  \bibfield  {author} {\bibinfo {author} {\bibfnamefont {K.}~\bibnamefont
  {Dasari}}, \bibinfo {author} {\bibfnamefont {S.}~\bibnamefont {Chhillar}},
  \bibinfo {author} {\bibfnamefont {R.}~\bibnamefont {Acharya}}, \bibinfo
  {author} {\bibfnamefont {D.}~\bibnamefont {Ray}}, \bibinfo {author}
  {\bibfnamefont {A.}~\bibnamefont {Behera}}, \bibinfo {author} {\bibfnamefont
  {N.~L.}\ \bibnamefont {Das}}, \ and\ \bibinfo {author} {\bibfnamefont
  {P.}~\bibnamefont {Pujari}},\ }\href {\doibase
  https://doi.org/10.1016/j.nimb.2014.08.017} {\bibfield  {journal} {\bibinfo
  {journal} {Nucl. Instrum. Methods in Phys. Res. Sect. B.}\ }\textbf {\bibinfo
  {volume} {339}},\ \bibinfo {pages} {37 } (\bibinfo {year}
  {2014})}\BibitemShut {NoStop}%
\bibitem [{\citenamefont {Tatischeff}(2003)}]{22}%
  \BibitemOpen
  \bibfield  {author} {\bibinfo {author} {\bibfnamefont {V.}~\bibnamefont
  {Tatischeff}},\ }\href {\doibase 10.1051/eas:2003038} {\bibfield  {journal}
  {\bibinfo  {journal} {EAS Publications Series}\ }\textbf {\bibinfo {volume}
  {7}},\ \bibinfo {pages} {79} (\bibinfo {year} {2003})}\BibitemShut {NoStop}%
\bibitem [{\citenamefont {Tatischeff}\ and\ \citenamefont {Kiener}(2004)}]{23}%
  \BibitemOpen
  \bibfield  {author} {\bibinfo {author} {\bibfnamefont {V.}~\bibnamefont
  {Tatischeff}}\ and\ \bibinfo {author} {\bibfnamefont {J.}~\bibnamefont
  {Kiener}},\ }\href {\doibase https://doi.org/10.1016/j.newar.2003.11.013}
  {\bibfield  {journal} {\bibinfo  {journal} {New Astron. Rev.}\ }\textbf
  {\bibinfo {volume} {48}},\ \bibinfo {pages} {99 } (\bibinfo {year}
  {2004})}\BibitemShut {NoStop}%
\bibitem [{\citenamefont {Benhabiles-Mezhoud}\ \emph
  {et~al.}(2013)\citenamefont {Benhabiles-Mezhoud}, \citenamefont {Kiener},
  \citenamefont {Tatischeff},\ and\ \citenamefont {Strong}}]{24}%
  \BibitemOpen
  \bibfield  {author} {\bibinfo {author} {\bibfnamefont {H.}~\bibnamefont
  {Benhabiles-Mezhoud}}, \bibinfo {author} {\bibfnamefont {J.}~\bibnamefont
  {Kiener}}, \bibinfo {author} {\bibfnamefont {V.}~\bibnamefont {Tatischeff}},
  \ and\ \bibinfo {author} {\bibfnamefont {A.~W.}\ \bibnamefont {Strong}},\
  }\href {\doibase 10.1088/0004-637x/763/2/98} {\bibfield  {journal} {\bibinfo
  {journal} {Astrophys. J. Supp. Ser.}\ }\textbf {\bibinfo {volume} {763}},\
  \bibinfo {pages} {98} (\bibinfo {year} {2013})}\BibitemShut {NoStop}%
\bibitem [{\citenamefont {Sharpey-Schafer}(2004)}]{25}%
  \BibitemOpen
  \bibfield  {author} {\bibinfo {author} {\bibfnamefont {J.}~\bibnamefont
  {Sharpey-Schafer}},\ }\href {\doibase 10.1080/10506890491034659} {\bibfield
  {journal} {\bibinfo  {journal} {Nucl. Phys. News}\ }\textbf {\bibinfo
  {volume} {14}},\ \bibinfo {pages} {3} (\bibinfo {year} {2004})}\BibitemShut
  {NoStop}%
\bibitem [{\citenamefont {Newman}\ \emph {et~al.}(1998)\citenamefont {Newman},
  \citenamefont {Lawrie}, \citenamefont {Babu}, \citenamefont {Fetea},
  \citenamefont {F{\"o}rtsch}, \citenamefont {Naguleswaran}, \citenamefont
  {Pilcher}, \citenamefont {Raavé}, \citenamefont {Rigollet}, \citenamefont
  {Sharpey-Schafer} \emph {et~al.}}]{25b}%
  \BibitemOpen
  \bibfield  {author} {\bibinfo {author} {\bibfnamefont {R.~T.}\ \bibnamefont
  {Newman}}, \bibinfo {author} {\bibfnamefont {J.~J.}\ \bibnamefont {Lawrie}},
  \bibinfo {author} {\bibfnamefont {B.~R.~S.}\ \bibnamefont {Babu}}, \bibinfo
  {author} {\bibfnamefont {M.~S.}\ \bibnamefont {Fetea}}, \bibinfo {author}
  {\bibfnamefont {S.~V.}\ \bibnamefont {F{\"o}rtsch}}, \bibinfo {author}
  {\bibfnamefont {S.}~\bibnamefont {Naguleswaran}}, \bibinfo {author}
  {\bibfnamefont {J.~V.}\ \bibnamefont {Pilcher}}, \bibinfo {author}
  {\bibfnamefont {D.~A.}\ \bibnamefont {Raavé}}, \bibinfo {author}
  {\bibfnamefont {C.}~\bibnamefont {Rigollet}}, \bibinfo {author}
  {\bibfnamefont {J.~F.}\ \bibnamefont {Sharpey-Schafer}},  \emph {et~al.},\
  }\href@noop {} {\bibfield  {journal} {\bibinfo  {journal} {Balkan Phys.
  Lett., Spec. Issue}\ ,\ \bibinfo {pages} {182}} (\bibinfo {year}
  {1998})}\BibitemShut {NoStop}%
\bibitem [{\citenamefont {Lipoglavšek}\ \emph {et~al.}(2006)\citenamefont
  {Lipoglavšek}, \citenamefont {Likar}, \citenamefont {Vencelj}, \citenamefont
  {Vidmar}, \citenamefont {Bark}, \citenamefont {Gueorguieva}, \citenamefont
  {Komati}, \citenamefont {Lawrie}, \citenamefont {Maliage}, \citenamefont
  {Mullins}, \citenamefont {Murray},\ and\ \citenamefont {Ramashidzha}}]{26}%
  \BibitemOpen
  \bibfield  {author} {\bibinfo {author} {\bibfnamefont {M.}~\bibnamefont
  {Lipoglavšek}}, \bibinfo {author} {\bibfnamefont {A.}~\bibnamefont {Likar}},
  \bibinfo {author} {\bibfnamefont {M.}~\bibnamefont {Vencelj}}, \bibinfo
  {author} {\bibfnamefont {T.}~\bibnamefont {Vidmar}}, \bibinfo {author}
  {\bibfnamefont {R.}~\bibnamefont {Bark}}, \bibinfo {author} {\bibfnamefont
  {E.}~\bibnamefont {Gueorguieva}}, \bibinfo {author} {\bibfnamefont
  {F.}~\bibnamefont {Komati}}, \bibinfo {author} {\bibfnamefont
  {J.}~\bibnamefont {Lawrie}}, \bibinfo {author} {\bibfnamefont
  {S.}~\bibnamefont {Maliage}}, \bibinfo {author} {\bibfnamefont
  {S.}~\bibnamefont {Mullins}}, \bibinfo {author} {\bibfnamefont
  {S.}~\bibnamefont {Murray}}, \ and\ \bibinfo {author} {\bibfnamefont
  {T.}~\bibnamefont {Ramashidzha}},\ }\href {\doibase
  https://doi.org/10.1016/j.nima.2005.11.067} {\bibfield  {journal} {\bibinfo
  {journal} {Nucl. Instrum. Methods in Phys. Res. Sect. A.}\ }\textbf {\bibinfo
  {volume} {557}},\ \bibinfo {pages} {523 } (\bibinfo {year}
  {2006})}\BibitemShut {NoStop}%
\bibitem [{\citenamefont {Duchêne}\ \emph {et~al.}(1999)\citenamefont
  {Duchêne}, \citenamefont {Beck}, \citenamefont {Twin}, \citenamefont
  {de~France}, \citenamefont {Curien}, \citenamefont {Han}, \citenamefont
  {Beausang}, \citenamefont {Bentley}, \citenamefont {Nolan},\ and\
  \citenamefont {Simpson}}]{27}%
  \BibitemOpen
  \bibfield  {author} {\bibinfo {author} {\bibfnamefont {G.}~\bibnamefont
  {Duchêne}}, \bibinfo {author} {\bibfnamefont {F.}~\bibnamefont {Beck}},
  \bibinfo {author} {\bibfnamefont {P.}~\bibnamefont {Twin}}, \bibinfo {author}
  {\bibfnamefont {G.}~\bibnamefont {de~France}}, \bibinfo {author}
  {\bibfnamefont {D.}~\bibnamefont {Curien}}, \bibinfo {author} {\bibfnamefont
  {L.}~\bibnamefont {Han}}, \bibinfo {author} {\bibfnamefont {C.}~\bibnamefont
  {Beausang}}, \bibinfo {author} {\bibfnamefont {M.}~\bibnamefont {Bentley}},
  \bibinfo {author} {\bibfnamefont {P.}~\bibnamefont {Nolan}}, \ and\ \bibinfo
  {author} {\bibfnamefont {J.}~\bibnamefont {Simpson}},\ }\href {\doibase
  https://doi.org/10.1016/S0168-9002(99)00277-6} {\bibfield  {journal}
  {\bibinfo  {journal} {Nucl. Instrum. Methods in Phys. Res. Sect. A.}\
  }\textbf {\bibinfo {volume} {432}},\ \bibinfo {pages} {90 } (\bibinfo {year}
  {1999})}\BibitemShut {NoStop}%
\bibitem [{\citenamefont {GEANT4}()}]{28}%
  \BibitemOpen
  \bibfield  {author} {\bibinfo {author} {\bibnamefont {GEANT4}},\ }\href
  {\doibase http://geant4.web.cern.ch/} {\
  http://geant4.web.cern.ch/}\BibitemShut {NoStop}%
\bibitem [{\citenamefont {ROOT-CERNLIB}()}]{29}%
  \BibitemOpen
  \bibfield  {author} {\bibinfo {author} {\bibnamefont {ROOT-CERNLIB}},\ }\href
  {\doibase https://root.cern.ch/} {\ https://root.cern.ch/}\BibitemShut
  {NoStop}%
\bibitem [{\citenamefont {Radford}()}]{30}%
  \BibitemOpen
  \bibfield  {author} {\bibinfo {author} {\bibfnamefont {R.~C.}\ \bibnamefont
  {Radford}},\ }\href {\doibase https://radware.phy.ornl.gov/} {\
  https://radware.phy.ornl.gov/}\BibitemShut {NoStop}%
\bibitem [{\citenamefont {Rose}(1953)}]{31}%
  \BibitemOpen
  \bibfield  {author} {\bibinfo {author} {\bibfnamefont {M.~E.}\ \bibnamefont
  {Rose}},\ }\href {\doibase 10.1103/PhysRev.91.610} {\bibfield  {journal}
  {\bibinfo  {journal} {Phys. Rev.}\ }\textbf {\bibinfo {volume} {91}},\
  \bibinfo {pages} {610} (\bibinfo {year} {1953})}\BibitemShut {NoStop}%
\bibitem [{\citenamefont {Iliadis}(2007)}]{32}%
  \BibitemOpen
  \bibfield  {author} {\bibinfo {author} {\bibfnamefont {C.}~\bibnamefont
  {Iliadis}},\ }\href@noop {} {\emph {\bibinfo {title} {{Nuclear Physics of
  Stars}}}}\ (\bibinfo  {publisher} {John Wiley \& Sons},\ \bibinfo {year}
  {2007})\BibitemShut {NoStop}%
\bibitem [{\citenamefont {Gilmore}(2008)}]{33}%
  \BibitemOpen
  \bibfield  {author} {\bibinfo {author} {\bibfnamefont {G.~R.}\ \bibnamefont
  {Gilmore}},\ }\href@noop {} {\emph {\bibinfo {title} {{Practical Gamma-ray
  Spectroscopy (2$^{nd}$ Edition)}}}}\ (\bibinfo  {publisher} {John Wiley \&
  Sons},\ \bibinfo {year} {2008})\BibitemShut {NoStop}%
\bibitem [{\citenamefont {Bunting}\ and\ \citenamefont {Kraushaar}(1974)}]{34}%
  \BibitemOpen
  \bibfield  {author} {\bibinfo {author} {\bibfnamefont {R.}~\bibnamefont
  {Bunting}}\ and\ \bibinfo {author} {\bibfnamefont {J.}~\bibnamefont
  {Kraushaar}},\ }\href {\doibase 10.1016/0029-554X(74)90667-3} {\bibfield
  {journal} {\bibinfo  {journal} {Nucl. Instrum. Methods}\ }\textbf {\bibinfo
  {volume} {118}},\ \bibinfo {pages} {565} (\bibinfo {year}
  {1974})}\BibitemShut {NoStop}%
\bibitem [{\citenamefont {Tatischeff}\ \emph {et~al.}(2006)\citenamefont
  {Tatischeff}, \citenamefont {Kozlovsky}, \citenamefont {Kiener},\ and\
  \citenamefont {Murphy}}]{35}%
  \BibitemOpen
  \bibfield  {author} {\bibinfo {author} {\bibfnamefont {V.}~\bibnamefont
  {Tatischeff}}, \bibinfo {author} {\bibfnamefont {B.}~\bibnamefont
  {Kozlovsky}}, \bibinfo {author} {\bibfnamefont {J.}~\bibnamefont {Kiener}}, \
  and\ \bibinfo {author} {\bibfnamefont {R.~J.}\ \bibnamefont {Murphy}},\
  }\href {\doibase 10.1086/505112} {\bibfield  {journal} {\bibinfo  {journal}
  {Astrophys. J. Supp. Ser.}\ }\textbf {\bibinfo {volume} {165}},\ \bibinfo
  {pages} {606} (\bibinfo {year} {2006})}\BibitemShut {NoStop}%
\bibitem [{\citenamefont {Soukhovitski}\ \emph {et~al.}()\citenamefont
  {Soukhovitski}, \citenamefont {Chiba}, \citenamefont {Capote}, \citenamefont
  {Quesada}, \citenamefont {Kunieda},\ and\ \citenamefont {Morogovskij}}]{36}%
  \BibitemOpen
  \bibfield  {author} {\bibinfo {author} {\bibfnamefont {E.~S.}\ \bibnamefont
  {Soukhovitski}}, \bibinfo {author} {\bibfnamefont {S.}~\bibnamefont {Chiba}},
  \bibinfo {author} {\bibfnamefont {R.}~\bibnamefont {Capote}}, \bibinfo
  {author} {\bibfnamefont {J.~M.}\ \bibnamefont {Quesada}}, \bibinfo {author}
  {\bibfnamefont {S.}~\bibnamefont {Kunieda}}, \ and\ \bibinfo {author}
  {\bibfnamefont {G.~B.}\ \bibnamefont {Morogovskij}},\ }\href
  {{https://www-nds.iaea.org/RIPL-3/codes/OPTMAN/}} {\bibinfo  {journal}
  {{OPTMAN code v10}}\ }\BibitemShut {NoStop}%
\bibitem [{\citenamefont {Plujko}\ \emph {et~al.}(2014)\citenamefont {Plujko},
  \citenamefont {Gorbachenko}, \citenamefont {Bondar},\ and\ \citenamefont
  {Rovenskykh}}]{38}%
  \BibitemOpen
\bibfield  {journal} {  }\bibfield  {author} {\bibinfo {author} {\bibfnamefont
  {V.}~\bibnamefont {Plujko}}, \bibinfo {author} {\bibfnamefont
  {O.}~\bibnamefont {Gorbachenko}}, \bibinfo {author} {\bibfnamefont
  {B.}~\bibnamefont {Bondar}}, \ and\ \bibinfo {author} {\bibfnamefont
  {E.}~\bibnamefont {Rovenskykh}},\ }\href {\doibase
  https://doi.org/10.1016/j.nds.2014.04.047} {\bibfield  {journal} {\bibinfo
  {journal} {Nucl. Data Sheets}\ }\textbf {\bibinfo {volume} {118}},\ \bibinfo
  {pages} {240 } (\bibinfo {year} {2014})}\BibitemShut {NoStop}%
\bibitem [{\citenamefont {RIPL-3}()}]{37}%
  \BibitemOpen
  \bibfield  {author} {\bibinfo {author} {\bibnamefont {RIPL-3}},\ }\href
  {\doibase https://www-nds.iaea.org/RIPL-3/} {\
  https://www-nds.iaea.org/RIPL-3/}\BibitemShut {NoStop}%
\bibitem [{\citenamefont {Reeves}\ \emph {et~al.}(1970)\citenamefont {Reeves},
  \citenamefont {Fowler},\ and\ \citenamefont {Hoyle}}]{39}%
  \BibitemOpen
  \bibfield  {author} {\bibinfo {author} {\bibfnamefont {H.}~\bibnamefont
  {Reeves}}, \bibinfo {author} {\bibfnamefont {W.~A.}\ \bibnamefont {Fowler}},
  \ and\ \bibinfo {author} {\bibfnamefont {F.}~\bibnamefont {Hoyle}},\ }\href
  {\doibase https://doi.org/10.1038/226727a0} {\bibfield  {journal} {\bibinfo
  {journal} {Nature}\ }\textbf {\bibinfo {volume} {226}},\ \bibinfo {pages}
  {727 } (\bibinfo {year} {1970})}\BibitemShut {NoStop}%
\bibitem [{\citenamefont {Rolfs}\ and\ \citenamefont {Rodney}(1988)}]{39b}%
  \BibitemOpen
  \bibfield  {author} {\bibinfo {author} {\bibfnamefont {C.~E.}\ \bibnamefont
  {Rolfs}}\ and\ \bibinfo {author} {\bibfnamefont {W.~S.}\ \bibnamefont
  {Rodney}},\ }\href@noop {} {\emph {\bibinfo {title} {{Cauldrons in the cosmos
  : nuclear astrophysics}}}}\ (\bibinfo  {publisher} {University of Chicago
  Press, Chicago.},\ \bibinfo {year} {1988})\BibitemShut {NoStop}%
\bibitem [{\citenamefont {Strong}\ \emph {et~al.}(2007)\citenamefont {Strong},
  \citenamefont {Moskalenko},\ and\ \citenamefont {Ptuskin}}]{40}%
  \BibitemOpen
  \bibfield  {author} {\bibinfo {author} {\bibfnamefont {A.~W.}\ \bibnamefont
  {Strong}}, \bibinfo {author} {\bibfnamefont {I.~V.}\ \bibnamefont
  {Moskalenko}}, \ and\ \bibinfo {author} {\bibfnamefont {V.~S.}\ \bibnamefont
  {Ptuskin}},\ }\href {\doibase 10.1146/annurev.nucl.57.090506.123011}
  {\bibfield  {journal} {\bibinfo  {journal} {Annu. Rev. Nucl. Part. Sci.}\
  }\textbf {\bibinfo {volume} {57}},\ \bibinfo {pages} {285} (\bibinfo {year}
  {2007})}\BibitemShut {NoStop}%
\bibitem [{\citenamefont {Grenier}\ \emph {et~al.}(2015)\citenamefont
  {Grenier}, \citenamefont {Black},\ and\ \citenamefont {Strong}}]{40b}%
  \BibitemOpen
  \bibfield  {author} {\bibinfo {author} {\bibfnamefont {I.~A.}\ \bibnamefont
  {Grenier}}, \bibinfo {author} {\bibfnamefont {J.~H.}\ \bibnamefont {Black}},
  \ and\ \bibinfo {author} {\bibfnamefont {A.~W.}\ \bibnamefont {Strong}},\
  }\href {\doibase 10.1146/annurev-astro-082214-122457} {\bibfield  {journal}
  {\bibinfo  {journal} {Annu. Rev. of Astron. Astrophys.}\ }\textbf {\bibinfo
  {volume} {53}},\ \bibinfo {pages} {199} (\bibinfo {year} {2015})}\BibitemShut
  {NoStop}%
\bibitem [{\citenamefont {Indriolo}\ \emph {et~al.}(2009)\citenamefont
  {Indriolo}, \citenamefont {Fields},\ and\ \citenamefont {McCall}}]{42}%
  \BibitemOpen
  \bibfield  {author} {\bibinfo {author} {\bibfnamefont {N.}~\bibnamefont
  {Indriolo}}, \bibinfo {author} {\bibfnamefont {B.~D.}\ \bibnamefont
  {Fields}}, \ and\ \bibinfo {author} {\bibfnamefont {B.~J.}\ \bibnamefont
  {McCall}},\ }\href {\doibase 10.1088/0004-637x/694/1/257} {\bibfield
  {journal} {\bibinfo  {journal} {Astrophys. J. Supp. Ser.}\ }\textbf {\bibinfo
  {volume} {694}},\ \bibinfo {pages} {257} (\bibinfo {year}
  {2009})}\BibitemShut {NoStop}%
\bibitem [{\citenamefont {Indriolo}\ and\ \citenamefont {McCall}(2012)}]{43}%
  \BibitemOpen
  \bibfield  {author} {\bibinfo {author} {\bibfnamefont {N.}~\bibnamefont
  {Indriolo}}\ and\ \bibinfo {author} {\bibfnamefont {B.~J.}\ \bibnamefont
  {McCall}},\ }\href {\doibase 10.1088/0004-637x/745/1/91} {\bibfield
  {journal} {\bibinfo  {journal} {Astrophys. J. Supp. Ser.}\ }\textbf {\bibinfo
  {volume} {745}},\ \bibinfo {pages} {91} (\bibinfo {year} {2012})}\BibitemShut
  {NoStop}%
\bibitem [{\citenamefont {Angelis}\ \emph {et~al.}(2018)\citenamefont
  {Angelis}, \citenamefont {Tatischeff}, \citenamefont {Grenier}, \citenamefont
  {McEnery}, \citenamefont {Mallamaci}, \citenamefont {Tavani}, \citenamefont
  {Oberlack}, \citenamefont {Hanlon}, \citenamefont {Walter}, \citenamefont
  {Argan} \emph {et~al.}}]{48}%
  \BibitemOpen
  \bibfield  {author} {\bibinfo {author} {\bibfnamefont {A.~D.}\ \bibnamefont
  {Angelis}}, \bibinfo {author} {\bibfnamefont {V.}~\bibnamefont {Tatischeff}},
  \bibinfo {author} {\bibfnamefont {I.}~\bibnamefont {Grenier}}, \bibinfo
  {author} {\bibfnamefont {J.}~\bibnamefont {McEnery}}, \bibinfo {author}
  {\bibfnamefont {M.}~\bibnamefont {Mallamaci}}, \bibinfo {author}
  {\bibfnamefont {M.}~\bibnamefont {Tavani}}, \bibinfo {author} {\bibfnamefont
  {U.}~\bibnamefont {Oberlack}}, \bibinfo {author} {\bibfnamefont
  {L.}~\bibnamefont {Hanlon}}, \bibinfo {author} {\bibfnamefont
  {R.}~\bibnamefont {Walter}}, \bibinfo {author} {\bibfnamefont
  {A.}~\bibnamefont {Argan}},  \emph {et~al.},\ }\href {\doibase
  https://doi.org/10.1016/j.jheap.2018.07.001} {\bibfield  {journal} {\bibinfo
  {journal} {J. High Energy Astrophys.}\ }\textbf {\bibinfo {volume} {19}},\
  \bibinfo {pages} {1 } (\bibinfo {year} {2018})}\BibitemShut {NoStop}%
\bibitem [{\citenamefont {Koning}\ and\ \citenamefont {Delaroche}(2003)}]{49}%
  \BibitemOpen
  \bibfield  {author} {\bibinfo {author} {\bibfnamefont {A.}~\bibnamefont
  {Koning}}\ and\ \bibinfo {author} {\bibfnamefont {J.}~\bibnamefont
  {Delaroche}},\ }\href {\doibase 10.1016/S0375-9474(02)01321-0} {\bibfield
  {journal} {\bibinfo  {journal} {Nucl. Phys. A}\ }\textbf {\bibinfo {volume}
  {713}},\ \bibinfo {pages} {231} (\bibinfo {year} {2003})}\BibitemShut
  {NoStop}%
\bibitem [{\citenamefont {Soukhovitski}\ \emph {et~al.}(2016)\citenamefont
  {Soukhovitski}, \citenamefont {Capote}, \citenamefont {Quesada},
  \citenamefont {Chiba},\ and\ \citenamefont {Martyanov}}]{50}%
  \BibitemOpen
  \bibfield  {author} {\bibinfo {author} {\bibfnamefont {E.~S.}\ \bibnamefont
  {Soukhovitski}}, \bibinfo {author} {\bibfnamefont {R.}~\bibnamefont
  {Capote}}, \bibinfo {author} {\bibfnamefont {J.~M.}\ \bibnamefont {Quesada}},
  \bibinfo {author} {\bibfnamefont {S.}~\bibnamefont {Chiba}}, \ and\ \bibinfo
  {author} {\bibfnamefont {D.~S.}\ \bibnamefont {Martyanov}},\ }\href {\doibase
  10.1103/PhysRevC.94.064605} {\bibfield  {journal} {\bibinfo  {journal} {Phys.
  Rev. C}\ }\textbf {\bibinfo {volume} {94}},\ \bibinfo {pages} {064605}
  (\bibinfo {year} {2016})}\BibitemShut {NoStop}%
\bibitem [{\citenamefont {Quesada}\ \emph {et~al.}(2003)\citenamefont
  {Quesada}, \citenamefont {Capote}, \citenamefont {Molina},\ and\
  \citenamefont {Lozano}}]{Quesada}%
  \BibitemOpen
  \bibfield  {author} {\bibinfo {author} {\bibfnamefont {J.}~\bibnamefont
  {Quesada}}, \bibinfo {author} {\bibfnamefont {R.}~\bibnamefont {Capote}},
  \bibinfo {author} {\bibfnamefont {A.}~\bibnamefont {Molina}}, \ and\ \bibinfo
  {author} {\bibfnamefont {M.}~\bibnamefont {Lozano}},\ }\href {\doibase
  https://doi.org/10.1016/S0010-4655(03)00157-7} {\bibfield  {journal}
  {\bibinfo  {journal} {Comput. Phys. Commun.}\ }\textbf {\bibinfo {volume}
  {153}},\ \bibinfo {pages} {97 } (\bibinfo {year} {2003})}\BibitemShut
  {NoStop}%
\bibitem [{\citenamefont {Davydov}\ and\ \citenamefont {Filippov}(1958)}]{51}%
  \BibitemOpen
  \bibfield  {author} {\bibinfo {author} {\bibfnamefont {A.}~\bibnamefont
  {Davydov}}\ and\ \bibinfo {author} {\bibfnamefont {G.}~\bibnamefont
  {Filippov}},\ }\href {\doibase https://doi.org/10.1016/0029-5582(58)90153-6}
  {\bibfield  {journal} {\bibinfo  {journal} {Nucl. Phys.}\ }\textbf {\bibinfo
  {volume} {8}},\ \bibinfo {pages} {237 } (\bibinfo {year} {1958})}\BibitemShut
  {NoStop}%
\bibitem [{\citenamefont {Esser}\ \emph {et~al.}(1997)\citenamefont {Esser},
  \citenamefont {Neuneyer}, \citenamefont {Casten},\ and\ \citenamefont {von
  Brentano}}]{52}%
  \BibitemOpen
  \bibfield  {author} {\bibinfo {author} {\bibfnamefont {L.}~\bibnamefont
  {Esser}}, \bibinfo {author} {\bibfnamefont {U.}~\bibnamefont {Neuneyer}},
  \bibinfo {author} {\bibfnamefont {R.~F.}\ \bibnamefont {Casten}}, \ and\
  \bibinfo {author} {\bibfnamefont {P.}~\bibnamefont {von Brentano}},\ }\href
  {\doibase 10.1103/PhysRevC.55.206} {\bibfield  {journal} {\bibinfo  {journal}
  {Phys. Rev. C}\ }\textbf {\bibinfo {volume} {55}},\ \bibinfo {pages} {206}
  (\bibinfo {year} {1997})}\BibitemShut {NoStop}%
\bibitem [{\citenamefont {Glatz}\ \emph {et~al.}(1986)\citenamefont {Glatz},
  \citenamefont {Norbert}, \citenamefont {Bitterwolf}, \citenamefont {Burkard},
  \citenamefont {Heidinger}, \citenamefont {Kern}, \citenamefont {Lehmann},
  \citenamefont {R{\"o}pke}, \citenamefont {Siefert}, \citenamefont
  {Schneider},\ and\ \citenamefont {Wildenthal}}]{53}%
  \BibitemOpen
  \bibfield  {author} {\bibinfo {author} {\bibfnamefont {F.}~\bibnamefont
  {Glatz}}, \bibinfo {author} {\bibfnamefont {S.}~\bibnamefont {Norbert}},
  \bibinfo {author} {\bibfnamefont {E.}~\bibnamefont {Bitterwolf}}, \bibinfo
  {author} {\bibfnamefont {A.}~\bibnamefont {Burkard}}, \bibinfo {author}
  {\bibfnamefont {F.}~\bibnamefont {Heidinger}}, \bibinfo {author}
  {\bibfnamefont {T.}~\bibnamefont {Kern}}, \bibinfo {author} {\bibfnamefont
  {R.}~\bibnamefont {Lehmann}}, \bibinfo {author} {\bibfnamefont
  {H.}~\bibnamefont {R{\"o}pke}}, \bibinfo {author} {\bibfnamefont
  {J.}~\bibnamefont {Siefert}}, \bibinfo {author} {\bibfnamefont
  {C.}~\bibnamefont {Schneider}}, \ and\ \bibinfo {author} {\bibfnamefont
  {B.~H.}\ \bibnamefont {Wildenthal}},\ }\href {\doibase 10.1007/BF01325130}
  {\bibfield  {journal} {\bibinfo  {journal} {Z. Phys. A.}\ }\textbf {\bibinfo
  {volume} {324}},\ \bibinfo {pages} {187} (\bibinfo {year}
  {1986})}\BibitemShut {NoStop}%
\bibitem [{\citenamefont {Alons}\ \emph {et~al.}(1981)\citenamefont {Alons},
  \citenamefont {Blok}, \citenamefont {Hienen},\ and\ \citenamefont
  {Blok}}]{54}%
  \BibitemOpen
  \bibfield  {author} {\bibinfo {author} {\bibfnamefont {P.}~\bibnamefont
  {Alons}}, \bibinfo {author} {\bibfnamefont {H.}~\bibnamefont {Blok}},
  \bibinfo {author} {\bibfnamefont {J.}~\bibnamefont {Hienen}}, \ and\ \bibinfo
  {author} {\bibfnamefont {J.}~\bibnamefont {Blok}},\ }\href {\doibase
  https://doi.org/10.1016/0375-9474(81)90277-3} {\bibfield  {journal} {\bibinfo
   {journal} {Nucl. Phys. A}\ }\textbf {\bibinfo {volume} {367}},\ \bibinfo
  {pages} {41 } (\bibinfo {year} {1981})}\BibitemShut {NoStop}%
\bibitem [{\citenamefont {Dybdal}\ \emph {et~al.}(1981)\citenamefont {Dybdal},
  \citenamefont {Forster}, \citenamefont {Hornshøj}, \citenamefont {Rud},\
  and\ \citenamefont {Straede}}]{55}%
  \BibitemOpen
  \bibfield  {author} {\bibinfo {author} {\bibfnamefont {K.}~\bibnamefont
  {Dybdal}}, \bibinfo {author} {\bibfnamefont {J.}~\bibnamefont {Forster}},
  \bibinfo {author} {\bibfnamefont {P.}~\bibnamefont {Hornshøj}}, \bibinfo
  {author} {\bibfnamefont {N.}~\bibnamefont {Rud}}, \ and\ \bibinfo {author}
  {\bibfnamefont {C.}~\bibnamefont {Straede}},\ }\href {\doibase
  https://doi.org/10.1016/0375-9474(81)90247-5} {\bibfield  {journal} {\bibinfo
   {journal} {Nucl. Phys. A}\ }\textbf {\bibinfo {volume} {359}},\ \bibinfo
  {pages} {431 } (\bibinfo {year} {1981})}\BibitemShut {NoStop}%
\bibitem [{\citenamefont {Kunieda}\ \emph {et~al.}(2007)\citenamefont
  {Kunieda}, \citenamefont {Chiba}, \citenamefont {Shibata}, \citenamefont
  {Ichihara},\ and\ \citenamefont {Sukhovitsk{\~i}}}]{56}%
  \BibitemOpen
  \bibfield  {author} {\bibinfo {author} {\bibfnamefont {S.}~\bibnamefont
  {Kunieda}}, \bibinfo {author} {\bibfnamefont {S.}~\bibnamefont {Chiba}},
  \bibinfo {author} {\bibfnamefont {K.}~\bibnamefont {Shibata}}, \bibinfo
  {author} {\bibfnamefont {A.}~\bibnamefont {Ichihara}}, \ and\ \bibinfo
  {author} {\bibfnamefont {E.~S.}\ \bibnamefont {Sukhovitsk{\~i}}},\
  }\href@noop {} {\bibfield  {journal} {\bibinfo  {journal} {J. Nucl. Sci.
  Technol.}\ }\textbf {\bibinfo {volume} {44}},\ \bibinfo {pages} {838}
  (\bibinfo {year} {2007})}\BibitemShut {NoStop}%
\bibitem [{\citenamefont {Olsson}\ \emph {et~al.}(1990)\citenamefont {Olsson},
  \citenamefont {Ramström},\ and\ \citenamefont {Trostell}}]{59}%
  \BibitemOpen
  \bibfield  {author} {\bibinfo {author} {\bibfnamefont {N.}~\bibnamefont
  {Olsson}}, \bibinfo {author} {\bibfnamefont {E.}~\bibnamefont {Ramström}}, \
  and\ \bibinfo {author} {\bibfnamefont {B.}~\bibnamefont {Trostell}},\ }\href
  {\doibase https://doi.org/10.1016/0375-9474(90)90096-5} {\bibfield  {journal}
  {\bibinfo  {journal} {Nucl. Phys. A}\ }\textbf {\bibinfo {volume} {513}},\
  \bibinfo {pages} {205 } (\bibinfo {year} {1990})}\BibitemShut {NoStop}%
\bibitem [{\citenamefont {Tailor}\ \emph {et~al.}(1983)\citenamefont {Tailor},
  \citenamefont {Rapaport}, \citenamefont {Finlay},\ and\ \citenamefont
  {Randers-Pehrson}}]{60}%
  \BibitemOpen
  \bibfield  {author} {\bibinfo {author} {\bibfnamefont {R.}~\bibnamefont
  {Tailor}}, \bibinfo {author} {\bibfnamefont {J.}~\bibnamefont {Rapaport}},
  \bibinfo {author} {\bibfnamefont {R.}~\bibnamefont {Finlay}}, \ and\ \bibinfo
  {author} {\bibfnamefont {G.}~\bibnamefont {Randers-Pehrson}},\ }\href
  {\doibase https://doi.org/10.1016/0375-9474(83)90528-6} {\bibfield  {journal}
  {\bibinfo  {journal} {Nucl. Phys. A}\ }\textbf {\bibinfo {volume} {401}},\
  \bibinfo {pages} {237 } (\bibinfo {year} {1983})}\BibitemShut {NoStop}%
\bibitem [{\citenamefont {Roy}\ \emph {et~al.}(1983)\citenamefont {Roy},
  \citenamefont {Lamontagne}, \citenamefont {Slobodrian}, \citenamefont
  {Arvieux}, \citenamefont {Birchall},\ and\ \citenamefont {Conzett}}]{78}%
  \BibitemOpen
  \bibfield  {author} {\bibinfo {author} {\bibfnamefont {R.}~\bibnamefont
  {Roy}}, \bibinfo {author} {\bibfnamefont {C.}~\bibnamefont {Lamontagne}},
  \bibinfo {author} {\bibfnamefont {R.}~\bibnamefont {Slobodrian}}, \bibinfo
  {author} {\bibfnamefont {J.}~\bibnamefont {Arvieux}}, \bibinfo {author}
  {\bibfnamefont {J.}~\bibnamefont {Birchall}}, \ and\ \bibinfo {author}
  {\bibfnamefont {H.}~\bibnamefont {Conzett}},\ }\href {\doibase
  https://doi.org/10.1016/0375-9474(83)90504-3} {\bibfield  {journal} {\bibinfo
   {journal} {Nucl. Phys. A}\ }\textbf {\bibinfo {volume} {411}},\ \bibinfo
  {pages} {1 } (\bibinfo {year} {1983})}\BibitemShut {NoStop}%
\bibitem [{\citenamefont {Fabrici}\ \emph {et~al.}(1980)\citenamefont
  {Fabrici}, \citenamefont {Micheletti}, \citenamefont {Pignanelli},
  \citenamefont {Resmini}, \citenamefont {{De Leo}}, \citenamefont {D'Erasmo},
  \citenamefont {Pantaleo}, \citenamefont {Escudi{\'e}},\ and\ \citenamefont
  {Tarrats}}]{79}%
  \BibitemOpen
  \bibfield  {author} {\bibinfo {author} {\bibfnamefont {E.}~\bibnamefont
  {Fabrici}}, \bibinfo {author} {\bibfnamefont {S.}~\bibnamefont {Micheletti}},
  \bibinfo {author} {\bibfnamefont {M.}~\bibnamefont {Pignanelli}}, \bibinfo
  {author} {\bibfnamefont {F.~G.}\ \bibnamefont {Resmini}}, \bibinfo {author}
  {\bibfnamefont {R.}~\bibnamefont {{De Leo}}}, \bibinfo {author}
  {\bibfnamefont {G.}~\bibnamefont {D'Erasmo}}, \bibinfo {author}
  {\bibfnamefont {A.}~\bibnamefont {Pantaleo}}, \bibinfo {author}
  {\bibfnamefont {J.~L.}\ \bibnamefont {Escudi{\'e}}}, \ and\ \bibinfo {author}
  {\bibfnamefont {A.}~\bibnamefont {Tarrats}},\ }\href {\doibase
  10.1103/PhysRevC.21.830} {\bibfield  {journal} {\bibinfo  {journal} {Phys.
  Rev. C}\ }\textbf {\bibinfo {volume} {21}},\ \bibinfo {pages} {830} (\bibinfo
  {year} {1980})}\BibitemShut {NoStop}%
\bibitem [{\citenamefont {Hasell}\ \emph {et~al.}(1983)\citenamefont {Hasell},
  \citenamefont {Davison}, \citenamefont {Nasr}, \citenamefont {Murdoch},
  \citenamefont {Sourkes},\ and\ \citenamefont {van Oers}}]{80}%
  \BibitemOpen
  \bibfield  {author} {\bibinfo {author} {\bibfnamefont {D.~K.}\ \bibnamefont
  {Hasell}}, \bibinfo {author} {\bibfnamefont {N.~E.}\ \bibnamefont {Davison}},
  \bibinfo {author} {\bibfnamefont {T.~N.}\ \bibnamefont {Nasr}}, \bibinfo
  {author} {\bibfnamefont {B.~T.}\ \bibnamefont {Murdoch}}, \bibinfo {author}
  {\bibfnamefont {A.~M.}\ \bibnamefont {Sourkes}}, \ and\ \bibinfo {author}
  {\bibfnamefont {W.~T.~H.}\ \bibnamefont {van Oers}},\ }\href {\doibase
  10.1103/PhysRevC.27.482} {\bibfield  {journal} {\bibinfo  {journal} {Phys.
  Rev. C}\ }\textbf {\bibinfo {volume} {27}},\ \bibinfo {pages} {482} (\bibinfo
  {year} {1983})}\BibitemShut {NoStop}%
\bibitem [{\citenamefont {Fulmer}(1962)}]{81}%
  \BibitemOpen
  \bibfield  {author} {\bibinfo {author} {\bibfnamefont {C.~B.}\ \bibnamefont
  {Fulmer}},\ }\href {\doibase 10.1103/PhysRev.125.631} {\bibfield  {journal}
  {\bibinfo  {journal} {Phys. Rev.}\ }\textbf {\bibinfo {volume} {125}},\
  \bibinfo {pages} {631} (\bibinfo {year} {1962})}\BibitemShut {NoStop}%
\bibitem [{\citenamefont {Eenmaa}\ \emph {et~al.}(1974)\citenamefont {Eenmaa},
  \citenamefont {Cole}, \citenamefont {Waddell}, \citenamefont {Sandhu},\ and\
  \citenamefont {Dittman}}]{82}%
  \BibitemOpen
  \bibfield  {author} {\bibinfo {author} {\bibfnamefont {J.}~\bibnamefont
  {Eenmaa}}, \bibinfo {author} {\bibfnamefont {R.}~\bibnamefont {Cole}},
  \bibinfo {author} {\bibfnamefont {C.}~\bibnamefont {Waddell}}, \bibinfo
  {author} {\bibfnamefont {H.}~\bibnamefont {Sandhu}}, \ and\ \bibinfo {author}
  {\bibfnamefont {R.}~\bibnamefont {Dittman}},\ }\href {\doibase
  https://doi.org/10.1016/0375-9474(74)90025-6} {\bibfield  {journal} {\bibinfo
   {journal} {Nucl. Phys. A}\ }\textbf {\bibinfo {volume} {218}},\ \bibinfo
  {pages} {125 } (\bibinfo {year} {1974})}\BibitemShut {NoStop}%
\bibitem [{\citenamefont {Rush}\ \emph {et~al.}(1967)\citenamefont {Rush},
  \citenamefont {Burge}, \citenamefont {Lewis}, \citenamefont {Smith},\ and\
  \citenamefont {Ganguly}}]{83}%
  \BibitemOpen
  \bibfield  {author} {\bibinfo {author} {\bibfnamefont {A.}~\bibnamefont
  {Rush}}, \bibinfo {author} {\bibfnamefont {E.}~\bibnamefont {Burge}},
  \bibinfo {author} {\bibfnamefont {V.}~\bibnamefont {Lewis}}, \bibinfo
  {author} {\bibfnamefont {D.}~\bibnamefont {Smith}}, \ and\ \bibinfo {author}
  {\bibfnamefont {N.}~\bibnamefont {Ganguly}},\ }\href {\doibase
  https://doi.org/10.1016/0375-9474(67)90561-1} {\bibfield  {journal} {\bibinfo
   {journal} {Nucl. Phys. A}\ }\textbf {\bibinfo {volume} {104}},\ \bibinfo
  {pages} {340 } (\bibinfo {year} {1967})}\BibitemShut {NoStop}%
\bibitem [{\citenamefont {Ohnuma}\ \emph {et~al.}(1980)\citenamefont {Ohnuma},
  \citenamefont {Kasagi}, \citenamefont {Kakimoto}, \citenamefont {Kubono},\
  and\ \citenamefont {Koyama}}]{84}%
  \BibitemOpen
  \bibfield  {author} {\bibinfo {author} {\bibfnamefont {H.}~\bibnamefont
  {Ohnuma}}, \bibinfo {author} {\bibfnamefont {J.}~\bibnamefont {Kasagi}},
  \bibinfo {author} {\bibfnamefont {F.}~\bibnamefont {Kakimoto}}, \bibinfo
  {author} {\bibfnamefont {S.}~\bibnamefont {Kubono}}, \ and\ \bibinfo {author}
  {\bibfnamefont {K.}~\bibnamefont {Koyama}},\ }\href {\doibase
  10.1143/JPSJ.48.1812} {\bibfield  {journal} {\bibinfo  {journal} {J. Phys.
  Soc. Jpn}\ }\textbf {\bibinfo {volume} {48}},\ \bibinfo {pages} {1812}
  (\bibinfo {year} {1980})}\BibitemShut {NoStop}%
\bibitem [{\citenamefont {Kato}\ \emph {et~al.}(1985)\citenamefont {Kato},
  \citenamefont {Okada}, \citenamefont {Kondo}, \citenamefont {Hosono},
  \citenamefont {Saito}, \citenamefont {Matsuoka}, \citenamefont {Hatanaka},
  \citenamefont {Noro}, \citenamefont {Nagamachi}, \citenamefont {Shimizu},
  \citenamefont {Ogino}, \citenamefont {Kadota}, \citenamefont {Matsuki},\ and\
  \citenamefont {Wakai}}]{85}%
  \BibitemOpen
  \bibfield  {author} {\bibinfo {author} {\bibfnamefont {S.}~\bibnamefont
  {Kato}}, \bibinfo {author} {\bibfnamefont {K.}~\bibnamefont {Okada}},
  \bibinfo {author} {\bibfnamefont {M.}~\bibnamefont {Kondo}}, \bibinfo
  {author} {\bibfnamefont {K.}~\bibnamefont {Hosono}}, \bibinfo {author}
  {\bibfnamefont {T.}~\bibnamefont {Saito}}, \bibinfo {author} {\bibfnamefont
  {N.}~\bibnamefont {Matsuoka}}, \bibinfo {author} {\bibfnamefont
  {K.}~\bibnamefont {Hatanaka}}, \bibinfo {author} {\bibfnamefont
  {T.}~\bibnamefont {Noro}}, \bibinfo {author} {\bibfnamefont {S.}~\bibnamefont
  {Nagamachi}}, \bibinfo {author} {\bibfnamefont {H.}~\bibnamefont {Shimizu}},
  \bibinfo {author} {\bibfnamefont {K.}~\bibnamefont {Ogino}}, \bibinfo
  {author} {\bibfnamefont {Y.}~\bibnamefont {Kadota}}, \bibinfo {author}
  {\bibfnamefont {S.}~\bibnamefont {Matsuki}}, \ and\ \bibinfo {author}
  {\bibfnamefont {M.}~\bibnamefont {Wakai}},\ }\href {\doibase
  10.1103/PhysRevC.31.1616} {\bibfield  {journal} {\bibinfo  {journal} {Phys.
  Rev. C}\ }\textbf {\bibinfo {volume} {31}},\ \bibinfo {pages} {1616}
  (\bibinfo {year} {1985})}\BibitemShut {NoStop}%
\bibitem [{\citenamefont {Sakaguchi}\ \emph {et~al.}(1982)\citenamefont
  {Sakaguchi}, \citenamefont {Nakamura}, \citenamefont {Hatanaka},
  \citenamefont {Goto}, \citenamefont {Noro}, \citenamefont {Ohtani},
  \citenamefont {Sakamoto}, \citenamefont {Ogawa},\ and\ \citenamefont
  {Kobayashi}}]{86}%
  \BibitemOpen
  \bibfield  {author} {\bibinfo {author} {\bibfnamefont {H.}~\bibnamefont
  {Sakaguchi}}, \bibinfo {author} {\bibfnamefont {M.}~\bibnamefont {Nakamura}},
  \bibinfo {author} {\bibfnamefont {K.}~\bibnamefont {Hatanaka}}, \bibinfo
  {author} {\bibfnamefont {A.}~\bibnamefont {Goto}}, \bibinfo {author}
  {\bibfnamefont {T.}~\bibnamefont {Noro}}, \bibinfo {author} {\bibfnamefont
  {F.}~\bibnamefont {Ohtani}}, \bibinfo {author} {\bibfnamefont
  {H.}~\bibnamefont {Sakamoto}}, \bibinfo {author} {\bibfnamefont
  {H.}~\bibnamefont {Ogawa}}, \ and\ \bibinfo {author} {\bibfnamefont
  {S.}~\bibnamefont {Kobayashi}},\ }\href {\doibase 10.1103/PhysRevC.26.944}
  {\bibfield  {journal} {\bibinfo  {journal} {Phys. Rev. C}\ }\textbf {\bibinfo
  {volume} {26}},\ \bibinfo {pages} {944} (\bibinfo {year} {1982})}\BibitemShut
  {NoStop}%
\bibitem [{\citenamefont {Hatanaka}\ \emph {et~al.}(1984)\citenamefont
  {Hatanaka}, \citenamefont {Fujiwara}, \citenamefont {Hosono}, \citenamefont
  {Matsuoka}, \citenamefont {Saito},\ and\ \citenamefont {Sakai}}]{87}%
  \BibitemOpen
  \bibfield  {author} {\bibinfo {author} {\bibfnamefont {K.}~\bibnamefont
  {Hatanaka}}, \bibinfo {author} {\bibfnamefont {M.}~\bibnamefont {Fujiwara}},
  \bibinfo {author} {\bibfnamefont {K.}~\bibnamefont {Hosono}}, \bibinfo
  {author} {\bibfnamefont {N.}~\bibnamefont {Matsuoka}}, \bibinfo {author}
  {\bibfnamefont {T.}~\bibnamefont {Saito}}, \ and\ \bibinfo {author}
  {\bibfnamefont {H.}~\bibnamefont {Sakai}},\ }\href {\doibase
  10.1103/PhysRevC.29.13} {\bibfield  {journal} {\bibinfo  {journal} {Phys.
  Rev. C}\ }\textbf {\bibinfo {volume} {29}},\ \bibinfo {pages} {13} (\bibinfo
  {year} {1984})}\BibitemShut {NoStop}%
\bibitem [{\citenamefont {Schwandt}\ \emph
  {et~al.}(1982{\natexlab{a}})\citenamefont {Schwandt}, \citenamefont {Meyer},
  \citenamefont {Jacobs}, \citenamefont {Bacher}, \citenamefont {Vigdor},
  \citenamefont {Kaitchuck},\ and\ \citenamefont {Donoghue}}]{88}%
  \BibitemOpen
  \bibfield  {author} {\bibinfo {author} {\bibfnamefont {P.}~\bibnamefont
  {Schwandt}}, \bibinfo {author} {\bibfnamefont {H.~O.}\ \bibnamefont {Meyer}},
  \bibinfo {author} {\bibfnamefont {W.~W.}\ \bibnamefont {Jacobs}}, \bibinfo
  {author} {\bibfnamefont {A.~D.}\ \bibnamefont {Bacher}}, \bibinfo {author}
  {\bibfnamefont {S.~E.}\ \bibnamefont {Vigdor}}, \bibinfo {author}
  {\bibfnamefont {M.~D.}\ \bibnamefont {Kaitchuck}}, \ and\ \bibinfo {author}
  {\bibfnamefont {T.~R.}\ \bibnamefont {Donoghue}},\ }\href {\doibase
  10.1103/PhysRevC.26.55} {\bibfield  {journal} {\bibinfo  {journal} {Phys.
  Rev. C}\ }\textbf {\bibinfo {volume} {26}},\ \bibinfo {pages} {55} (\bibinfo
  {year} {1982}{\natexlab{a}})}\BibitemShut {NoStop}%
\bibitem [{\citenamefont {Hicks}\ \emph {et~al.}(1988)\citenamefont {Hicks},
  \citenamefont {Jeppesen}, \citenamefont {Lin}, \citenamefont {Abegg},
  \citenamefont {Jackson}, \citenamefont {H{\"a}usser}, \citenamefont
  {Lisantti}, \citenamefont {Miller}, \citenamefont {Rost}, \citenamefont
  {Sawafta}, \citenamefont {Vetterli},\ and\ \citenamefont {Yen}}]{89}%
  \BibitemOpen
  \bibfield  {author} {\bibinfo {author} {\bibfnamefont {K.~H.}\ \bibnamefont
  {Hicks}}, \bibinfo {author} {\bibfnamefont {R.~G.}\ \bibnamefont {Jeppesen}},
  \bibinfo {author} {\bibfnamefont {C.~C.~K.}\ \bibnamefont {Lin}}, \bibinfo
  {author} {\bibfnamefont {R.}~\bibnamefont {Abegg}}, \bibinfo {author}
  {\bibfnamefont {K.~P.}\ \bibnamefont {Jackson}}, \bibinfo {author}
  {\bibfnamefont {O.}~\bibnamefont {H{\"a}usser}}, \bibinfo {author}
  {\bibfnamefont {J.}~\bibnamefont {Lisantti}}, \bibinfo {author}
  {\bibfnamefont {C.~A.}\ \bibnamefont {Miller}}, \bibinfo {author}
  {\bibfnamefont {E.}~\bibnamefont {Rost}}, \bibinfo {author} {\bibfnamefont
  {R.}~\bibnamefont {Sawafta}}, \bibinfo {author} {\bibfnamefont {M.~C.}\
  \bibnamefont {Vetterli}}, \ and\ \bibinfo {author} {\bibfnamefont
  {S.}~\bibnamefont {Yen}},\ }\href {\doibase 10.1103/PhysRevC.38.229}
  {\bibfield  {journal} {\bibinfo  {journal} {Phys. Rev. C}\ }\textbf {\bibinfo
  {volume} {38}},\ \bibinfo {pages} {229} (\bibinfo {year} {1988})}\BibitemShut
  {NoStop}%
\bibitem [{\citenamefont {Herman}\ \emph {et~al.}(1976)\citenamefont {Herman},
  \citenamefont {Marcinkowski}, \citenamefont {Zwieglinski}, \citenamefont
  {Augustyniak}, \citenamefont {Bielewicz},\ and\ \citenamefont {Zych}}]{90}%
  \BibitemOpen
  \bibfield  {author} {\bibinfo {author} {\bibfnamefont {M.}~\bibnamefont
  {Herman}}, \bibinfo {author} {\bibfnamefont {A.}~\bibnamefont
  {Marcinkowski}}, \bibinfo {author} {\bibfnamefont {B.}~\bibnamefont
  {Zwieglinski}}, \bibinfo {author} {\bibfnamefont {W.}~\bibnamefont
  {Augustyniak}}, \bibinfo {author} {\bibfnamefont {J.}~\bibnamefont
  {Bielewicz}}, \ and\ \bibinfo {author} {\bibfnamefont {W.}~\bibnamefont
  {Zych}},\ }\href {\doibase 10.1088/0305-4616/2/11/008} {\bibfield  {journal}
  {\bibinfo  {journal} {J. Phys. G}\ }\textbf {\bibinfo {volume} {2}},\
  \bibinfo {pages} {831} (\bibinfo {year} {1976})}\BibitemShut {NoStop}%
\bibitem [{\citenamefont {Crawley}\ and\ \citenamefont {Garvey}(1967)}]{91}%
  \BibitemOpen
  \bibfield  {author} {\bibinfo {author} {\bibfnamefont {G.~M.}\ \bibnamefont
  {Crawley}}\ and\ \bibinfo {author} {\bibfnamefont {G.~T.}\ \bibnamefont
  {Garvey}},\ }\href {\doibase 10.1103/PhysRev.160.981} {\bibfield  {journal}
  {\bibinfo  {journal} {Phys. Rev.}\ }\textbf {\bibinfo {volume} {160}},\
  \bibinfo {pages} {981} (\bibinfo {year} {1967})}\BibitemShut {NoStop}%
\bibitem [{\citenamefont {Blair}\ \emph {et~al.}(1970)\citenamefont {Blair},
  \citenamefont {Glashausser}, \citenamefont {de~Swiniarski}, \citenamefont
  {Goudergues}, \citenamefont {Lombard}, \citenamefont {Mayer}, \citenamefont
  {Thirion},\ and\ \citenamefont {Vaganov}}]{64}%
  \BibitemOpen
  \bibfield  {author} {\bibinfo {author} {\bibfnamefont {A.~G.}\ \bibnamefont
  {Blair}}, \bibinfo {author} {\bibfnamefont {C.}~\bibnamefont {Glashausser}},
  \bibinfo {author} {\bibfnamefont {R.}~\bibnamefont {de~Swiniarski}}, \bibinfo
  {author} {\bibfnamefont {J.}~\bibnamefont {Goudergues}}, \bibinfo {author}
  {\bibfnamefont {R.}~\bibnamefont {Lombard}}, \bibinfo {author} {\bibfnamefont
  {B.}~\bibnamefont {Mayer}}, \bibinfo {author} {\bibfnamefont
  {J.}~\bibnamefont {Thirion}}, \ and\ \bibinfo {author} {\bibfnamefont
  {P.}~\bibnamefont {Vaganov}},\ }\href {\doibase 10.1103/PhysRevC.1.444}
  {\bibfield  {journal} {\bibinfo  {journal} {Phys. Rev. C}\ }\textbf {\bibinfo
  {volume} {1}},\ \bibinfo {pages} {444} (\bibinfo {year} {1970})}\BibitemShut
  {NoStop}%
\bibitem [{\citenamefont {Zwi\ifmmode \mbox{\c{e}}\else
  \c{e}\fi{}gli\ifmmode~\acute{n}\else \'{n}\fi{}ski}\ \emph
  {et~al.}(1983)\citenamefont {Zwi\ifmmode \mbox{\c{e}}\else
  \c{e}\fi{}gli\ifmmode~\acute{n}\else \'{n}\fi{}ski}, \citenamefont {Crawley},
  \citenamefont {Nolen},\ and\ \citenamefont {Ronningen}}]{92}%
  \BibitemOpen
  \bibfield  {author} {\bibinfo {author} {\bibfnamefont {B.}~\bibnamefont
  {Zwi\ifmmode \mbox{\c{e}}\else \c{e}\fi{}gli\ifmmode~\acute{n}\else
  \'{n}\fi{}ski}}, \bibinfo {author} {\bibfnamefont {G.~M.}\ \bibnamefont
  {Crawley}}, \bibinfo {author} {\bibfnamefont {J.~A.}\ \bibnamefont {Nolen}},
  \ and\ \bibinfo {author} {\bibfnamefont {R.~M.}\ \bibnamefont {Ronningen}},\
  }\href {\doibase 10.1103/PhysRevC.28.542} {\bibfield  {journal} {\bibinfo
  {journal} {Phys. Rev. C}\ }\textbf {\bibinfo {volume} {28}},\ \bibinfo
  {pages} {542} (\bibinfo {year} {1983})}\BibitemShut {NoStop}%
\bibitem [{\citenamefont {Kliczewski}\ and\ \citenamefont
  {Lewandowski}(1978)}]{61}%
  \BibitemOpen
  \bibfield  {author} {\bibinfo {author} {\bibfnamefont {S.}~\bibnamefont
  {Kliczewski}}\ and\ \bibinfo {author} {\bibfnamefont {Z.}~\bibnamefont
  {Lewandowski}},\ }\href {\doibase 10.1016/0375-9474(78)90237-3} {\bibfield
  {journal} {\bibinfo  {journal} {Nucl. Phys. A}\ }\textbf {\bibinfo {volume}
  {304}},\ \bibinfo {pages} {269} (\bibinfo {year} {1978})}\BibitemShut
  {NoStop}%
\bibitem [{\citenamefont {Höhn}\ \emph {et~al.}(1969)\citenamefont {Höhn},
  \citenamefont {Pose}, \citenamefont {Seeliger},\ and\ \citenamefont
  {Reif}}]{62}%
  \BibitemOpen
  \bibfield  {author} {\bibinfo {author} {\bibfnamefont {J.}~\bibnamefont
  {Höhn}}, \bibinfo {author} {\bibfnamefont {H.}~\bibnamefont {Pose}},
  \bibinfo {author} {\bibfnamefont {D.}~\bibnamefont {Seeliger}}, \ and\
  \bibinfo {author} {\bibfnamefont {R.}~\bibnamefont {Reif}},\ }\href {\doibase
  https://doi.org/10.1016/0375-9474(69)91053-7} {\bibfield  {journal} {\bibinfo
   {journal} {Nucl. Phys. A}\ }\textbf {\bibinfo {volume} {134}},\ \bibinfo
  {pages} {289 } (\bibinfo {year} {1969})}\BibitemShut {NoStop}%
\bibitem [{\citenamefont {Al-Ohali}\ \emph {et~al.}(2012)\citenamefont
  {Al-Ohali}, \citenamefont {Delaroche}, \citenamefont {Howell}, \citenamefont
  {Nagadi}, \citenamefont {Naqvi}, \citenamefont {Tornow}, \citenamefont
  {Walter},\ and\ \citenamefont {Weisel}}]{63}%
  \BibitemOpen
  \bibfield  {author} {\bibinfo {author} {\bibfnamefont {M.~A.}\ \bibnamefont
  {Al-Ohali}}, \bibinfo {author} {\bibfnamefont {J.~P.}\ \bibnamefont
  {Delaroche}}, \bibinfo {author} {\bibfnamefont {C.~R.}\ \bibnamefont
  {Howell}}, \bibinfo {author} {\bibfnamefont {M.~M.}\ \bibnamefont {Nagadi}},
  \bibinfo {author} {\bibfnamefont {A.~A.}\ \bibnamefont {Naqvi}}, \bibinfo
  {author} {\bibfnamefont {W.}~\bibnamefont {Tornow}}, \bibinfo {author}
  {\bibfnamefont {R.~L.}\ \bibnamefont {Walter}}, \ and\ \bibinfo {author}
  {\bibfnamefont {G.~J.}\ \bibnamefont {Weisel}},\ }\href {\doibase
  10.1103/PhysRevC.86.034603} {\bibfield  {journal} {\bibinfo  {journal} {Phys.
  Rev. C}\ }\textbf {\bibinfo {volume} {86}},\ \bibinfo {pages} {034603}
  (\bibinfo {year} {2012})}\BibitemShut {NoStop}%
\bibitem [{\citenamefont {Alarcon}\ and\ \citenamefont {Rapaport}(1986)}]{65}%
  \BibitemOpen
  \bibfield  {author} {\bibinfo {author} {\bibfnamefont {R.}~\bibnamefont
  {Alarcon}}\ and\ \bibinfo {author} {\bibfnamefont {J.}~\bibnamefont
  {Rapaport}},\ }\href {\doibase 10.1016/0375-9474(86)90048-5} {\bibfield
  {journal} {\bibinfo  {journal} {Nucl. Phys. A}\ }\textbf {\bibinfo {volume}
  {458}},\ \bibinfo {pages} {502} (\bibinfo {year} {1986})}\BibitemShut
  {NoStop}%
\bibitem [{\citenamefont {DeVito}\ \emph {et~al.}(1983)\citenamefont {DeVito},
  \citenamefont {Austin}, \citenamefont {Berg}, \citenamefont {{De Leo}},\ and\
  \citenamefont {Sterrenburg}}]{66}%
  \BibitemOpen
  \bibfield  {author} {\bibinfo {author} {\bibfnamefont {R.~P.}\ \bibnamefont
  {DeVito}}, \bibinfo {author} {\bibfnamefont {S.~M.}\ \bibnamefont {Austin}},
  \bibinfo {author} {\bibfnamefont {U.~E.~P.}\ \bibnamefont {Berg}}, \bibinfo
  {author} {\bibfnamefont {R.}~\bibnamefont {{De Leo}}}, \ and\ \bibinfo
  {author} {\bibfnamefont {W.~A.}\ \bibnamefont {Sterrenburg}},\ }\href
  {\doibase 10.1103/PhysRevC.28.2530} {\bibfield  {journal} {\bibinfo
  {journal} {Phys. Rev. C}\ }\textbf {\bibinfo {volume} {28}},\ \bibinfo
  {pages} {2530} (\bibinfo {year} {1983})}\BibitemShut {NoStop}%
\bibitem [{\citenamefont {{M. Ibaraki}}\ \emph {et~al.}(2000)\citenamefont {{M.
  Ibaraki}}, \citenamefont {Miura}, \citenamefont {Nauchi}, \citenamefont
  {Hirasawa}, \citenamefont {Hirakawa}, \citenamefont {Nakashima},
  \citenamefont {Meigo}, \citenamefont {Iwamoto},\ and\ \citenamefont
  {Tanaka}}]{67}%
  \BibitemOpen
  \bibfield  {author} {\bibinfo {author} {\bibfnamefont {M.~B.}\ \bibnamefont
  {{M. Ibaraki}}}, \bibinfo {author} {\bibfnamefont {T.}~\bibnamefont {Miura}},
  \bibinfo {author} {\bibfnamefont {Y.}~\bibnamefont {Nauchi}}, \bibinfo
  {author} {\bibfnamefont {Y.}~\bibnamefont {Hirasawa}}, \bibinfo {author}
  {\bibfnamefont {N.}~\bibnamefont {Hirakawa}}, \bibinfo {author}
  {\bibfnamefont {H.}~\bibnamefont {Nakashima}}, \bibinfo {author}
  {\bibfnamefont {S.}~\bibnamefont {Meigo}}, \bibinfo {author} {\bibfnamefont
  {O.}~\bibnamefont {Iwamoto}}, \ and\ \bibinfo {author} {\bibfnamefont
  {S.}~\bibnamefont {Tanaka}},\ }\href {\doibase
  10.1080/00223131.2000.10874976} {\bibfield  {journal} {\bibinfo  {journal}
  {J. Nucl. Sci. Technol.}\ }\textbf {\bibinfo {volume} {37}},\ \bibinfo
  {pages} {683} (\bibinfo {year} {2000})}\BibitemShut {NoStop}%
\bibitem [{\citenamefont {Ibaraki}\ \emph {et~al.}(1999)\citenamefont
  {Ibaraki}, \citenamefont {Baba}, \citenamefont {Miura}, \citenamefont
  {Nauchi}, \citenamefont {Hirasawa}, \citenamefont {Hirakawa}, \citenamefont
  {Nakashima}, \citenamefont {ichiro Meigo}, \citenamefont {Iwamoto},\ and\
  \citenamefont {Tanaka}}]{68}%
  \BibitemOpen
  \bibfield  {author} {\bibinfo {author} {\bibfnamefont {M.}~\bibnamefont
  {Ibaraki}}, \bibinfo {author} {\bibfnamefont {M.}~\bibnamefont {Baba}},
  \bibinfo {author} {\bibfnamefont {T.}~\bibnamefont {Miura}}, \bibinfo
  {author} {\bibfnamefont {Y.}~\bibnamefont {Nauchi}}, \bibinfo {author}
  {\bibfnamefont {Y.}~\bibnamefont {Hirasawa}}, \bibinfo {author}
  {\bibfnamefont {N.}~\bibnamefont {Hirakawa}}, \bibinfo {author}
  {\bibfnamefont {H.}~\bibnamefont {Nakashima}}, \bibinfo {author}
  {\bibfnamefont {S.}~\bibnamefont {ichiro Meigo}}, \bibinfo {author}
  {\bibfnamefont {O.}~\bibnamefont {Iwamoto}}, \ and\ \bibinfo {author}
  {\bibfnamefont {S.}~\bibnamefont {Tanaka}},\ }\href@noop {} {\bibfield
  {journal} {\bibinfo  {journal} {Proceeding of the Ninth International
  Conference on Radiation Shielding, Tsukuba, Japan}\ } (\bibinfo {year}
  {October 17-22 1999})}\BibitemShut {NoStop}%
\bibitem [{\citenamefont {Hjort}\ \emph {et~al.}(1994)\citenamefont {Hjort},
  \citenamefont {Brady}, \citenamefont {Romero}, \citenamefont {Drummond},
  \citenamefont {Sorenson}, \citenamefont {Osborne}, \citenamefont
  {McEachern},\ and\ \citenamefont {Hansen}}]{69}%
  \BibitemOpen
  \bibfield  {author} {\bibinfo {author} {\bibfnamefont {E.~L.}\ \bibnamefont
  {Hjort}}, \bibinfo {author} {\bibfnamefont {F.~P.}\ \bibnamefont {Brady}},
  \bibinfo {author} {\bibfnamefont {J.~L.}\ \bibnamefont {Romero}}, \bibinfo
  {author} {\bibfnamefont {J.~R.}\ \bibnamefont {Drummond}}, \bibinfo {author}
  {\bibfnamefont {D.~S.}\ \bibnamefont {Sorenson}}, \bibinfo {author}
  {\bibfnamefont {J.~H.}\ \bibnamefont {Osborne}}, \bibinfo {author}
  {\bibfnamefont {B.}~\bibnamefont {McEachern}}, \ and\ \bibinfo {author}
  {\bibfnamefont {L.~F.}\ \bibnamefont {Hansen}},\ }\href {\doibase
  10.1103/PhysRevC.50.275} {\bibfield  {journal} {\bibinfo  {journal} {Phys.
  Rev. C}\ }\textbf {\bibinfo {volume} {50}},\ \bibinfo {pages} {275} (\bibinfo
  {year} {1994})}\BibitemShut {NoStop}%
\bibitem [{\citenamefont {Cohen}\ and\ \citenamefont {Cookson}(1961)}]{93}%
  \BibitemOpen
  \bibfield  {author} {\bibinfo {author} {\bibfnamefont {A.}~\bibnamefont
  {Cohen}}\ and\ \bibinfo {author} {\bibfnamefont {J.}~\bibnamefont
  {Cookson}},\ }\href {\doibase https://doi.org/10.1016/0029-5582(61)90428-X}
  {\bibfield  {journal} {\bibinfo  {journal} {Nucl. Phys.}\ }\textbf {\bibinfo
  {volume} {24}},\ \bibinfo {pages} {529 } (\bibinfo {year}
  {1961})}\BibitemShut {NoStop}%
\bibitem [{\citenamefont {Baugh}\ \emph {et~al.}(1965)\citenamefont {Baugh},
  \citenamefont {Greenless}, \citenamefont {Lilley},\ and\ \citenamefont
  {Roman}}]{94}%
  \BibitemOpen
  \bibfield  {author} {\bibinfo {author} {\bibfnamefont {D.}~\bibnamefont
  {Baugh}}, \bibinfo {author} {\bibfnamefont {G.}~\bibnamefont {Greenless}},
  \bibinfo {author} {\bibfnamefont {J.}~\bibnamefont {Lilley}}, \ and\ \bibinfo
  {author} {\bibfnamefont {S.}~\bibnamefont {Roman}},\ }\href {\doibase
  https://doi.org/10.1016/0029-5582(65)90292-0} {\bibfield  {journal} {\bibinfo
   {journal} {Nucl. Phys.}\ }\textbf {\bibinfo {volume} {65}},\ \bibinfo
  {pages} {33 } (\bibinfo {year} {1965})}\BibitemShut {NoStop}%
\bibitem [{\citenamefont {Put}\ \emph {et~al.}(1971)\citenamefont {Put},
  \citenamefont {Urone},\ and\ \citenamefont {Paans}}]{95}%
  \BibitemOpen
  \bibfield  {author} {\bibinfo {author} {\bibfnamefont {L.}~\bibnamefont
  {Put}}, \bibinfo {author} {\bibfnamefont {P.}~\bibnamefont {Urone}}, \ and\
  \bibinfo {author} {\bibfnamefont {A.}~\bibnamefont {Paans}},\ }\href
  {\doibase 10.1016/0370-2693(71)90266-8} {\bibfield  {journal} {\bibinfo
  {journal} {Phys. Lett. B}\ }\textbf {\bibinfo {volume} {35}},\ \bibinfo
  {pages} {311} (\bibinfo {year} {1971})}\BibitemShut {NoStop}%
\bibitem [{\citenamefont {Pignanelli}\ \emph {et~al.}(1986)\citenamefont
  {Pignanelli}, \citenamefont {Micheletti}, \citenamefont {{De Leo}},
  \citenamefont {Brandenburg},\ and\ \citenamefont {Harakeh}}]{96}%
  \BibitemOpen
  \bibfield  {author} {\bibinfo {author} {\bibfnamefont {M.}~\bibnamefont
  {Pignanelli}}, \bibinfo {author} {\bibfnamefont {S.}~\bibnamefont
  {Micheletti}}, \bibinfo {author} {\bibfnamefont {R.}~\bibnamefont {{De
  Leo}}}, \bibinfo {author} {\bibfnamefont {S.}~\bibnamefont {Brandenburg}}, \
  and\ \bibinfo {author} {\bibfnamefont {M.~N.}\ \bibnamefont {Harakeh}},\
  }\href {\doibase 10.1103/PhysRevC.33.40} {\bibfield  {journal} {\bibinfo
  {journal} {Phys. Rev. C}\ }\textbf {\bibinfo {volume} {33}},\ \bibinfo
  {pages} {40} (\bibinfo {year} {1986})}\BibitemShut {NoStop}%
\bibitem [{\citenamefont {Fricke}\ \emph {et~al.}(1967)\citenamefont {Fricke},
  \citenamefont {Gross},\ and\ \citenamefont {Zucker}}]{118}%
  \BibitemOpen
  \bibfield  {author} {\bibinfo {author} {\bibfnamefont {M.~P.}\ \bibnamefont
  {Fricke}}, \bibinfo {author} {\bibfnamefont {E.~E.}\ \bibnamefont {Gross}}, \
  and\ \bibinfo {author} {\bibfnamefont {A.}~\bibnamefont {Zucker}},\ }\href
  {\doibase 10.1103/PhysRev.163.1153} {\bibfield  {journal} {\bibinfo
  {journal} {Phys. Rev.}\ }\textbf {\bibinfo {volume} {163}},\ \bibinfo {pages}
  {1153} (\bibinfo {year} {1967})}\BibitemShut {NoStop}%
\bibitem [{\citenamefont {Nakamura}\ \emph {et~al.}(1983)\citenamefont
  {Nakamura}, \citenamefont {Sakaguchi}, \citenamefont {Sakamoto},
  \citenamefont {Ogawa}, \citenamefont {Cynshi}, \citenamefont {Kobayashi},
  \citenamefont {Kato}, \citenamefont {Matsuoka}, \citenamefont {Hatanaka},\
  and\ \citenamefont {Noro}}]{97}%
  \BibitemOpen
  \bibfield  {author} {\bibinfo {author} {\bibfnamefont {M.}~\bibnamefont
  {Nakamura}}, \bibinfo {author} {\bibfnamefont {H.}~\bibnamefont {Sakaguchi}},
  \bibinfo {author} {\bibfnamefont {H.}~\bibnamefont {Sakamoto}}, \bibinfo
  {author} {\bibfnamefont {H.}~\bibnamefont {Ogawa}}, \bibinfo {author}
  {\bibfnamefont {O.}~\bibnamefont {Cynshi}}, \bibinfo {author} {\bibfnamefont
  {S.}~\bibnamefont {Kobayashi}}, \bibinfo {author} {\bibfnamefont
  {S.}~\bibnamefont {Kato}}, \bibinfo {author} {\bibfnamefont {N.}~\bibnamefont
  {Matsuoka}}, \bibinfo {author} {\bibfnamefont {K.}~\bibnamefont {Hatanaka}},
  \ and\ \bibinfo {author} {\bibfnamefont {T.}~\bibnamefont {Noro}},\ }\href
  {\doibase 10.1016/0167-5087(83)90689-0} {\bibfield  {journal} {\bibinfo
  {journal} {Nucl. Instrum. Methods in Phys. Res.}\ }\textbf {\bibinfo {volume}
  {212}},\ \bibinfo {pages} {173} (\bibinfo {year} {1983})}\BibitemShut
  {NoStop}%
\bibitem [{\citenamefont {Toba}\ \emph {et~al.}(1978)\citenamefont {Toba},
  \citenamefont {Sakaguchi}, \citenamefont {Goto}, \citenamefont {Ohtani},
  \citenamefont {Nakanishi}, \citenamefont {Kishida}, \citenamefont {Yasue},\
  and\ \citenamefont {Hasegawa}}]{98}%
  \BibitemOpen
  \bibfield  {author} {\bibinfo {author} {\bibfnamefont {Y.}~\bibnamefont
  {Toba}}, \bibinfo {author} {\bibfnamefont {H.}~\bibnamefont {Sakaguchi}},
  \bibinfo {author} {\bibfnamefont {A.}~\bibnamefont {Goto}}, \bibinfo {author}
  {\bibfnamefont {F.}~\bibnamefont {Ohtani}}, \bibinfo {author} {\bibfnamefont
  {N.}~\bibnamefont {Nakanishi}}, \bibinfo {author} {\bibfnamefont
  {N.}~\bibnamefont {Kishida}}, \bibinfo {author} {\bibfnamefont
  {M.}~\bibnamefont {Yasue}}, \ and\ \bibinfo {author} {\bibfnamefont
  {T.}~\bibnamefont {Hasegawa}},\ }\href {\doibase 10.1143/JPSJ.45.367}
  {\bibfield  {journal} {\bibinfo  {journal} {J. Phys. Soc. Jpn}\ }\textbf
  {\bibinfo {volume} {45}},\ \bibinfo {pages} {367} (\bibinfo {year}
  {1978})}\BibitemShut {NoStop}%
\bibitem [{\citenamefont {Schwandt}\ \emph
  {et~al.}(1982{\natexlab{b}})\citenamefont {Schwandt}, \citenamefont {Meyer},
  \citenamefont {Jacobs}, \citenamefont {Bacher}, \citenamefont {Vigdor},
  \citenamefont {Kaitchuck},\ and\ \citenamefont {Donoghue}}]{99}%
  \BibitemOpen
  \bibfield  {author} {\bibinfo {author} {\bibfnamefont {P.}~\bibnamefont
  {Schwandt}}, \bibinfo {author} {\bibfnamefont {H.~O.}\ \bibnamefont {Meyer}},
  \bibinfo {author} {\bibfnamefont {W.~W.}\ \bibnamefont {Jacobs}}, \bibinfo
  {author} {\bibfnamefont {A.~D.}\ \bibnamefont {Bacher}}, \bibinfo {author}
  {\bibfnamefont {S.~E.}\ \bibnamefont {Vigdor}}, \bibinfo {author}
  {\bibfnamefont {M.~D.}\ \bibnamefont {Kaitchuck}}, \ and\ \bibinfo {author}
  {\bibfnamefont {T.~R.}\ \bibnamefont {Donoghue}},\ }\href {\doibase
  10.1103/PhysRevC.26.55} {\bibfield  {journal} {\bibinfo  {journal} {Phys.
  Rev. C}\ }\textbf {\bibinfo {volume} {26}},\ \bibinfo {pages} {55} (\bibinfo
  {year} {1982}{\natexlab{b}})}\BibitemShut {NoStop}%
\bibitem [{\citenamefont {Chen}\ \emph {et~al.}(1990)\citenamefont {Chen},
  \citenamefont {Kelly}, \citenamefont {Singh}, \citenamefont {Radhakrishna},
  \citenamefont {Jones},\ and\ \citenamefont {Nann}}]{100}%
  \BibitemOpen
  \bibfield  {author} {\bibinfo {author} {\bibfnamefont {Q.}~\bibnamefont
  {Chen}}, \bibinfo {author} {\bibfnamefont {J.~J.}\ \bibnamefont {Kelly}},
  \bibinfo {author} {\bibfnamefont {P.~P.}\ \bibnamefont {Singh}}, \bibinfo
  {author} {\bibfnamefont {M.~C.}\ \bibnamefont {Radhakrishna}}, \bibinfo
  {author} {\bibfnamefont {W.~P.}\ \bibnamefont {Jones}}, \ and\ \bibinfo
  {author} {\bibfnamefont {H.}~\bibnamefont {Nann}},\ }\href {\doibase
  10.1103/PhysRevC.41.2514} {\bibfield  {journal} {\bibinfo  {journal} {Phys.
  Rev. C}\ }\textbf {\bibinfo {volume} {41}},\ \bibinfo {pages} {2514}
  (\bibinfo {year} {1990})}\BibitemShut {NoStop}%
\bibitem [{\citenamefont {Sundberg}\ \emph {et~al.}(1967)\citenamefont
  {Sundberg}, \citenamefont {Johansson}, \citenamefont {Tibell}, \citenamefont
  {Dahlgren}, \citenamefont {Hasselgren}, \citenamefont {H{\"o}istad},
  \citenamefont {Ingemarsson},\ and\ \citenamefont {Renberg}}]{101}%
  \BibitemOpen
  \bibfield  {author} {\bibinfo {author} {\bibfnamefont {O.}~\bibnamefont
  {Sundberg}}, \bibinfo {author} {\bibfnamefont {A.}~\bibnamefont {Johansson}},
  \bibinfo {author} {\bibfnamefont {G.}~\bibnamefont {Tibell}}, \bibinfo
  {author} {\bibfnamefont {S.}~\bibnamefont {Dahlgren}}, \bibinfo {author}
  {\bibfnamefont {D.}~\bibnamefont {Hasselgren}}, \bibinfo {author}
  {\bibfnamefont {B.}~\bibnamefont {H{\"o}istad}}, \bibinfo {author}
  {\bibfnamefont {A.}~\bibnamefont {Ingemarsson}}, \ and\ \bibinfo {author}
  {\bibfnamefont {P.-U.}\ \bibnamefont {Renberg}},\ }\href {\doibase
  10.1016/0375-9474(67)90647-1} {\bibfield  {journal} {\bibinfo  {journal}
  {Nucl. Phys. A}\ }\textbf {\bibinfo {volume} {101}},\ \bibinfo {pages} {481}
  (\bibinfo {year} {1967})}\BibitemShut {NoStop}%
\bibitem [{\citenamefont {Pellegrini}\ \emph {et~al.}(1978)\citenamefont
  {Pellegrini}, \citenamefont {Calvelli}, \citenamefont {Guazzoni},\ and\
  \citenamefont {Micheletti}}]{102}%
  \BibitemOpen
  \bibfield  {author} {\bibinfo {author} {\bibfnamefont {F.}~\bibnamefont
  {Pellegrini}}, \bibinfo {author} {\bibfnamefont {G.}~\bibnamefont
  {Calvelli}}, \bibinfo {author} {\bibfnamefont {P.}~\bibnamefont {Guazzoni}},
  \ and\ \bibinfo {author} {\bibfnamefont {S.}~\bibnamefont {Micheletti}},\
  }\href {\doibase 10.1103/PhysRevC.18.613} {\bibfield  {journal} {\bibinfo
  {journal} {Phys. Rev. C}\ }\textbf {\bibinfo {volume} {18}},\ \bibinfo
  {pages} {613} (\bibinfo {year} {1978})}\BibitemShut {NoStop}%
\bibitem [{\citenamefont {Kelly}\ \emph {et~al.}(1990)\citenamefont {Kelly},
  \citenamefont {Chen}, \citenamefont {Singh}, \citenamefont {Radhakrishna},
  \citenamefont {Jones},\ and\ \citenamefont {Nann}}]{103}%
  \BibitemOpen
  \bibfield  {author} {\bibinfo {author} {\bibfnamefont {J.~J.}\ \bibnamefont
  {Kelly}}, \bibinfo {author} {\bibfnamefont {Q.}~\bibnamefont {Chen}},
  \bibinfo {author} {\bibfnamefont {P.~P.}\ \bibnamefont {Singh}}, \bibinfo
  {author} {\bibfnamefont {M.~C.}\ \bibnamefont {Radhakrishna}}, \bibinfo
  {author} {\bibfnamefont {W.~P.}\ \bibnamefont {Jones}}, \ and\ \bibinfo
  {author} {\bibfnamefont {H.}~\bibnamefont {Nann}},\ }\href {\doibase
  10.1103/PhysRevC.41.2525} {\bibfield  {journal} {\bibinfo  {journal} {Phys.
  Rev. C}\ }\textbf {\bibinfo {volume} {41}},\ \bibinfo {pages} {2525}
  (\bibinfo {year} {1990})}\BibitemShut {NoStop}%
\bibitem [{\citenamefont {Boschung}\ \emph {et~al.}(1971)\citenamefont
  {Boschung}, \citenamefont {Lindow},\ and\ \citenamefont {Shrader}}]{70}%
  \BibitemOpen
  \bibfield  {author} {\bibinfo {author} {\bibfnamefont {P.}~\bibnamefont
  {Boschung}}, \bibinfo {author} {\bibfnamefont {J.}~\bibnamefont {Lindow}}, \
  and\ \bibinfo {author} {\bibfnamefont {E.}~\bibnamefont {Shrader}},\ }\href
  {\doibase https://doi.org/10.1016/0375-9474(71)90388-5} {\bibfield  {journal}
  {\bibinfo  {journal} {Nucl. Phys. A}\ }\textbf {\bibinfo {volume} {161}},\
  \bibinfo {pages} {593 } (\bibinfo {year} {1971})}\BibitemShut {NoStop}%
\bibitem [{\citenamefont {El-Kadi}\ \emph {et~al.}(1982)\citenamefont
  {El-Kadi}, \citenamefont {Nelson}, \citenamefont {Purser}, \citenamefont
  {Walter}, \citenamefont {Beyerle}, \citenamefont {Gould},\ and\ \citenamefont
  {Seagondollar}}]{71}%
  \BibitemOpen
  \bibfield  {author} {\bibinfo {author} {\bibfnamefont {S.}~\bibnamefont
  {El-Kadi}}, \bibinfo {author} {\bibfnamefont {C.}~\bibnamefont {Nelson}},
  \bibinfo {author} {\bibfnamefont {F.}~\bibnamefont {Purser}}, \bibinfo
  {author} {\bibfnamefont {R.}~\bibnamefont {Walter}}, \bibinfo {author}
  {\bibfnamefont {A.}~\bibnamefont {Beyerle}}, \bibinfo {author} {\bibfnamefont
  {C.}~\bibnamefont {Gould}}, \ and\ \bibinfo {author} {\bibfnamefont
  {L.}~\bibnamefont {Seagondollar}},\ }\href {\doibase
  https://doi.org/10.1016/0375-9474(82)90281-0} {\bibfield  {journal} {\bibinfo
   {journal} {Nucl. Phys. A}\ }\textbf {\bibinfo {volume} {390}},\ \bibinfo
  {pages} {509 } (\bibinfo {year} {1982})}\BibitemShut {NoStop}%
\bibitem [{\citenamefont {Mellema}\ \emph {et~al.}(1983)\citenamefont
  {Mellema}, \citenamefont {Finlay}, \citenamefont {Dietrich},\ and\
  \citenamefont {Petrovich}}]{72}%
  \BibitemOpen
  \bibfield  {author} {\bibinfo {author} {\bibfnamefont {S.}~\bibnamefont
  {Mellema}}, \bibinfo {author} {\bibfnamefont {R.~W.}\ \bibnamefont {Finlay}},
  \bibinfo {author} {\bibfnamefont {F.~S.}\ \bibnamefont {Dietrich}}, \ and\
  \bibinfo {author} {\bibfnamefont {F.}~\bibnamefont {Petrovich}},\ }\href
  {\doibase 10.1103/PhysRevC.28.2267} {\bibfield  {journal} {\bibinfo
  {journal} {Phys. Rev. C}\ }\textbf {\bibinfo {volume} {28}},\ \bibinfo
  {pages} {2267} (\bibinfo {year} {1983})}\BibitemShut {NoStop}%
\bibitem [{\citenamefont {Tutubalin}\ and\ \citenamefont {al}(1973)}]{73}%
  \BibitemOpen
  \bibfield  {author} {\bibinfo {author} {\bibfnamefont {A.~I.}\ \bibnamefont
  {Tutubalin}}\ and\ \bibinfo {author} {\bibnamefont {al}},\ }\href@noop {}
  {\bibfield  {journal} {\bibinfo  {journal} {2nd Conference on Neutron
  Physics, Kiev.}\ }\textbf {\bibinfo {volume} {3}},\ \bibinfo {pages} {62}
  (\bibinfo {year} {1973})}\BibitemShut {NoStop}%
\bibitem [{\citenamefont {Pedroni}\ \emph {et~al.}(1988)\citenamefont
  {Pedroni}, \citenamefont {Howell}, \citenamefont {Honor\'e}, \citenamefont
  {Pfutzner}, \citenamefont {Byrd}, \citenamefont {Walter},\ and\ \citenamefont
  {Delaroche}}]{74}%
  \BibitemOpen
  \bibfield  {author} {\bibinfo {author} {\bibfnamefont {R.~S.}\ \bibnamefont
  {Pedroni}}, \bibinfo {author} {\bibfnamefont {C.~R.}\ \bibnamefont {Howell}},
  \bibinfo {author} {\bibfnamefont {G.~M.}\ \bibnamefont {Honor\'e}}, \bibinfo
  {author} {\bibfnamefont {H.~G.}\ \bibnamefont {Pfutzner}}, \bibinfo {author}
  {\bibfnamefont {R.~C.}\ \bibnamefont {Byrd}}, \bibinfo {author}
  {\bibfnamefont {R.~L.}\ \bibnamefont {Walter}}, \ and\ \bibinfo {author}
  {\bibfnamefont {J.~P.}\ \bibnamefont {Delaroche}},\ }\href {\doibase
  10.1103/PhysRevC.38.2052} {\bibfield  {journal} {\bibinfo  {journal} {Phys.
  Rev. C}\ }\textbf {\bibinfo {volume} {38}},\ \bibinfo {pages} {2052}
  (\bibinfo {year} {1988})}\BibitemShut {NoStop}%
\bibitem [{\citenamefont {Ramirez}\ \emph {et~al.}(2017)\citenamefont
  {Ramirez}, \citenamefont {Vanhoy}, \citenamefont {Hicks}, \citenamefont
  {McEllistrem}, \citenamefont {Peters}, \citenamefont {Mukhopadhyay},
  \citenamefont {Harrison}, \citenamefont {Howard}, \citenamefont {Jackson},
  \citenamefont {Lenzen} \emph {et~al.}}]{75}%
  \BibitemOpen
  \bibfield  {author} {\bibinfo {author} {\bibfnamefont {A.~P.~D.}\
  \bibnamefont {Ramirez}}, \bibinfo {author} {\bibfnamefont {J.~R.}\
  \bibnamefont {Vanhoy}}, \bibinfo {author} {\bibfnamefont {S.~F.}\
  \bibnamefont {Hicks}}, \bibinfo {author} {\bibfnamefont {M.~T.}\ \bibnamefont
  {McEllistrem}}, \bibinfo {author} {\bibfnamefont {E.~E.}\ \bibnamefont
  {Peters}}, \bibinfo {author} {\bibfnamefont {S.}~\bibnamefont
  {Mukhopadhyay}}, \bibinfo {author} {\bibfnamefont {T.~D.}\ \bibnamefont
  {Harrison}}, \bibinfo {author} {\bibfnamefont {T.~J.}\ \bibnamefont
  {Howard}}, \bibinfo {author} {\bibfnamefont {D.~T.}\ \bibnamefont {Jackson}},
  \bibinfo {author} {\bibfnamefont {P.~D.}\ \bibnamefont {Lenzen}},  \emph
  {et~al.},\ }\href {\doibase 10.1103/PhysRevC.95.064605} {\bibfield  {journal}
  {\bibinfo  {journal} {Phys. Rev. C}\ }\textbf {\bibinfo {volume} {95}},\
  \bibinfo {pages} {064605} (\bibinfo {year} {2017})}\BibitemShut {NoStop}%
\bibitem [{\citenamefont {\"Ohrn}\ \emph {et~al.}(2008)\citenamefont {\"Ohrn},
  \citenamefont {Blomgren}, \citenamefont {Andersson}, \citenamefont
  {Ata\ifmmode~\mbox{\c{c}}\else \c{c}\fi{}}, \citenamefont {Gustavsson},
  \citenamefont {Klug}, \citenamefont {Mermod}, \citenamefont {Pomp},
  \citenamefont {Wolniewicz}, \citenamefont {\"Osterlund} \emph {et~al.}}]{77}%
  \BibitemOpen
  \bibfield  {author} {\bibinfo {author} {\bibfnamefont {A.}~\bibnamefont
  {\"Ohrn}}, \bibinfo {author} {\bibfnamefont {J.}~\bibnamefont {Blomgren}},
  \bibinfo {author} {\bibfnamefont {P.}~\bibnamefont {Andersson}}, \bibinfo
  {author} {\bibfnamefont {A.}~\bibnamefont {Ata\ifmmode~\mbox{\c{c}}\else
  \c{c}\fi{}}}, \bibinfo {author} {\bibfnamefont {C.}~\bibnamefont
  {Gustavsson}}, \bibinfo {author} {\bibfnamefont {J.}~\bibnamefont {Klug}},
  \bibinfo {author} {\bibfnamefont {P.}~\bibnamefont {Mermod}}, \bibinfo
  {author} {\bibfnamefont {S.}~\bibnamefont {Pomp}}, \bibinfo {author}
  {\bibfnamefont {P.}~\bibnamefont {Wolniewicz}}, \bibinfo {author}
  {\bibfnamefont {M.}~\bibnamefont {\"Osterlund}},  \emph {et~al.},\ }\href
  {\doibase 10.1103/PhysRevC.77.024605} {\bibfield  {journal} {\bibinfo
  {journal} {Phys. Rev. C}\ }\textbf {\bibinfo {volume} {77}},\ \bibinfo
  {pages} {024605} (\bibinfo {year} {2008})}\BibitemShut {NoStop}%
\bibitem [{\citenamefont {Boukharouba}\ \emph {et~al.}(1992)\citenamefont
  {Boukharouba}, \citenamefont {Brient}, \citenamefont {Grimes}, \citenamefont
  {Mishra},\ and\ \citenamefont {Pedroni}}]{104}%
  \BibitemOpen
  \bibfield  {author} {\bibinfo {author} {\bibfnamefont {N.}~\bibnamefont
  {Boukharouba}}, \bibinfo {author} {\bibfnamefont {C.~E.}\ \bibnamefont
  {Brient}}, \bibinfo {author} {\bibfnamefont {S.~M.}\ \bibnamefont {Grimes}},
  \bibinfo {author} {\bibfnamefont {V.}~\bibnamefont {Mishra}}, \ and\ \bibinfo
  {author} {\bibfnamefont {R.~S.}\ \bibnamefont {Pedroni}},\ }\href {\doibase
  10.1103/PhysRevC.46.2375} {\bibfield  {journal} {\bibinfo  {journal} {Phys.
  Rev. C}\ }\textbf {\bibinfo {volume} {46}},\ \bibinfo {pages} {2375}
  (\bibinfo {year} {1992})}\BibitemShut {NoStop}%
\bibitem [{\citenamefont {Greenlees}\ \emph {et~al.}(1971)\citenamefont
  {Greenlees}, \citenamefont {Poppe}, \citenamefont {Sievers},\ and\
  \citenamefont {Watson}}]{105}%
  \BibitemOpen
  \bibfield  {author} {\bibinfo {author} {\bibfnamefont {G.~W.}\ \bibnamefont
  {Greenlees}}, \bibinfo {author} {\bibfnamefont {C.~H.}\ \bibnamefont
  {Poppe}}, \bibinfo {author} {\bibfnamefont {J.~A.}\ \bibnamefont {Sievers}},
  \ and\ \bibinfo {author} {\bibfnamefont {D.~L.}\ \bibnamefont {Watson}},\
  }\href {\doibase 10.1103/PhysRevC.3.1231} {\bibfield  {journal} {\bibinfo
  {journal} {Phys. Rev. C}\ }\textbf {\bibinfo {volume} {3}},\ \bibinfo {pages}
  {1231} (\bibinfo {year} {1971})}\BibitemShut {NoStop}%
\bibitem [{\citenamefont {Ahmed}\ \emph {et~al.}(1970)\citenamefont {Ahmed},
  \citenamefont {Lowe}, \citenamefont {Rolph},\ and\ \citenamefont
  {Karban}}]{106}%
  \BibitemOpen
  \bibfield  {author} {\bibinfo {author} {\bibfnamefont {M.}~\bibnamefont
  {Ahmed}}, \bibinfo {author} {\bibfnamefont {J.}~\bibnamefont {Lowe}},
  \bibinfo {author} {\bibfnamefont {P.}~\bibnamefont {Rolph}}, \ and\ \bibinfo
  {author} {\bibfnamefont {O.}~\bibnamefont {Karban}},\ }\href {\doibase
  https://doi.org/10.1016/0375-9474(70)90267-8} {\bibfield  {journal} {\bibinfo
   {journal} {Nucl. Phys. A}\ }\textbf {\bibinfo {volume} {147}},\ \bibinfo
  {pages} {273 } (\bibinfo {year} {1970})}\BibitemShut {NoStop}%
\bibitem [{\citenamefont {Lombardi}\ \emph {et~al.}(1972)\citenamefont
  {Lombardi}, \citenamefont {Boyd}, \citenamefont {Arking},\ and\ \citenamefont
  {Robbins}}]{107}%
  \BibitemOpen
  \bibfield  {author} {\bibinfo {author} {\bibfnamefont {J.}~\bibnamefont
  {Lombardi}}, \bibinfo {author} {\bibfnamefont {R.}~\bibnamefont {Boyd}},
  \bibinfo {author} {\bibfnamefont {R.}~\bibnamefont {Arking}}, \ and\ \bibinfo
  {author} {\bibfnamefont {A.}~\bibnamefont {Robbins}},\ }\href {\doibase
  https://doi.org/10.1016/0375-9474(72)90095-4} {\bibfield  {journal} {\bibinfo
   {journal} {Nucl. Phys. A}\ }\textbf {\bibinfo {volume} {192}},\ \bibinfo
  {pages} {641 } (\bibinfo {year} {1972})}\BibitemShut {NoStop}%
\bibitem [{\citenamefont {Perey}\ \emph {et~al.}(1968)\citenamefont {Perey},
  \citenamefont {Perey}, \citenamefont {Dickens},\ and\ \citenamefont
  {Silva}}]{108}%
  \BibitemOpen
  \bibfield  {author} {\bibinfo {author} {\bibfnamefont {C.~M.}\ \bibnamefont
  {Perey}}, \bibinfo {author} {\bibfnamefont {F.~G.}\ \bibnamefont {Perey}},
  \bibinfo {author} {\bibfnamefont {J.~K.}\ \bibnamefont {Dickens}}, \ and\
  \bibinfo {author} {\bibfnamefont {R.~J.}\ \bibnamefont {Silva}},\ }\href
  {\doibase 10.1103/PhysRev.175.1460} {\bibfield  {journal} {\bibinfo
  {journal} {Phys. Rev.}\ }\textbf {\bibinfo {volume} {175}},\ \bibinfo {pages}
  {1460} (\bibinfo {year} {1968})}\BibitemShut {NoStop}%
\bibitem [{\citenamefont {Lee}\ and\ \citenamefont {Schiffer}(1964)}]{109}%
  \BibitemOpen
  \bibfield  {author} {\bibinfo {author} {\bibfnamefont {L.~L.}\ \bibnamefont
  {Lee}}\ and\ \bibinfo {author} {\bibfnamefont {J.~P.}\ \bibnamefont
  {Schiffer}},\ }\href {\doibase 10.1103/PhysRev.134.B765} {\bibfield
  {journal} {\bibinfo  {journal} {Phys. Rev.}\ }\textbf {\bibinfo {volume}
  {134}},\ \bibinfo {pages} {B765} (\bibinfo {year} {1964})}\BibitemShut
  {NoStop}%
\bibitem [{\citenamefont {Hall}\ \emph {et~al.}(1977)\citenamefont {Hall},
  \citenamefont {Melssen}, \citenamefont {Wassenaar}, \citenamefont {Poppema},
  \citenamefont {Klein},\ and\ \citenamefont {Nijgh}}]{110}%
  \BibitemOpen
  \bibfield  {author} {\bibinfo {author} {\bibfnamefont {P.~V.}\ \bibnamefont
  {Hall}}, \bibinfo {author} {\bibfnamefont {J.}~\bibnamefont {Melssen}},
  \bibinfo {author} {\bibfnamefont {S.}~\bibnamefont {Wassenaar}}, \bibinfo
  {author} {\bibfnamefont {O.}~\bibnamefont {Poppema}}, \bibinfo {author}
  {\bibfnamefont {S.}~\bibnamefont {Klein}}, \ and\ \bibinfo {author}
  {\bibfnamefont {G.}~\bibnamefont {Nijgh}},\ }\href {\doibase
  https://doi.org/10.1016/0375-9474(77)90199-3} {\bibfield  {journal} {\bibinfo
   {journal} {Nucl. Phys. A}\ }\textbf {\bibinfo {volume} {291}},\ \bibinfo
  {pages} {63 } (\bibinfo {year} {1977})}\BibitemShut {NoStop}%
\bibitem [{\citenamefont {Gray}\ \emph {et~al.}(1965)\citenamefont {Gray},
  \citenamefont {Kenefick},\ and\ \citenamefont {Kraushaar}}]{111}%
  \BibitemOpen
  \bibfield  {author} {\bibinfo {author} {\bibfnamefont {W.}~\bibnamefont
  {Gray}}, \bibinfo {author} {\bibfnamefont {R.}~\bibnamefont {Kenefick}}, \
  and\ \bibinfo {author} {\bibfnamefont {J.}~\bibnamefont {Kraushaar}},\ }\href
  {\doibase https://doi.org/10.1016/0029-5582(65)90547-X} {\bibfield  {journal}
  {\bibinfo  {journal} {Nucl. Phys.}\ }\textbf {\bibinfo {volume} {67}},\
  \bibinfo {pages} {565 } (\bibinfo {year} {1965})}\BibitemShut {NoStop}%
\bibitem [{\citenamefont {Eccles}\ \emph {et~al.}(1966)\citenamefont {Eccles},
  \citenamefont {Lutz},\ and\ \citenamefont {Madsen}}]{112}%
  \BibitemOpen
  \bibfield  {author} {\bibinfo {author} {\bibfnamefont {S.~F.}\ \bibnamefont
  {Eccles}}, \bibinfo {author} {\bibfnamefont {H.~F.}\ \bibnamefont {Lutz}}, \
  and\ \bibinfo {author} {\bibfnamefont {V.~A.}\ \bibnamefont {Madsen}},\
  }\href {\doibase 10.1103/PhysRev.141.1067} {\bibfield  {journal} {\bibinfo
  {journal} {Phys. Rev.}\ }\textbf {\bibinfo {volume} {141}},\ \bibinfo {pages}
  {1067} (\bibinfo {year} {1966})}\BibitemShut {NoStop}%
\bibitem [{\citenamefont {Tesmer}\ and\ \citenamefont {Schmidt}(1972)}]{113}%
  \BibitemOpen
  \bibfield  {author} {\bibinfo {author} {\bibfnamefont {J.~R.}\ \bibnamefont
  {Tesmer}}\ and\ \bibinfo {author} {\bibfnamefont {F.~H.}\ \bibnamefont
  {Schmidt}},\ }\href {\doibase 10.1103/PhysRevC.5.864} {\bibfield  {journal}
  {\bibinfo  {journal} {Phys. Rev. C}\ }\textbf {\bibinfo {volume} {5}},\
  \bibinfo {pages} {864} (\bibinfo {year} {1972})}\BibitemShut {NoStop}%
\bibitem [{\citenamefont {Bertrand}\ and\ \citenamefont {Peelle}(1973)}]{114}%
  \BibitemOpen
  \bibfield  {author} {\bibinfo {author} {\bibfnamefont {F.~E.}\ \bibnamefont
  {Bertrand}}\ and\ \bibinfo {author} {\bibfnamefont {R.~W.}\ \bibnamefont
  {Peelle}},\ }\href {\doibase 10.1103/PhysRevC.8.1045} {\bibfield  {journal}
  {\bibinfo  {journal} {Phys. Rev. C}\ }\textbf {\bibinfo {volume} {8}},\
  \bibinfo {pages} {1045} (\bibinfo {year} {1973})}\BibitemShut {NoStop}%
\bibitem [{\citenamefont {Greaves}\ \emph {et~al.}(1972)\citenamefont
  {Greaves}, \citenamefont {Hnizdo}, \citenamefont {Lowe},\ and\ \citenamefont
  {Karban}}]{115}%
  \BibitemOpen
  \bibfield  {author} {\bibinfo {author} {\bibfnamefont {P.}~\bibnamefont
  {Greaves}}, \bibinfo {author} {\bibfnamefont {V.}~\bibnamefont {Hnizdo}},
  \bibinfo {author} {\bibfnamefont {J.}~\bibnamefont {Lowe}}, \ and\ \bibinfo
  {author} {\bibfnamefont {O.}~\bibnamefont {Karban}},\ }\href {\doibase
  https://doi.org/10.1016/0375-9474(72)90101-7} {\bibfield  {journal} {\bibinfo
   {journal} {Nucl. Phys. A}\ }\textbf {\bibinfo {volume} {179}},\ \bibinfo
  {pages} {1 } (\bibinfo {year} {1972})}\BibitemShut {NoStop}%
\bibitem [{\citenamefont {Stovall}\ and\ \citenamefont {Hintz}(1964)}]{116}%
  \BibitemOpen
  \bibfield  {author} {\bibinfo {author} {\bibfnamefont {T.}~\bibnamefont
  {Stovall}}\ and\ \bibinfo {author} {\bibfnamefont {N.~M.}\ \bibnamefont
  {Hintz}},\ }\href {\doibase 10.1103/PhysRev.135.B330} {\bibfield  {journal}
  {\bibinfo  {journal} {Phys. Rev.}\ }\textbf {\bibinfo {volume} {135}},\
  \bibinfo {pages} {B330} (\bibinfo {year} {1964})}\BibitemShut {NoStop}%
\bibitem [{\citenamefont {Brussel}\ and\ \citenamefont {Williams}(1959)}]{117}%
  \BibitemOpen
  \bibfield  {author} {\bibinfo {author} {\bibfnamefont {M.~K.}\ \bibnamefont
  {Brussel}}\ and\ \bibinfo {author} {\bibfnamefont {J.~H.}\ \bibnamefont
  {Williams}},\ }\href {\doibase 10.1103/PhysRev.114.525} {\bibfield  {journal}
  {\bibinfo  {journal} {Phys. Rev.}\ }\textbf {\bibinfo {volume} {114}},\
  \bibinfo {pages} {525} (\bibinfo {year} {1959})}\BibitemShut {NoStop}%
\bibitem [{\citenamefont {Noro}\ \emph {et~al.}(1981)\citenamefont {Noro},
  \citenamefont {Sakaguchi}, \citenamefont {Nakamura}, \citenamefont
  {Hatanaka}, \citenamefont {Ohtani}, \citenamefont {Sakamoto},\ and\
  \citenamefont {Kobayashi}}]{119}%
  \BibitemOpen
  \bibfield  {author} {\bibinfo {author} {\bibfnamefont {T.}~\bibnamefont
  {Noro}}, \bibinfo {author} {\bibfnamefont {H.}~\bibnamefont {Sakaguchi}},
  \bibinfo {author} {\bibfnamefont {M.}~\bibnamefont {Nakamura}}, \bibinfo
  {author} {\bibfnamefont {K.}~\bibnamefont {Hatanaka}}, \bibinfo {author}
  {\bibfnamefont {F.}~\bibnamefont {Ohtani}}, \bibinfo {author} {\bibfnamefont
  {H.}~\bibnamefont {Sakamoto}}, \ and\ \bibinfo {author} {\bibfnamefont
  {S.}~\bibnamefont {Kobayashi}},\ }\href {\doibase
  https://doi.org/10.1016/0375-9474(81)90283-9} {\bibfield  {journal} {\bibinfo
   {journal} {Nucl. Phys. A}\ }\textbf {\bibinfo {volume} {366}},\ \bibinfo
  {pages} {189 } (\bibinfo {year} {1981})}\BibitemShut {NoStop}%
\bibitem [{\citenamefont {Funsten}\ \emph {et~al.}(1964)\citenamefont
  {Funsten}, \citenamefont {Roberson},\ and\ \citenamefont {Rost}}]{120}%
  \BibitemOpen
  \bibfield  {author} {\bibinfo {author} {\bibfnamefont {H.~O.}\ \bibnamefont
  {Funsten}}, \bibinfo {author} {\bibfnamefont {N.~R.}\ \bibnamefont
  {Roberson}}, \ and\ \bibinfo {author} {\bibfnamefont {E.}~\bibnamefont
  {Rost}},\ }\href {\doibase 10.1103/PhysRev.134.B117} {\bibfield  {journal}
  {\bibinfo  {journal} {Phys. Rev.}\ }\textbf {\bibinfo {volume} {134}},\
  \bibinfo {pages} {B117} (\bibinfo {year} {1964})}\BibitemShut {NoStop}%
\bibitem [{\citenamefont {Kossanyi-Demay}\ \emph {et~al.}(1967)\citenamefont
  {Kossanyi-Demay}, \citenamefont {de~Swiniarski},\ and\ \citenamefont
  {Glashausser}}]{121}%
  \BibitemOpen
  \bibfield  {author} {\bibinfo {author} {\bibfnamefont {P.}~\bibnamefont
  {Kossanyi-Demay}}, \bibinfo {author} {\bibfnamefont {R.}~\bibnamefont
  {de~Swiniarski}}, \ and\ \bibinfo {author} {\bibfnamefont {C.}~\bibnamefont
  {Glashausser}},\ }\href {\doibase
  https://doi.org/10.1016/0375-9474(67)90428-9} {\bibfield  {journal} {\bibinfo
   {journal} {Nucl. Phys. A}\ }\textbf {\bibinfo {volume} {94}},\ \bibinfo
  {pages} {513 } (\bibinfo {year} {1967})}\BibitemShut {NoStop}%
\bibitem [{\citenamefont {Kumar}\ \emph {et~al.}(2001)\citenamefont {Kumar},
  \citenamefont {Avasthi}, \citenamefont {Tripathi}, \citenamefont {Datta},\
  and\ \citenamefont {Govil}}]{122}%
  \BibitemOpen
  \bibfield  {author} {\bibinfo {author} {\bibfnamefont {A.}~\bibnamefont
  {Kumar}}, \bibinfo {author} {\bibfnamefont {D.~K.}\ \bibnamefont {Avasthi}},
  \bibinfo {author} {\bibfnamefont {A.}~\bibnamefont {Tripathi}}, \bibinfo
  {author} {\bibfnamefont {S.~K.}\ \bibnamefont {Datta}}, \ and\ \bibinfo
  {author} {\bibfnamefont {I.~M.}\ \bibnamefont {Govil}},\ }\href {\doibase
  10.1103/PhysRevC.65.014305} {\bibfield  {journal} {\bibinfo  {journal} {Phys.
  Rev. C}\ }\textbf {\bibinfo {volume} {65}},\ \bibinfo {pages} {014305}
  (\bibinfo {year} {2001})}\BibitemShut {NoStop}%
\bibitem [{\citenamefont {Ridley}\ and\ \citenamefont {Turner}(1964)}]{123}%
  \BibitemOpen
  \bibfield  {author} {\bibinfo {author} {\bibfnamefont {B.}~\bibnamefont
  {Ridley}}\ and\ \bibinfo {author} {\bibfnamefont {J.}~\bibnamefont
  {Turner}},\ }\href {\doibase https://doi.org/10.1016/0029-5582(64)90561-9}
  {\bibfield  {journal} {\bibinfo  {journal} {Nucl. Phys.}\ }\textbf {\bibinfo
  {volume} {58}},\ \bibinfo {pages} {497 } (\bibinfo {year}
  {1964})}\BibitemShut {NoStop}%
\bibitem [{\citenamefont {Mani}(1971)}]{124}%
  \BibitemOpen
  \bibfield  {author} {\bibinfo {author} {\bibfnamefont {G.}~\bibnamefont
  {Mani}},\ }\href {\doibase https://doi.org/10.1016/0375-9474(71)90758-5}
  {\bibfield  {journal} {\bibinfo  {journal} {Nucl. Phys. A}\ }\textbf
  {\bibinfo {volume} {165}},\ \bibinfo {pages} {225 } (\bibinfo {year}
  {1971})}\BibitemShut {NoStop}%
\bibitem [{\citenamefont {{De Leo}}\ \emph {et~al.}(1996)\citenamefont {{De
  Leo}}, \citenamefont {Akimune}, \citenamefont {Blasi}, \citenamefont {Daito},
  \citenamefont {Fujita}, \citenamefont {Fujiwara}, \citenamefont {Hayakawa},
  \citenamefont {Hatori}, \citenamefont {Hosono}, \citenamefont {Ikegami} \emph
  {et~al.}}]{125}%
  \BibitemOpen
  \bibfield  {author} {\bibinfo {author} {\bibfnamefont {R.}~\bibnamefont {{De
  Leo}}}, \bibinfo {author} {\bibfnamefont {H.}~\bibnamefont {Akimune}},
  \bibinfo {author} {\bibfnamefont {N.}~\bibnamefont {Blasi}}, \bibinfo
  {author} {\bibfnamefont {I.}~\bibnamefont {Daito}}, \bibinfo {author}
  {\bibfnamefont {Y.}~\bibnamefont {Fujita}}, \bibinfo {author} {\bibfnamefont
  {M.}~\bibnamefont {Fujiwara}}, \bibinfo {author} {\bibfnamefont {S.~I.}\
  \bibnamefont {Hayakawa}}, \bibinfo {author} {\bibfnamefont {S.}~\bibnamefont
  {Hatori}}, \bibinfo {author} {\bibfnamefont {K.}~\bibnamefont {Hosono}},
  \bibinfo {author} {\bibfnamefont {H.}~\bibnamefont {Ikegami}},  \emph
  {et~al.},\ }\href {\doibase 10.1103/PhysRevC.53.2718} {\bibfield  {journal}
  {\bibinfo  {journal} {Phys. Rev. C}\ }\textbf {\bibinfo {volume} {53}},\
  \bibinfo {pages} {2718} (\bibinfo {year} {1996})}\BibitemShut {NoStop}%
\bibitem [{\citenamefont {Takamatsu}(1990)}]{126}%
  \BibitemOpen
  \bibfield  {author} {\bibinfo {author} {\bibfnamefont {J.}~\bibnamefont
  {Takamatsu}},\ }\href@noop {} {\bibfield  {journal} {\bibinfo  {journal} {J.
  Phys. (Paris)}\ }\textbf {\bibinfo {volume} {55}},\ \bibinfo {pages} {423}
  (\bibinfo {year} {1990})}\BibitemShut {NoStop}%
\bibitem [{\citenamefont {Comparat}\ \emph {et~al.}(1974)\citenamefont
  {Comparat}, \citenamefont {Frascaria}, \citenamefont {Marty}, \citenamefont
  {Morlet},\ and\ \citenamefont {Willis}}]{127}%
  \BibitemOpen
  \bibfield  {author} {\bibinfo {author} {\bibfnamefont {V.}~\bibnamefont
  {Comparat}}, \bibinfo {author} {\bibfnamefont {R.}~\bibnamefont {Frascaria}},
  \bibinfo {author} {\bibfnamefont {N.}~\bibnamefont {Marty}}, \bibinfo
  {author} {\bibfnamefont {M.}~\bibnamefont {Morlet}}, \ and\ \bibinfo {author}
  {\bibfnamefont {A.}~\bibnamefont {Willis}},\ }\href {\doibase
  https://doi.org/10.1016/0375-9474(74)90327-3} {\bibfield  {journal} {\bibinfo
   {journal} {Nucl. Phys. A}\ }\textbf {\bibinfo {volume} {221}},\ \bibinfo
  {pages} {403 } (\bibinfo {year} {1974})}\BibitemShut {NoStop}%
\bibitem [{\citenamefont {{Experimental Nuclear Reaction Data
  (EXFOR)}}()}]{57}%
  \BibitemOpen
  \bibfield  {author} {\bibinfo {author} {\bibnamefont {{Experimental Nuclear
  Reaction Data (EXFOR)}}},\ }\href {\doibase
  https://www-nds.iaea.org/exfor/exfor.htm} {\
  https://www-nds.iaea.org/exfor/exfor.htm}\BibitemShut {NoStop}%
\end{thebibliography}%

\end{document}